\documentclass[a4paper,12pt]{article}
\usepackage{amsfonts,amsmath,amssymb}
\usepackage{graphicx}
\usepackage{color,here}
\usepackage{cite}
\usepackage{hyperref}
\usepackage{finite_corr}
\usepackage{simplewick}
\usepackage{enumitem}
\usepackage{mathrsfs}
\usepackage[vcentermath]{youngtab}
\usepackage{tocloft}

\usepackage{pgf}
\usepackage{ytableau,varwidth}
\definecolor{xgray}{rgb}{0.75, 0.75, 0.75}

\usepackage{xparse}
\usepackage{tikz}
\usetikzlibrary{arrows,arrows.meta,backgrounds,calc,cd,decorations.markings,decorations.pathmorphing,decorations.shapes,external,graphs,math,matrix,positioning,shapes.geometric}

\tikzset{
perm/.style={trapezium, trapezium angle=90, minimum size=3mm, thick, draw=black},
perm2/.style={trapezium, trapezium angle=90, shape border rotate=180, minimum size=3mm, thick, draw=black},
midperm/.style={trapezium, shape 
border uses incircle, trapezium angle=90, minimum size=15mm, thick, draw=black},
bigperm/.style={trapezium, shape 
border uses incircle, trapezium angle=90, minimum size=3mm, thick, draw=black},
branch/.style={circle, thick, draw=black},
branch2/.style={circle, thick, draw=black, fill=black!10},
branch3/.style={circle, thick, draw=black, fill=black!30},
wavy/.style={thick, decorate, decoration={coil, aspect=0, segment length=2mm, amplitude=.5mm}},
highwavy/.style={thick, decorate, decoration={coil, aspect=0, segment length=1mm, amplitude=.5mm}},
ziggy/.style={thick, decorate, decoration={zigzag, aspect=0, segment length=2mm, amplitude=1mm}},
crossing/.style={thick, decorate, decoration={crosses, segment length=2mm}},
triangling/.style={thick, decorate, decoration={triangles, segment length=2mm}},
lineR/.style={thick, double distance=1pt, postaction={decorate}},
lineS/.style={thick, double=black!50, double distance=1.2pt, postaction={decorate}},
lineT/.style={thick, double=black, double distance=1.2pt, postaction={decorate}},
liner1/.style={wavy, -{Straight Barb[]}},
liner2/.style={thick, -{Straight Barb[]}},
liner3/.style={triangling, -{Straight Barb[]}},
lines1/.style={highwavy, -{Straight Barb[]}},
lines2/.style={thick, dashed, -{Straight Barb[]}},
lines3/.style={crossing, -{Straight Barb[]}}
}

\numberwithin{equation}{section}

\let\OLDthebibliography\thebibliography
\renewcommand\thebibliography[1]{
  \OLDthebibliography{#1}
  \setlength{\parskip}{2pt}
  \setlength{\itemsep}{2pt plus 2pt}
}

\makeatletter
\renewcommand*\env@matrix[1][*\c@MaxMatrixCols c]{%
  \hskip -\arraycolsep
  \let\@ifnextchar\new@ifnextchar
  \array{#1}}
\makeatother

\begin{document}
\thispagestyle{empty}

\ \vskip 30mm

{\LARGE 
\centerline{\bf Three-point Functions in $\cN=4$ SYM}

\vspace{3mm}
\centerline{\bf at Finite $N_c$ and Background Independence}

}

\vskip 25mm

\centerline{
{\large \bf Ryo Suzuki}
}

{\let\thefootnote\relax\footnotetext{{\tt rsuzuki.mp\_at\_gmail.com}}}

\vskip 15mm

\centerline{{\it Shing-Tung Yau Center of Southeast University,}}
\centerline{{\it 15th Floor, Yifu Architecture Building, No.2 Sipailou,}}
\centerline{{\it Xuanwu district, Nanjing, Jiangsu, 210096, China}}

\vskip 25mm


\centerline{\bf Abstract}

\vskip 10mm 

We compute non-extremal three-point functions of scalar operators in $\cN=4$ super Yang-Mills at tree-level in $g_{\rm YM}$ and at finite $N_c$\,, using the operator basis of the restricted Schur characters.
We make use of the diagrammatic methods called quiver calculus to simplify the three-point functions.
The results involve an invariant product of the generalized Racah-Wigner tensors ($6j$ symbols).
Assuming that the invariant product is written by the Littlewood-Richardson coefficients, we show that the non-extremal three-point functions satisfy the large $N_c$ background independence; correspondence between the string excitations on \AdSxS\ and those in the LLM geometry.

\newpage
\tableofcontents

\section{Introduction}

Recently we have seen remarkable progress in the computation of the correlation functions of $\cN=4$ super Yang-Mills theory (SYM) in the hope of establishing the AdS/CFT correspondence \cite{Maldacena:1997re}.
There are two complementary approaches to this problem.

The first approach is based on the integrability of $\cN=4$ SYM in the planar limit. 
The planar three-point functions of single-trace operators are regarded as a pair of hexagons glued together, where each hexagon form-factor is severely constrained by the centrally-extended $\alg{su}(2|2)$ symmetry \cite{Basso:2015zoa}.
The $n$-point functions of BPS operators can be studied by {\it hexagonization}.
The gluing of four hexagons give us the planar four-point functions \cite{Fleury:2016ykk,Eden:2016xvg,Fleury:2017eph}, and the gluing of $2n-4+4g$ hexagons should give the $g$-th non-planar corrections \cite{Bargheer:2017nne,Eden:2017ozn,Bargheer:2018jvq}.
Furthermore, certain four-point functions in the large charge limit decompose into a pair of octagons \cite{Coronado:2018ypq,Coronado:2018cxj}, which can be resummed \cite{Bargheer:2019kxb,Bargheer:2019exp}.

The integrability approach tells us how single-trace correlation functions depend on the 't Hooft coupling $\lambda = N_c \, g_{\rm YM}^2$. However, only the non-extremal correlation functions have been studied, because the non-extremality is related to the so-called bridge length (the number of Wick contractions between a pair of operators), which suppresses the complicated wrapping corrections to the asymptotic formula \cite{Basso:2015eqa,Eden:2015ija,Basso:2017muf,Eden:2018vug,deLeeuw:2019qvz}.

The second approach is based on the finite-group theory.
In this approach, one obtains the results valid for any values of $N_c$\,, though most results are limited to tree-level or a few orders of small $\lambda$ expansion.
In the finite-group approach, extremal correlation functions are often studied, because they are roughly equal to the two-point functions at tree level.

Quite recently the author studied the $n$-point functions of multi-trace scalar operators at tree-level of $\cN=4$ SYM with $U(N_c)$ gauge group, based on the finite group methods \cite{Suzuki:2018aep}.
Those results are written in terms of permutations, meaning that they are valid to any orders of $1/N_c$ expansions, but not at any values of $N_c$ because the finite-$N_c$ constraints are not taken into consideration.
The primary purpose of this paper is to generalize the permutation-based results to finite $N_c$\,, by taking a Fourier transform of symmetric groups.

Two types of operator bases of $\cN=4$ SYM are well-known, which carry a set of Young diagrams as the operator label, diagonalize tree-level two-point functions at finite $N_c$\,, generalizing the pioneering work of \cite{Corley:2001zk}. 
The covariant basis (also called BHR basis) introduced in \cite{Brown:2007xh,Brown:2008ij} respects the global (or flavor) symmetry of the operator. As such, one can construct $O(N_f)$ singlets for general $N_f$ \cite{Kimura:2016bzo}.
The restricted Schur basis was introduced in a series of papers \cite{deMelloKoch:2007rqf,deMelloKoch:2007nbd,Bekker:2007ea} and related to multi-matrix models in \cite{Bhattacharyya:2008rb,Bhattacharyya:2008xy}.\footnote{Note that the restricted Schur basis can compute the observables of a multi-matrix model, which are not the function of the multi-matrix eigenvalues only.}
The restricted Schur basis respects the permutation symmetry of the operator, and suitable for explicit calculation. In other words, one has to specify a state inside the irreducible representation of the global (or flavor) symmetry, like the highest weight state.
Here is a brief comparison of the two representation bases \cite{Collins:2008gc}:
\begin{center}
\begin{tabular}{c|cc}
Operator basis & Symmetry respected & Analogy \\\hline
Covariant & Global symmetry & Spherical coordinates \\
Restricted Schur & Permutation of constituents & Cartesian coordinates
\end{tabular}
\end{center}

In this paper, we consider general non-extremal three-point functions of the scalar operators in the restricted Schur basis. 
There are several important ideas in this computation.
The first idea is the Schur-Weyl duality between $U(N_c)$ and $S_L$, which converts powers of $N_c$ into the irreducible characters of the symmetric group $S_L$\,.
The second idea is the {\it quiver calculus} initiated by \cite{Pasukonis:2013ts}. This is a set of diagrammatic rules which enormously simplify the manipulation of representation-theoretical objects.
The third idea is the generalized Racah-Wigner tensor. 
Since the three-point function is non-extremal, we need to compute a non-trivial overlap between the states under different subgroup decompositions of $S_L$\,.
The invariant products we encounter are more general than Wigner's $6j$ symbols.\footnote{The $6j$ symbol is also called Racah's $W$ coefficient or recoupling coefficient. The $6j$ symbols of symmetrical groups are called $6f$ symbols in \cite{Kramer67}, and they are related to the $6j$ symbols of unitary groups by the through the duality factor \cite{HB64b}.}

Let us summarize the main results. Our notation is explained in Appendix \ref{app:notation}.
We are particularly interested in two types of the non-extremal three-point functions (or equivalently non-extremal OPE coefficients).
The first type is the super-protected three-point functions \cite{Drukker:2009sf} in the restricted Schur basis, given by \eqref{Cooo comp I123}
\begin{multline}
\text{Fourier transform of }
\Vev{ 
\tr_{\! L_1} \bigl( \alpha_1 \, Z^{\otimes L_1} \bigr) \ 
\tr_{\! L_2} \bigl( \alpha_2 \, \tilde Z^{L_2} \bigr) \ 
\tr_{\! L_3} \bigl( \alpha_3 \, \olZ^{L_3} \bigr) }
\\
= \( \prod_{i=1}^3 \frac{L_i!}{\olL_i!} \)
\sum_{\hat R \, \vdash \hL} 
\frac{\Dim_{N_c} (\hat R)}{d_{R_1} d_{R_2} d_{R_3}} \ 
\sum_{Q_1 \vdash \olL_2 } 
\sum_{Q_2 \vdash \olL_3 }
\sum_{Q_3 \vdash \olL_1 } 
\( \prod_{i=1}^3 d_{Q_i} \) \, \cG_{123} \,.
\label{OPE1st}
\end{multline}
The second type is the three-point functions of the scalar operators made of three pairs of complex scalars in $\cN=4$ SYM, given by \eqref{CXYZ comp I123}
\begin{multline}
\text{Fourier transform of } \Big\langle 
\tr_{\! L_1} \Bigl( \alpha_1 \, \olX^{\otimes (\ell_{31}-h_2)} \, \olY^{\otimes h_3} \, Z^{\otimes (\ell_{12} - h_3 + h_2) } \Bigr) \ \times
\\[1mm]
\tr_{\! L_2} \Bigl( \alpha_2 \, \olX^{\otimes h_1} \, Y^{\otimes (\ell_{23}-h_1+h_3)} \, \olZ^{\otimes (\ell_{12}-h_3)} \Bigr) \,
\tr_{\! L_3} \Bigl( \alpha_3 \, X^{\otimes (\ell_{31}-h_2+h_1)} \, \olY^{\otimes (\ell_{23}-h_1)} \, \olZ^{\otimes h_2} \Bigr) 
\Big\rangle
\\
= \( \prod_{i=1}^3 \frac{L_i!}{\olL_i!} \)
\sum_{\hat R \, \vdash \hL} 
\frac{\Dim_{N_c} (\hat R)}{d_{R_1} d_{R_2} d_{R_3}} 
\( d_{q_1} \, d_{q_2} \, d_{r_1} \, d_{r_3} \, d_{s_2} \, d_{s_3} \)
\oldelta^{\, \nu_{1-} \, \nu_{2+} } \,
\oldelta^{\, \nu_{2-} \, \nu_{3+} } \,
\oldelta^{\, \nu_{3-} \, \nu_{1+} } \, 
\cG'_{123} \,.
\label{OPE2nd}
\end{multline}
The objects $\cG_{123}$ and $\cG'_{123}$ are related to the invariant products of the generalized Racah-Wigner tensors.

Mathematically, the branching coefficient of $R = \mathop \oplus \limits_{r,s} (r \otimes s)$ is the building block of the restricted Schur character and the generalized Racah-Wigner tensor. 
In the literature, the orthonormal basis of $r \otimes s$ is called the split basis \cite{ABH98}, and the branching coefficients are called fractional parentage coefficients \cite{EHJ53}, subduction coefficients \cite{CCG83,PC93} or the split-standard transformation coefficients \cite{ABH98,AB99,MH02}. 
In general, explicit computation of the branching coefficients is a hard problem.
See \cite{Chilla05,Chilla06,deMelloKoch:2011qz} for the recent results on the branching coefficients, and on the construction of the restricted Schur basis \cite{Koch:2011hb}.

Likewise, it is difficult to compute $\cG_{123} \,, \cG'_{123}$ explicitly. 
We conjecture that they can be written by the Littlewood-Richardson coefficients, based on the fact that they satisfy certain sum rules.

From \eqref{OPE1st} and \eqref{OPE2nd}, it is straightforward to show the large $N_c$ background independence in $\cN=4$ SYM \cite{deMelloKoch:2018ert}.
The background independence is a conjectured correspondence between the operators with $\cO(N_c^0)$ canonical dimensions and those with $\cO(N_c^2)$ canonical dimensions, where the latter is constructed from the former by ``attaching'' a large number of background boxes.
By AdS/CFT, this conjecture implies that the stringy excitations in \AdSxS\ and those in the (concentric circle configuration of) LLM geometry \cite{Lin:2004nb}.

On the gauge theory side, the large $N_c$ background independence has been checked for the case of two-point functions and extremal $n$-point functions.
On the gravity side, some string spectrum of in the $SL(2)$ sector has been studied in \cite{Kim:2018gwx}.
We find that the non-extremal OPE coefficients in the LLM background are essentially given by the rescaling of $N_c$ in \eqref{OPE1st}, \eqref{OPE2nd}.
Our results provide strong support that the large $N_c$ background independence can be found also in the string interactions.

\section{Two-point functions in the representation basis}\label{sec:2pt}

We review the construction of the restricted Schur basis, and introduce the diagrammatic computation methods called quiver calculus.

\subsection{Set-up}

We consider $\cN=4$ SYM of $U(N_c)$ gauge group at tree-level. This theory has three complex scalars $(X, Y, Z)$, which satisfy the $U(N_c)$ Wick rule,
\begin{equation}
\contraction{}{X}{_a^b (x) \, }{\olX}%
\contraction{X_a^b (x) \, \olX_c^d (0) = }{Y}{_a^b (x) \, }{\olY}
\contraction{X_a^b (x) \, \olX_c^d (0) = Y_a^b (x) \, \olY_c^d (0) = }{Z}{_a^b (x) \, }{\olZ}
X_a^b (x) \, \olX_c^d (0) = Y_a^b (x) \, \olY_c^d (0) = Z_a^b (x) \, \olZ_c^d (0) 
= |x|^{-2}  \, \delta^d_a \, \delta^b_c \,.
\label{UNc Wick}
\end{equation}

With $\alpha \in S_{l+m+n}$\,, we define a multi-trace operator in the permutation basis
\begin{equation}
\begin{aligned}
\cO_\alpha^{(l, m,n)} 
&= \tr_{\! l+m+n} \( \alpha \, X^{\otimes l} \, Y^{\otimes m} \, Z^{\otimes n} \)
\\
&\equiv \sum_{i_1 , i_2 , \dots , i_{l+m+n} =1}^{N_c}
X^{i_1}_{i_{\alpha (1)}} \dots X^{i_l}_{i_{\alpha(l)}} \, 
Y^{i_{m+1}}_{i_{\alpha(l+1)}} \dots Y^{i_{l+m}}_{i_{\alpha(l+m)}} \,
Z^{i_{l+m+1}}_{i_{\alpha(l+m+1)}} \dots Z^{i_{l+m+n}}_{i_{\alpha(l+m+n)}} \,.
\end{aligned}
\label{def:perm multi-trace}
\end{equation}
The usual single-trace operators can be expressed in the permutation basis as
\begin{equation}
\tr ( X^l \, Y^m \, Z^n )
\ \to \ \trb{\,L} ( \alpha \, X^{\otimes l} \, Y^{\otimes m} \, Z^{\otimes n} ), \qquad
(\alpha_i \in \bb{Z}_{l+m+n}).
\end{equation}
The correspondence between a multi-trace operator and $\alpha \in S_{\hL}$ is not one-to-one, because $\alpha$ is defined modulo conjugation,
\begin{equation}
\cO_\alpha^{(l,m,n)} = \cO_{\gamma \alpha \gamma^{-1}}^{(l,m,n)} , \qquad
\gamma \in S_l \otimes S_m \otimes S_n 
\label{def:symm multi-trace}
\end{equation}
which we call the flavor symmetry (or global symmetry).
For example,
\begin{equation}
\begin{aligned}
\tr (XXZZ ) &= \trb{\,L=4} ( (1234) \, X^{\otimes 2} Z^{\otimes 2} )
= \trb{\,L=4} ( (2143) \, X^{\otimes 2} Z^{\otimes 2} ) = \dots
\\
\tr (XZXZ ) &= \trb{\,L=4} ( (1324) \, X^{\otimes 2} Z^{\otimes 2} )
= \trb{\,L=4} ( (3142) \, X^{\otimes 2} Z^{\otimes 2} ) = \dots
\end{aligned}
\end{equation}
where $\dots$ represents the other permutations generated by the flavor symmetry \eqref{def:symm multi-trace}.

We define the complex conjugate operator by
\begin{equation}
\olcO_\alpha^{(l,m,n)} = \tr_{\! l+m+n} \( \alpha \, \olX^{\otimes l} \, \olY^{\otimes m} \, \olZ^{\otimes n} \)
\label{def:perm multi-trace2}
\end{equation}
The two-point function between $\cO_{\alpha_1}^{(l,m,n)}$ and $\olcO_{\alpha_2}^{(l,m,n)}$ at tree-level is given by
\begin{equation}
\vev{ \cO_{\alpha_1}^{(l,m,n)} (x) \, \olcO_{\alpha_2}^{(l,m,n)} (0) }
=|x|^{-2(l+m+n)} \sum_{\gamma \in S_l \otimes S_m \otimes S_n} 
N_c^{C (\alpha_1 \gamma \alpha_2 \gamma^{-1}) }
\label{two-point su2}
\end{equation}
where $C(\omega)$ counts the number of cycles in $\omega \in S_{l+m+n}$\,.
We write $\vev{ \cO_1 \, \olcO_2 } \equiv \vev{ \cO_1 (1) \, \olcO_2 (0) }$.

\subsection{Diagonalizing the tree-level two-point}\label{sec:diag 2pt}

Following \cite{Pasukonis:2013ts}, we show how to ``derive'' the representation basis of operators starting from the two-point functions on the permutation basis \eqref{two-point su2}.
The resulting tree-level two-point functions are diagonal at any $N_c$\,.
The readers familiar with the restricted Schur basis can skip this subsection. The basic formulae are summarized in Appendix \ref{app:RSB}.

First, we rewrite the equation \eqref{two-point su2} by using \eqref{decompose Nc(sigma)} as
\begin{equation}
\begin{aligned}
\vev{ \cO_{\alpha_1}^{(l,m,n)} \, \olcO_{\alpha_2}^{(l,m,n)} }
&= \sum_{\gamma \in S_l \otimes S_m \otimes S_n} \sum_{R \, \vdash (l+m+n)} \Dim_{N_c} (R)
\chi^R (\alpha_1 \gamma \alpha_2 \gamma^{-1}) 
\\
&= \sum_{R \, \vdash (l+m+n)} \Dim_{N_c} (R) \sum_{\gamma \in S_l \otimes S_m \otimes S_n} \ 
\begin{tikzpicture}[node distance=10mm, baseline=(current bounding box.center)]
\begin{scope}[decoration={markings, mark=at position 0.7 with {\arrow{Straight Barb[]}}}]
 \node (al1) [perm] {$\alpha_1$};
 \node (gam1) [perm, above right of=al1, xshift=5mm] {$\gamma^{-1}$};
 \node (gam2) [perm, below right of=al1, xshift=5mm] {$\gamma$}; 
 \node (al2) [perm2, right of=al1, xshift=16mm] {$\alpha_2$};
 \draw [lineR] (gam1) -- (al1);
 \draw [lineR] (al1) -- (gam2);
 \draw [lineR] (gam2) -- (al2);
 \draw [lineR] (al2) -- (gam1);
\end{scope}
\end{tikzpicture}
\end{aligned}
\label{two-point su2 comp}
\end{equation}
where we used the quiver calculus notation of Appendix \ref{app:quiver calc} in the second line.
We introduce $\gamma = \gamma_1 \circ \gamma_2 \circ \gamma_3 \in S_l \otimes S_m \otimes S_n$ and the branching coefficients for $S_{l+m+n} \downarrow (S_l \otimes S_m \otimes S_n)$ to make use of the identity \eqref{gen:perm matrix after branching} for $\ell=3$.
The equation \eqref{two-point su2 comp} becomes
\begin{equation}
\vev{ \cO_{\alpha_1}^{(l,m,n)} \, \olcO_{\alpha_2}^{(l,m,n)} } = 
\sum_{R \, \vdash (l+m+n)} \Dim_{N_c} (R) 
\sum_{\substack{\gamma_1 \in S_l \\[.6mm] \gamma_2 \in S_m \\[.6mm] \gamma_3 \in S_n}} \ 
\sum_{\substack{r_1, r_2, r_3 ,\nu_- \\[.6mm] s_1, s_2, s_3, \nu_+}} \ 
\begin{tikzpicture}[node distance=10mm, baseline=(current bounding box.center)]
\begin{scope}[decoration={markings, mark=at position 0.7 with {\arrow{Straight Barb[]}}}]
 \node (al1) [perm] {$\alpha_1$};
 \node (br1) [branch, above right of=al1, xshift=3mm, yshift=8mm] {$\nu_+$};
 \node (br2) [branch, below right of=al1, xshift=3mm, yshift=-8mm] {$\nu_-$};  
 \node (gam1) [perm, shape border rotate=270, right of=br1, xshift=6mm, yshift=10mm] {$\gamma_1^{-1}$};
 \node (gam1b) [perm, shape border rotate=270, right of=br1, xshift=6mm] {$\gamma_2^{-1}$};
 \node (gam1c) [perm, shape border rotate=270, right of=br1, xshift=6mm, yshift=-10mm] {$\gamma_3^{-1}$};
 \node (gam2) [perm, shape border rotate=90, right of=br2, xshift=6mm, yshift=10mm] {$\gamma_1$}; 
 \node (gam2b) [perm, shape border rotate=90, right of=br2, xshift=6mm] {$\gamma_2$}; 
 \node (gam2c) [perm, shape border rotate=90, right of=br2, xshift=6mm, yshift=-10mm] {$\gamma_3$}; 
 \node (br3) [branch, right of=gam1b, xshift=6mm] {$\nu_+$};
 \node (br4) [branch, right of=gam2b, xshift=6mm] {$\nu_-$};  
 \node (al2) [perm2, right of=al1, xshift=40mm] {$\alpha_2$};
 \draw [lineR] (br1) -- (al1);
 \draw [lineR] (al1) -- (br2);
 \draw [lineR] (al2) -- (br3);
 \draw [lineR] (br4) -- (al2);
 \draw [liner1] (gam1.west) -- (br1.north east);
 \draw [liner2] (gam1b.west) -- (br1.east);
 \draw [liner3] (gam1c.west) -- (br1.south east);
 \draw [lines1] (br2.north east) -- (gam2.west);
 \draw [lines2] (br2.east) -- (gam2b.west);
 \draw [lines3] (br2.south east) -- (gam2c.west);
 \draw [liner1] (br3.north west) -- (gam1.east);
 \draw [liner2] (br3.west) -- (gam1b.east);
 \draw [liner3] (br3.south west) -- (gam1c.east);
 \draw [lines1] (gam2.east) -- (br4.north west);
 \draw [lines2] (gam2b.east) -- (br4.west);
 \draw [lines3] (gam2c.east) -- (br4.south west);
\end{scope}
\end{tikzpicture}
\end{equation}
We apply the grand orthogonality \eqref{grand orthogonality quiver2} to the matrix elements of $\gamma_1 \,, \gamma_2$ and $\gamma_3$ to obtain
\begin{equation}
\begin{aligned}
\vev{ \cO_{\alpha_1}^{(l,m,n)} \, \olcO_{\alpha_2}^{(l,m,n)} } &=  
\sum_{R \, \vdash (l+m+n)} \Dim_{N_c} (R) \, 
\sum_{r_1, r_2, r_3, \nu_- , \nu_+} \frac{l! \, m! \, n!}{d_{r_1} d_{r_2} d_{r_3}} \ \ 
\begin{tikzpicture}[node distance=10mm, baseline=(current bounding box.center)]
\begin{scope}[decoration={markings, mark=at position 0.7 with {\arrow{Straight Barb[]}}}]
 \node (al1) [perm, xshift=62mm] {$\alpha_1$};
 \node (br1) [branch, above of=al1, yshift=4mm] {$\nu_+$};
 \node (br2) [branch, below of=al1, yshift=-4mm] {$\nu_-$}; 
 \draw [lineR] (br1) -- (al1);
 \draw [lineR] (al1) -- (br2);
 \draw [liner1] (br2.west) .. controls (5.2,-0.5) and (5.2,0.5) .. (br1.west);
 \draw [liner2] (br2.east) .. controls (7.2,-0.5) and (7.2,0.5) .. (br1.east);
 \draw [liner3] (br2.south) .. controls (5.1,-2) and (4.8,-0.5) .. (4.8,0) .. controls (4.8,0.5) and (5.1,2) .. (br1.north);
 \node (al2) [perm2, right of=al1, xshift=16mm] {$\alpha_2$};
 \node (br3) [branch, above of=al2, yshift=4mm] {$\nu_+$};
 \node (br4) [branch, below of=al2, yshift=-4mm] {$\nu_-$}; 
 \draw [lineR] (al2) -- (br3);
 \draw [lineR] (br4) -- (al2);
 \draw [liner1] (br3.west) .. controls (7.8,.5) and (7.8,-0.5) .. (br4.west);
 \draw [liner2] (br3.east) .. controls (9.8,0.5) and (9.8,-0.5) .. (br4.east);
 \draw [liner3] (br3.north) .. controls (7.7,2) and (7.4,0.5) .. (7.4,0) .. controls (7.4,-0.5) and (7.7,-2) .. (br4.south);
\end{scope}
\end{tikzpicture}
\\
&= \sum_{R, r_1, r_2, r_3, \nu_- , \nu_+}
\Dim_{N_c} (R) \, \frac{l! \, m! \, n!}{d_{r_1} d_{r_2} d_{r_3}} \,
\chi^{R, (r_1,r_2, r_3), (\nu_+,\nu_-)} (\alpha_1) \,
\chi^{R, (r_1,r_2, r_3), (\nu_-,\nu_+)} (\alpha_2) 
\end{aligned}
\notag
\end{equation}
where $\chi^{R, (r_1,r_2,r_3), (\nu_+,\nu_-)} (\alpha)$ is the restricted characters defined through branching coefficients,
\begin{equation}
\chi^{R,(r_1,r_2,r_3),\nu_+,\nu_-} (\sigma) 
\equiv \sum_{I,J} \sum_{i,j} B^{R \to (r_1, r_2,r_3) \nu_+}_{I \to (i,j,k)} \, 
\brT^{R \to (r_1, r_2, r_3) \nu_-}_{J \to (i,j,k)} \, 
D^R_{IJ} (\sigma) .
\label{def:restricted character}
\end{equation}

The restricted characters satisfy the orthogonality relations \eqref{restricted row orthogonality}.
It is straightforward to find a linear combination of operators which diagonalizes the two-point function;
\begin{equation}
\begin{aligned}
\cO^{S,(s_1,s_2,s_3),\mu_+,\mu_-} (x) = 
\frac{1}{l! \, m! \, n!} \sum_{\alpha \in S_{l+m+n}}
\chi^{S,(s_1,s_2,s_3),\mu_+,\mu_-} (\alpha) \, \cO_{\alpha}^{(l,m,n)} (x)
\\
\olcO^{T,(t_1,t_2,t_3),\eta_+,\eta_-} (y) = 
\frac{1}{l! \, m! \, n!} \sum_{\alpha \in S_{l+m+n}}
\chi^{T,(t_1,t_2,t_3),\eta_+,\eta_-} (\alpha) \, \olcO_{\alpha}^{(l,m,n)} (y).
\end{aligned}
\end{equation}
It follows that                      
\begin{align}
&\Vev{ \cO^{S,(s_1,s_2,s_3),\mu_+,\mu_-} \, \olcO^{T,(t_1,t_2,t_3),\eta_+,\eta_-} }
= \( \frac{1}{l! \, m! \, n!}\)^2 \sum_{R, r_1, r_2, r_3 ,\nu_- , \nu_+} 
\Dim_{N_c} (R) \, \frac{l! m! n!}{d_{r_1} d_{r_2} d_{r_3}} \ \times
\notag \\[1mm]
&\qquad
\sum_{\alpha_1 \,, \alpha_2 \in S_{l+m+n}} 
\chi^{S,(s_1,s_2,s_3),\mu_+,\mu_-} (\alpha_1) \, \chi^{T,(t_1,t_2,t_3),\eta_+,\eta_-} (\alpha_2) \,
\chi^{R, (r_1,r_2,r_3), (\nu_+,\nu_-)} (\alpha_1) \,
\chi^{R, (r_1,r_2,r_3), (\nu_-,\nu_+)} (\alpha_2) 
\notag \\[1mm]
&\qquad
= \Dim_{N_c} (S) \, \frac{(l+m+n)!^2}{l! \, m! \, n!} \, \frac{d_{s_1} d_{s_2} d_{s_3}}{d_S^2} \,
\delta^{ST} \delta^{s_1 t_1} \delta^{s_2 t_2} \delta^{s_3 t_3} \delta^{\mu_+ \eta_-} \delta^{\mu_- \eta_+} 
\notag \\[1mm]
&\qquad
= \wt_{N_c} (S) \, \frac{\hook_S}{\hook_{s_1} \hook_{s_2} \hook_{s_3}} \,
\delta^{ST} \delta^{s_1 t_1} \delta^{s_2 t_2} \delta^{s_3 t_3} \delta^{\mu_+ \eta_-} \delta^{\mu_- \eta_+}
\label{two-point su2 comp2}
\end{align}
where we used \eqref{def:DimNR wtNR}.

Recall that $\cO_\alpha^{(l, m,n)}$ in \eqref{def:perm multi-trace} becomes half-BPS when $l=m=0$, and the restricted character \eqref{def:restricted character} reduces to the usual irreducible characters of $S_n$\,.
The two-point function \eqref{two-point su2 comp2} becomes
\begin{equation}
\Vev{ \cO^{S} \, \olcO^{T} }
= \wt_{N_c} (S) \, \delta^{ST} 
\label{BPS hw normalization}
\end{equation}
which gives the same normalization of half-BPS operators as in \cite{Corley:2001zk}.

\section{Three-point functions in the representation basis}\label{sec:3pt}

In \cite{Suzuki:2018aep}, tree-level formulae of the $n$-point functions of general scalar operators in the permutation basis have been derived.
We consider three-point functions of scalar operators in the restricted Schur basis below. 
The three-point functions of $\cN=4$ SYM are related to the OPE coefficients by
\begin{equation}
\Vev{\cO_1(x_1) \cO_2(x_2) \cO_3(x_3)} = \frac{C_{123}}{\abs{x_1-x_2}^{\Delta_1+\Delta_2-\Delta_3}
\abs{x_2-x_3}^{\Delta_2+\Delta_3-\Delta_1}
\abs{x_3-x_1}^{\Delta_3+\Delta_1-\Delta_2}} 
\label{def:OPE coeff}
\end{equation}
thanks to the conformal symmetry. By abuse of notation, we write \eqref{def:OPE coeff} as
\begin{equation}
\vev{\cO_1 \cO_2 \cO_3} = C_{123} \,.
\end{equation}

\subsection{Set-up}\label{sec:three-point setup}

Let us recall the tree-level permutation formula for three-point functions in \cite{Suzuki:2018aep}.
That formula has been derived based on the following idea.
Consider a non-extremal three-point function of the operators labeled by $\alpha_i \in S_{L_i}$ for $i=1,2,3$. 
We expect that the tree-level Wick contractions give the quantity like $N_c^{C (\alpha_1 \alpha_2 \alpha_3)}$.
However, we cannot define the multiplication of elements in $S_{L_1}$ and $S_{L_2}$ if $L_1 \neq L_2$\,.
This problem can be solved by extending $\alpha_i$ to $\hat \alpha_i \in S_{\hL}$ for some $\hL$\,, which makes the quantity $N_c^{C (\hat \alpha_1 \hat \alpha_2 \hat \alpha_3)}$ well-defined.

\bigskip
Let us explain how this idea works.
First, we extend the operator $\cO_i$ by adding identity fields,
\begin{equation}
\hat \cO_i \equiv \cO_{\alpha_i} \times \tr ({\bf 1})^{\olL_i} 
\equiv \prod_{p=1}^{\hL}  (\Phi^{\hat A_p^{(i)}})^{a_p}_{a_{\hat \alpha_i(p)}} \,, \qquad
\hat \alpha_i = \alpha_i \circ {\bf 1}_{\olL_i} \in S_{L_i} \times S_{\olL_i} \subset S_{\hL} 
\label{def:Ohat}
\end{equation}
where
\begin{equation}
\hL= \frac{L_1+L_2+L_3}{2} \,, \qquad
\olL_i = \hL-L_i \,, \qquad
\Phi^{\hat A_p^{(i)}} \in (X, \olX, Y, \olY, Z, \olZ, {\bf 1}). 
\label{def:olL}
\end{equation}
The permutation $\hat \alpha_i$ acts as the identity at the position $p$ at which $\Phi^{\hat A_p^{(i)}} = {\bf 1}$.
The (edge-type) permutation formula reads
\begin{equation}
C_{123} = \frac{1}{\prod_{i=1}^3 \olL_i !} \, \frac{1}{\hL!} \ \sum_{\{U_i\} \in S_L^{\otimes 3} } 
\( \prod_{p=1}^{\hL} \, h^{\chA^{(1)}_{p} \chA^{(2)}_{p} \chA^{(3)}_{p} } \) 
N_c^{C (\chalpha_1 \, \chalpha_2 \, \chalpha_3 ) }
\label{SL tree n-pt}
\end{equation}
where $\chA^{(i)}_p \equiv \hat A^{(i)}_{U_i(p)}$, $\check \alpha_i \equiv U_i^{-1} \, \hat \alpha_i \, U_i$ and
\begin{equation}
h^{ABC} = h^{AB} \, \delta^C_{\bf 1} 
+ h^{BC} \, \delta^A_{\bf 1} + h^{CA} \, \delta^B_{\bf 1} \,, \qquad
h^{AB} = \begin{cases}
g^{AB} \equiv \vev{ \Phi^A (1) \Phi^B (0) }
&\ ({\rm both} \ \Phi^A , \Phi^B \neq {\bf 1}) \\
0 &\ ({\rm otherwise}).
\end{cases}
\label{def:hABC}
\end{equation}
We call $h^{ABC}$ a triple Wick contraction.

We will consider two types of three-point functions.
The first type is the three-point functions of half-BPS multi-trace operators,
\begin{equation}
C_{\circ \circ \circ} = \Vev{ 
\tr_{\! L_1} \bigl( \alpha_1 \, Z^{\otimes L_1} \bigr) \ 
\tr_{\! L_2} \bigl( \alpha_2 \, \tilde Z^{\otimes L_2} \bigr) \ 
\tr_{\! L_3} \bigl( \alpha_3 \, \olZ^{\otimes L_3} \bigr) }, \qquad
\tilde Z = (Z + \olZ + Y - \olY).
\label{def:Cooo}
\end{equation}
The field $\tilde Z$ belongs to the one-parameter family of operators used in \cite{Drukker:2009sf,Basso:2015zoa},
\begin{equation}
\mathfrak{Z}_i (a) = (Z + a_i \, (Y-\olY) + a_i^2 \, \olZ) (x_i), \qquad
x_i = (0,a_i,0,0).
\end{equation}
The second type is general three-point functions of the scalar multi-trace operator \eqref{def:perm multi-trace},
\begin{multline}
C^{XYZ}_{\vec h} = \Big\langle 
\tr_{\! L_1} \Bigl( \alpha_1 \, \olX^{\otimes (\ell_{31}-h_2)} \, \olY^{\otimes h_3} \, Z^{\otimes (\ell_{12} - h_3 + h_2) } \Bigr) \ \times
\\[1mm]
\tr_{\! L_2} \Bigl( \alpha_2 \, \olX^{\otimes h_1} \, Y^{\otimes (\ell_{23}-h_1+h_3)} \, \olZ^{\otimes (\ell_{12}-h_3)} \Bigr) \,
\tr_{\! L_3} \Bigl( \alpha_3 \, X^{\otimes (\ell_{31}-h_2+h_1)} \, \olY^{\otimes (\ell_{23}-h_1)} \, \olZ^{\otimes h_2} \Bigr) 
\Big\rangle
\label{def:CXYZ vec}
\end{multline}
where $\ell_{ij}$ is the number of tree-level Wick contractions between $\cO_i$ and $\cO_j$ (called the bridge length), given by
\begin{equation}
\ell_{12} = \frac{L_1 + L_2 -L_3}{2} \,, \qquad
\ell_{23} = \frac{L_2 + L_3 -L_1}{2} \,, \qquad
\ell_{31} = \frac{L_3 + L_1 -L_2}{2} 
\end{equation}
and $h_i$ is an integer inside the range
\begin{equation}
0 \le h_1 \le \ell_{23} \,, \qquad
0 \le h_2 \le \ell_{31} \,, \qquad
0 \le h_3 \le \ell_{12} \,.
\label{ranges oh h}
\end{equation}

\subsection{Partial Fourier transform}\label{sec:Fourier}

We construct the three-point functions in the restricted Schur basis by taking the Fourier transform of $C_{\circ \circ \circ}$ in \eqref{def:Cooo} and $C^{XYZ}_{\vec h}$ \eqref{def:CXYZ vec}. 
Recall that the usual Fourier transform of the delta function is a constant. In the Fourier transform over a finite group, the Fourier transform of the identity permutation should be a sum over all representations. In other words, if we write
\begin{equation}
R_i \, \vdash L_i \ \ \leftrightarrow \ \ \text{FT of} \ \alpha_i \in S_{L_i} \,, \qquad
t_i \, \vdash \olL_i \ \ \leftrightarrow \ \ \text{FT of} \ {\bf 1}^{\olL_i} \in S_{\olL_i}
\label{def:partial Fourier transform}
\end{equation}
then we should sum $t_i$ over all possible partitions of $\olL_i$\,.
In fact, $t_i$ is an unphysical parameter, and we can perform a calculation without using $t_i$\,.
Thus we call the procedure \eqref{def:partial Fourier transform} a partial Fourier transform.

\bigskip
In order to treat $C_{\circ \circ \circ}$ and $C^{XYZ}_{\vec h}$ simultaneously, we extend the multi-trace operator \eqref{def:perm multi-trace} as in \eqref{def:Ohat}, 
\begin{equation}
\begin{gathered}
\cO_{\hat \alpha_i}^{(l_i, m_i,n_i, \olL_i)} [X,Y,Z,{\bf 1}]
= \tr_{\! L_i} \Bigl( \alpha_i \, X^{\otimes l_i} \, Y^{\otimes m_i} \, Z^{\otimes n_i} \Bigr) \times 
\tr ({\bf 1})^{\olL_i}
\\[1mm]
l_i + m_i + n_i = L_i \,, \qquad
L_i + \olL_i = \hL \,, \qquad
\hat \alpha_i = \alpha_i \circ {\bf 1}_{\olL_i} \in S_{\hL}
\end{gathered}
\label{def:hat cOk XYZ}
\end{equation}
and define the partial Fourier transform by
\begin{equation}
\begin{gathered}
\hat \cO^{\bsR_i \, (\olL_i )} [X,Y,Z,{\bf 1}] = 
\frac{1}{l_i! \, m_i! \, n_i!} \sum_{\alpha_i \in S_{L_i}}
\chi^{\bsR_i} (\alpha_i) \, 
\cO_{\hat \alpha_i}^{(l_i, m_i,n_i, \olL_i)} [X,Y,Z,{\bf 1}]
\\[1mm]
\bsR_i = \pare{ R_i , (q_i, r_i, s_i), \nu_{i-}, \nu_{i+} }, \qquad
\( R_i \vdash L_i \,,\, q_i \vdash l_i \,,\, r_i \vdash m_i \,,\, s_i \vdash n_i \).
\end{gathered}
\label{extended rep basis}
\end{equation}
The partial Fourier transform can be rewritten as a linear combination of the complete Fourier transform. 
To see this, we recall \eqref{delta fn2} and
\begin{equation}
\chi^{\bsR_i \otimes t_i} (\alpha_i \circ {\bf 1}_{\olL_i} ) 
= \chi^{\bsR_i} (\alpha_i) \, d_{t_i} \,,
\qquad
\sum_{t_i \vdash \olL_i} d_{t_i}^2 = \olL_i
\label{def:extended character}
\end{equation}
giving us a dummy representation $t_i$ to be summed over the partitions of $\olL_i$\,.
It follows that
\begin{equation}
\hat \cO^{\bsR_i \, (\olL_i )} [X,Y,Z,{\bf 1}]
= \frac{1}{l_i! \, m_i! \, n_i! \, \olL_i!} 
\sum_{t_i \, \vdash \olL_i} \, \sum_{\hat \alpha_i \in S_{L_i} \times {\bf 1}_{\olL_i} }
d_{t_i} \, \chi^{\bsR_i \otimes t_i} (\hat \alpha_i) \, 
\cO_{\hat \alpha_i}^{(l_i,m_i,n_i, \olL_i)} [X,Y,Z,{\bf 1}] .
\end{equation}

As for $C_{\circ \circ \circ}$\,, we introduce the Fourier transform of the half-BPS operators as
\begin{equation}
\widetilde{\cO}_1 = \hat \cO_1^{R_1 (\olL_1)} [Z,{\bf 1}] , \qquad 
\widetilde{\cO}_2 = \hat \cO_2^{R_2 (\olL_2)} [\tilde Z, {\bf 1}] , \qquad
\widetilde{\cO}_3 = \hat \cO_3^{R_3 (\olL_3)} [\olZ, {\bf 1}] ,\qquad
\bsR_i = R_i \vdash L_i 
\end{equation}
and define
\begin{equation}
\tilde C_{\circ\circ\circ} = \Vev{
\hat \cO_1^{R_1 (\olL_1)} [Z,{\bf 1}] \,
\hat \cO_2^{R_2 (\olL_2)} [\tilde Z, {\bf 1}] \,
\hat \cO_3^{R_3 (\olL_3)} [\olZ, {\bf 1}] }.
\label{def:tilde Cooo}
\end{equation}
As for $C^{XYZ}_{\vec h}$\,, we take the Fourier transform of the operators in \eqref{def:CXYZ vec} as
\begin{alignat}{9}
\widetilde{\cO}_1 &= \hat \cO_1^{\bsR_1 (\olL_1)} [\olX,\olY,Z,{\bf 1}] &\qquad
(l_1,m_1,n_1) &= (\ell_{31}-h_2, \ h_3, \ \ell_{12} - h_3 + h_2)
\notag \\[1mm]
\widetilde{\cO}_2 &= \hat \cO_2^{\bsR_2 (\olL_2)} [\olX,Y,\olZ, {\bf 1}] &\qquad
(l_2, m_2, n_2) &= (h_1, \ \ell_{23}-h_1+h_3, \ \ell_{12}-h_3)
\notag \\[1mm]
\widetilde{\cO}_3 &= \hat \cO_3^{\bsR_3 (\olL_3)} [X,\olY,\olZ, {\bf 1}] &\qquad
(l_3, m_3, n_3) &= (\ell_{31}-h_2+h_1, \ \ell_{23}-h_1, \ h_2)
\end{alignat}
and define
\begin{equation}
\tilde C^{XYZ}_{\vec h} =  
\Vev{
\hat \cO_1^{\bsR_1 (\olL_1)} [\olX,\olY,Z,{\bf 1}] \,
\hat \cO_2^{\bsR_2 (\olL_2)} [\olX,Y,\olZ, {\bf 1}] \,
\hat \cO_3^{\bsR_3 (\olL_3)} [X,\olY,\olZ, {\bf 1}] }.
\label{def:tilde CXYZ}
\end{equation}
We collectively denote the three-point functions of the operators in the representation basis by
\begin{equation}
\tilde{C}_{123} \equiv \Vev{ \widetilde{\cO}_1 \, \widetilde{\cO}_2 \, \widetilde{\cO}_3 } .
\end{equation}
From \eqref{SL tree n-pt} we get
\begin{multline}
\tilde{C}_{123} = 
\frac{1}{\prod_{i=1}^3 l_i! \, m_i! \, n_i! \, (\olL_i !)^2 } \, \frac{1}{\hL!} \ 
\sum_{\{U_i \} \in S_L^{\otimes 3} }
\( \prod_{p=1}^{\hL} \, 
h^{\hat A^{(1)}_{U_1(p)} \hat A^{(2)}_{U_2(p)} \hat A^{(3)}_{U_3(p)} } \)
\sum_{\{ t_i \vdash \olL_i \}} \( \prod_{i=1}^3 d_{t_i} \) \times
\\
\sum_{ \{ \hat \alpha_i \in S_{L_i} \times {\bf 1}_{\olL_i} \} }
\( \prod_{i=1}^3 \chi^{\bsR_i \otimes t_i} (\hat \alpha_i) \)
N_c^{C ( U_1^{-1} \, \hat \alpha_1 \, U_1 U_2^{-1} \, \hat \alpha_2 \, U_2 U_3^{-1} \, \hat \alpha_3 \, U_3 ) } .
\label{CXZm comp}
\end{multline}

\bigskip
Consider the second line of \eqref{CXZm comp}.
We use the identity \eqref{decompose Nc(sigma)} and \eqref{def:irrep char} to obtain
\begin{align}
&\sum_{ \{ \hat \alpha_i \in S_{L_i} \times {\bf 1}_{\olL_i} \} }
\( \prod_{i=1}^3 \chi^{\bsR_i \otimes t_i} (\hat \alpha_i) \)
N_c^{C ( U_1^{-1} \, \hat \alpha_1 \, U_1 U_2^{-1} \, \hat \alpha_2 \, U_2 U_3^{-1} \, \hat \alpha_3 \, U_3 ) } 
\label{rep 3pt comp} \\[1mm]
&=\sum_{ \{ \hat \alpha_i \in S_{L_i} \times {\bf 1}_{\olL_i} \} }
\sum_{\hat R \, \vdash \hL} \Dim_{N_c} (\hat R) \, 
\( \prod_{i=1}^3 \chi^{\bsR_i \otimes t_i} (\hat \alpha_i) 
D^{\hat R} _{\hat I_i \hat J_i}(\hat\alpha_i) \) 
D^{\hat R}_{\hat J_1\hat I_2} (U_1 U_2^{-1})
D^{\hat R}_{\hat J_2\hat I_3} (U_2 U_3^{-1})
D^{\hat R}_{\hat J_3\hat I_1} ( U_3 U_1^{-1}) .
\notag
\end{align}
We simplify the sum over $\{ \hat \alpha_i \}$ in the last line.
The character is given by \eqref{def:extended character}.
We decompose the matrix elements $D^{\hat R} _{\hat I_i \hat J_i}( \hat\alpha_i )$ according to the restriction
\begin{equation}
S_{\hL} \downarrow (S_{L_i} \otimes S_{\olL_i}), \qquad
\hat R = \bigoplus_{R'_i \, \vdash L_i} \bigoplus_{T_i \, \vdash \olL_i} \bigoplus_{\mu_i=1}^{g (R'_i , t'_i ; \hat R)} \, (R'_i \otimes T_i)_{\mu_i} \,.
\label{alpha-decomp}
\end{equation}

When $\tilde C_{123} = \tilde C_{\circ \circ \circ}$\,, we have $\bsR_i = R_i$\,. 
From \eqref{alpha-decomp} we get
\begin{align}
\sum_{\hat \alpha_i}
\chi^{\bsR_i \otimes t_i} (\hat \alpha_i) D^{\hat R} _{\hat I_i \hat J_i}( \hat\alpha_i )
&= \sum_{\alpha_i \in S_{L_i}} \, \sum_{R'_i \, \vdash L_i} \,
\sum_{T_i \, \vdash \olL_i} \, \sum_{\mu_i=1}^{g(R'_i, T_i;\hat R)} \,
\chi^{R_i} (\alpha_i) \, d_{t_i} \,
B^{\hat R \to (R'_i, T_i), \mu_i}_{\hat I_i \to (I_i, c_i)} \, 
(B^T)^{\hat R \to (R'_i, T_i), \mu_i}_{\hat J_i \to (J_i, c_i)} \, 
D^{R'_i} _{I_i J_i} (\alpha_i)
\notag \\[1mm]
&= \sum_{R'_i, T_i, \mu_i} \, \Big\{ \sum_{\alpha_i \in S_{L_i}}\, 
\chi^{R_i} (\alpha_i) 
D^{R'_i} _{I_i J_i} (\alpha_i) \Big\}
d_{t_i} \,
B^{\hat R \to (R'_i, T_i), \mu_i}_{\hat I_i \to (I_i, c_i)} \, 
(B^T)^{\hat R \to (R'_i, T_i), \mu_i}_{\hat J_i \to (J_i, c_i)} \, 
\notag \\[1mm]
&= \sum_{T_i , \mu_i} \, \frac{L_i! \, d_{t_i} \,}{d_{R_i}} \, 
B^{\hat R \to (R_i, T_i), \mu_i}_{\hat I_i \to (I_i, c_i)} \, 
(B^T)^{\hat R \to (R_i, T_i), \mu_i}_{\hat J_i \to (I_i, c_i)} 
\notag \\[1mm]
&= \sum_{T_i \, \vdash \olL_i} \, \sum_{\mu_i=1}^{g(R_i ,T_i \,;\hat R)} \,
\frac{L_i! \, d_{t_i} \,}{d_{R_i}} \, \proj_{\hat I_i \hat J_i}^{\hat R \to (R_i, T_i), \mu_i, \mu_i}
\label{rep Cooo comp}
\end{align}
where we used \eqref{def:extended character}, \eqref{perm matrix after branching}, \eqref{orthogonality_representation} and \eqref{def:restricted projector R}.
When $\tilde C_{123} = \tilde C^{XYZ}_{\vec h}$\,, by using the definition of the restricted character \eqref{app:restricted character} we find
\begin{align}
&\sum_{\hat \alpha_i \in S_{L_i} \times {\bf 1}_{\olL_i}}
\chi^{\bsR_i \otimes t_i} (\hat \alpha_i) D^{\hat R} _{\hat I_i \hat J_i}( \hat\alpha_i)
\notag \\[1mm]
&= \sum_{R'_i, T_i, \mu_i} \, \Big\{ \sum_{\alpha_i \in S_{L_i}}\, 
D^{R_i}_{I' J'} (\alpha_i) 
D^{R'_i} _{I_i J_i} (\alpha_i) \Big\}
d_{t_i} \,
B^{R_i \to (q_i,r_i, s_i) \nu_{i-}}_{I' \to (j',k',l')} \, 
(B^T)^{R_i \to (q_i,r_i, s_i) \nu_{i+}}_{J' \to (j',k',l')} \, 
B^{\hat R \to (R'_i, T_i), \mu_i}_{\hat I_i \to (I_i, c_i)} \, 
(B^T)^{\hat R \to (R'_i, T_i), \mu_i}_{\hat J_i \to (J_i, c_i)} \, 
\notag \\[1mm]
&= \sum_{T_i , \mu_i} \, \frac{L_i! \, d_{t_i} \,}{d_{R_i}} \, 
B^{\hat R \to (R_i, T_i), \mu_i}_{\hat I_i \to (I_i, c_i)} \, 
B^{R_i \to (q_i,r_i, s_i) \nu_{i-}}_{I_i \to (j',k',l')} \, 
(B^T)^{\hat R \to (R_i, T_i), \mu_i}_{\hat J_i \to (J_i, c_i)} \, 
(B^T)^{R_i \to (q_i,r_i, s_i) \nu_{i+}}_{J_i \to (j',k',l')} 
\notag \\[1mm]
&\equiv \sum_{T_i , \mu_i} \, \frac{L_i! \, d_{t_i} \,}{d_{R_i}} \, 
\proj_{\hat I \hat J}^{\hat R \to \bs{RT}_{i-,i+}}
\label{rep CXYZ comp}
\end{align}
where we introduced the double projector
\begin{align}
\proj^{\hat R \to \bs{RT}_{i-,i+}}_{\hat I \hat J} &= \sum_{j,k,l,c}
\cB^{\hat R \to \bs{RT}_{i-}}_{\hat I \longrightarrow (j,k,l,c)} \,
(\cB^T)^{\hat R \to \bs{RT}_{i +}}_{\hat J \longrightarrow (j,k,l,c)}
\label{def:proj hat bsR}
\\[1mm]
\cB^{\hat R \to \bs{RT}_{i \mp}}_{\hat I \longrightarrow (j,k,l,c)}
&\equiv \sum_{I=1}^{d_{R_i} }
B^{\hat R \to (R_i , T_i), \mu_i}_{\hat I \to (I, c)} \, 
B^{R_i \to (q_i,r_i,s_i), \nu_{i \mp} }_{I \to (j,k,l)}  \,.
\label{def:double branching coeff}
\\[1mm]
\pare{ \hat R \to \bs{RT}_{i \mp} } &= \pare{ \hat R \to (R_i, T_i), \mu_i \, \to (q_i,r_i,s_i, T_i), (\mu_i,\nu_{i \mp}) } 
\label{def:hat bsR}
\end{align}
which come from the double restriction $S_{\hL} \downarrow (S_{L_i} \otimes S_{\olL_i}) \downarrow (S_{l_i} \otimes S_{m_i} \otimes S_{n_i} \otimes S_{\olL_i})$. 
Here we should keep in mind that the restriction to the subgroup of $S_{\hL}$ is different for each $i=1,2,3$.
We will revisit this issue in Section \ref{sec:sum projectors}.

Now the equation \eqref{rep 3pt comp} is simplified as
\begin{multline}
\sum_{ \{ \hat \alpha_i \in S_{L_i} \times {\bf 1}_{\olL_i} \} }
\( \prod_{i=1}^3 \chi^{\bsR_i \otimes t_i} (\hat \alpha_i) \)
N_c^{C ( U_1^{-1} \, \hat \alpha_1 \, U_1 U_2^{-1} \, \hat \alpha_2 \, U_2 U_3^{-1} \, \hat \alpha_3 \, U_3 ) } 
\\
= \sum_{ \{ T_i , \mu_i \} } \( \prod_{i=1}^3 
\proj^{\hat R \to \, {\rm sub}}_{\hat I_i \hat J_i} \) 
D^{\hat R}_{\hat J_1\hat I_2} (U_1 U_2^{-1}) \,
D^{\hat R}_{\hat J_2\hat I_3} (U_2 U_3^{-1}) \,
D^{\hat R}_{\hat J_3\hat I_1} ( U_3 U_1^{-1}) 
\end{multline}
where the projector $\proj^{\hat R \to {\rm sub}}_{\hat I_i \hat J_i}$ is given by
\begin{equation}
\proj^{\hat R \to \, {\rm sub}}_{\hat I_i \hat J_i} \equiv
\begin{cases}
\proj^{\hat R \to (R_i,T_i) \mu_i, \mu_i}_{\hat I_i \hat J_i} 
= B^{\hat R \to (R_i, T_i), \mu_i}_{\hat I_i \to (I_i, c_i)} \, 
(B^T)^{\hat R \to (R_i, T_i), \mu_i}_{\hat J_i \to (I_i, c_i)} 
&\quad \( {\rm for} \ \ \tilde C_{\circ \circ \circ} \)
\\[3mm]
\proj^{\hat R \to \bs{RT}_{i-,i+}}_{\hat I_i \hat J_i} 
= \cB^{\hat R \to \bs{RT}_{i-}}_{\hat I \longrightarrow (j,k,l,c)} \,
(\cB^T)^{\hat R \to \bs{RT}_{i +}}_{\hat J \longrightarrow (j,k,l,c)}
&\quad \( {\rm for} \ \ \tilde C^{XYZ}_{\vec h} \).
\end{cases}
\label{def:proj sub}
\end{equation}
The three-point function \eqref{CXZm comp} becomes
\begin{multline}
\tilde C_{123} = 
\( \prod_{i=1}^3 \frac{L_i!}{l_i! \, m_i! \, n_i! \, \olL_i ! } \) \frac{1}{\hL!} \ 
\sum_{\{U_i \} \in S_L^{\otimes 3} }
\( \prod_{p=1}^{\hL} \, h^{\hat A^{(1)}_{U_1(p)} \hat A^{(2)}_{U_2(p)} \hat A^{(3)}_{U_3(p)} } \)
\sum_{\hat R \, \vdash \hL} 
\frac{\Dim_{N_c} (\hat R)}{d_{R_1} d_{R_2} d_{R_3}} \ \times
\\[1mm]
\sum_{ \{ T_i , \mu_i \} } \( \prod_{i=1}^3 
\proj^{\hat R \to \, {\rm sub}}_{\hat I_i \hat J_i} \) 
D^{\hat R}_{\hat J_1\hat I_2} (U_1 U_2^{-1}) \,
D^{\hat R}_{\hat J_2\hat I_3} (U_2 U_3^{-1}) \,
D^{\hat R}_{\hat J_3\hat I_1} ( U_3 U_1^{-1}) 
\label{CXZm comp2}
\end{multline}
where \eqref{def:extended character} is used to sum over $t_i$\,.

\subsection{Sum over Wick contractions}\label{sec:sum Wick}

We simplify the sum over the Wick contractions, denoted by $\{U_i \} \in S_L^{\otimes 3}$ in \eqref{CXZm comp2}.

\subsubsection{Symmetry of the permutation formula}

To begin with, let us review the symmetry in the permutation formula \eqref{SL tree n-pt} for a fixed $\{ U_i \}$,
\begin{equation}
C_{123} ( \{ U_i\} ) =
\frac{1}{\olL_1 ! \, \olL_2 ! \, \olL_3 ! \, \hL!} \ 
\( \prod_{p=1}^{\hL} \, h^{\hat A^{(1)}_{U_1(p)} \hat A^{(2)}_{U_2(p)} \hat A^{(3)}_{U_3(p)} } \) 
N_c^{C ( U_1^{-1} \, \hat \alpha_1 \, U_1 U_2^{-1} \, \hat \alpha_2 \, U_2 U_3^{-1} \, \hat \alpha_3 \, U_3 ) } \,.
\label{SL tree n-pt summand}
\end{equation}
Since $\tilde C_{123}$ is a linear combination of $C_{123}$\,, the equation \eqref{CXZm comp2} should inherit the same symmetry.

First, $C_{123} ( \{ U_i\} )$ is invariant under the simultaneous transformation 
\begin{equation}
\( U_1 \,, U_2 \,, U_3 \) \ \mapsto \ \(U_1 V_0 \,,\, U_2 V_0 \,,\, U_3 V_0 \), \qquad
\forall V_0 \in S_{\hL} 
\label{V0 redundancy}
\end{equation}
which corresponds to the relabeling $p \mapsto V_0 (p)$ in \eqref{SL tree n-pt summand}.
Second, $C_{123} ( \{ U_i\} )$ is invariant under the permutation of identity fields
\begin{equation}
\begin{aligned}
\( U_1 \,, U_2 \,, U_3 \) \ &\mapsto \ \( V_1 U_1 , \, V_2 U_2, \, V_3 U_3 \)
\\[1mm]
\( V_1 \,, V_2 \,, V_3 \) \ &\in \ \( 
{\bf 1}_{L_1} \otimes S_{\olL_1} \,,
{\bf 1}_{L_2} \otimes S_{\olL_2} \,,
{\bf 1}_{L_3} \otimes S_{\olL_3} \) \ \subset \ S_{\hL}^{\otimes 3} 
\end{aligned}
\label{identity redundancy}
\end{equation}
which follows from the definition $\hat \alpha_i = \alpha_i \circ {\bf 1}_{\olL_i}$\,.
Third, $C_{123} ( \{ U_i\} )$ is invariant under the flavor symmetry \eqref{def:symm multi-trace},
\begin{equation}
\begin{aligned}
\( U_1 \,, U_2 \,, U_3 \) \ &\mapsto \ \( W_1 U_1 , \, W_2 U_2, \, W_3 U_3 \),
\\[1mm]
\( W_1 \,, W_2 \,, W_3 \) \ &\in \ \( 
S_{l_1} \otimes S_{m_1} \otimes S_{n_1} \otimes {\bf 1}_{\olL_1} \,,
S_{l_2} \otimes S_{m_2} \otimes S_{n_2} \otimes {\bf 1}_{\olL_2} \,,
S_{l_3} \otimes S_{m_3} \otimes S_{n_3} \otimes {\bf 1}_{\olL_3} \) 
\end{aligned}
\label{flavor symmetry C123}
\end{equation}
The redundancy \eqref{V0 redundancy} and \eqref{identity redundancy} are unphysical, which should be canceled by the numerical factors $\olL!$ and $\prod_i \olL_i!$ in \eqref{SL tree n-pt summand}.
The last operation \eqref{flavor symmetry C123} is the symmetry of the external operators, and interchanges different Wick contractions.

\subsubsection{Fixing redundancy}

Let us rewrite the flavor factor $\prod_p h^{ABC}$ in \eqref{SL tree n-pt summand} as
\begin{equation}
\fH \[ \hA^{(i)}_{U_i (p)} \] \equiv
\prod_{p=1}^{\hL} \, h^{\hat A^{(1)}_{U_1(p)} \hat A^{(2)}_{U_2(p)} \hat A^{(3)}_{U_3(p)} } 
\label{def:fH}
\end{equation}
where $\[ \hA^{(i)}_{U_i (p)} \]$ is the $3 \times \hL$ Wick-contraction matrix,\footnote{Each element of this matrix represents the flavor data. Note that this notation is slightly different from \cite{Suzuki:2018aep}, where the Wick-contraction matrix is defined by the color data.}
\begin{equation}
\[ \hA^{(i)}_{U_i (p)} \] = 
\begin{bmatrix}
\hA^{(1)}_{U_1(1)} & \hA^{(1)}_{U_1(2)} & \dots  & \hA^{(1)}_{U_1(\hL)} \\[2mm]
\hA^{(2)}_{U_2(1)} & \hA^{(2)}_{U_2(2)} & \dots  & \hA^{(2)}_{U_2(\hL)} \\[2mm]
\hA^{(3)}_{U_3(1)} & \hA^{(3)}_{U_3(2)} & \dots  & \hA^{(3)}_{U_3(\hL)} 
\end{bmatrix} .
\label{def:Wick mat}
\end{equation}
Note that the position of each column is unimportant for computing the flavor factor \eqref{def:fH},
\begin{equation}
\[ \hA^{(i)}_{U_i (p)} \] \ \simeq \ \[ \hA^{(i)}_{U_i (\sigma(p))} \] , \qquad \forall \sigma \in S_{\hL} \,.
\label{equivalence Wick mat}
\end{equation}

We fix the redundancy of $V_0$ in \eqref{V0 redundancy} as follows. 
Let us choose the position of the identity fields for each operator as
\begin{equation}
\begin{aligned}
\Phi^{\hat A_p^{(1)}} &= {\bf 1}_p \,, \qquad (p=1,2,\dots, \olL_1)
\\
\Phi^{\hat A_p^{(2)}} &= {\bf 1}_p \,, \qquad (p=\olL_1+1,\olL_1+2,\dots, \olL_1+\olL_2)
\\
\Phi^{\hat A_p^{(3)}} &= {\bf 1}_p \,, \qquad (p=\olL_1+\olL_2+1,\olL_1+\olL_2+2,\dots, \hL) .
\end{aligned}
\label{gauge fix partial}
\end{equation}
Here the subscript of ${\bf 1}$ is a dummy index, which will disappear after the identification \eqref{equivalence Wick mat}.
The Wick-contraction matrix becomes
\begin{equation}
\[ \hA^{(i)}_{U_i (p)} \] = 
\begin{bmatrix}
{\bf 1}_1 & \dots  & {\bf 1}_{\olL_1} & \hA^{(1)}_{U_1(\olL_1+1)} & \dots & \hA^{(1)}_{U_1(L_3)}  & 
\hA^{(1)}_{U_1(L_3+1)} & \dots & \hA^{(1)}_{U_1(\hL)} \\[2mm]
\hA^{(2)}_{U_2(1)} & \dots  & \hA^{(2)}_{U_2(\olL_1)} & {\bf 1}_{\olL_1+1} & \dots & {\bf 1}_{L_3} &
\hA^{(2)}_{U_2(L_3+1)} & \dots & \hA^{(2)}_{U_2(\hL)} \\[2mm]
\hA^{(3)}_{U_3(1)} & \dots  & \hA^{(3)}_{U_3(\olL_1)} & 
\hA^{(3)}_{U_3(\olL_1+1)} & \dots & \hA^{(3)}_{U_3(L_3)} & {\bf 1}_{L_3+1} & \dots & {\bf 1}_{\hL}
\end{bmatrix} .
\label{Wick mat part}
\end{equation}
The residual redundancy of $V_0$ is now $V'_0 \in S_{\olL_1} \otimes S_{\olL_2} \otimes S_{\olL_3}$\,.

After the partial gauge fixing \eqref{gauge fix partial}, $\{ U_i \}$ permute the non-identity fields only,
\begin{equation}
U_1 \ \in \ S_{L_1} \otimes {\bf 1}_{\olL_1} \,, \qquad
U_2 \ \in \ S_{L_2} \otimes {\bf 1}_{\olL_2} \,, \qquad
U_3 \ \in \ S_{L_3} \otimes {\bf 1}_{\olL_3} \,.
\label{partial Ui ranges}
\end{equation}
There is still residual redundancy generated by a combination of $V'_0$ and $V_i$ in \eqref{identity redundancy},
\begin{equation}
\tilde V \, : \, \{ U_i \} \ \mapsto \ \{ U'_i \}, \qquad
\hA^{(i)}_{U'_i (p)}  =
\begin{cases}
{\bf 1}_p &\quad ({\rm if} \ \ \hA^{(i)}_{U_i(p)} = {\bf 1}_p) \\[1mm]
\hA^{(i)}_{\tilde V^{-1} U_i \tilde V (p)} &\quad ({\rm if} \ \ \hA^{(i)}_{U_i(p)} \neq {\bf 1}_p)
\end{cases}
\label{residual redundancy}
\end{equation}
for any $\tilde V \in S_{\olL_1} \otimes S_{\olL_2} \otimes S_{\olL_3}$\,.
This map does not permute identity fields, but permutes the non-identity fields sitting in the same column.

\subsubsection{Counting inequivalent Wick contractions}

We pick up one set of partially gauge-fixed permutations $\{ U_i^\bullet \}$ such that $\prod_{p=1}^{\hL} \, h^{\hat A^{(1)}_{U_1^\bullet(p)} \hat A^{(2)}_{U_2^\bullet(p)} \hat A^{(3)}_{U_3^\bullet(p)} } \neq  0$.
We generate other $\{ U_i \}$ by applying the flavor symmetry, $U_i^\bullet \to W_i U_i^\bullet$ in \eqref{flavor symmetry C123}.

\bigskip
This procedure generates all non-vanishing Wick pairings.
To show this, consider two sets of permutations $\{ U_i^\bullet \}$ and $\{ U_i^\circ \}$, both of which are subject to the partial gauge fixing \eqref{partial Ui ranges} and giving the non-vanishing flavor factor \eqref{def:fH}.
Define
\begin{equation}
U_i^\bullet \equiv W_i^{\bullet \circ} U_i^\circ \,, \qquad
W_i^{\bullet \circ} \in S_{L_i} \otimes {\bf 1}_{\olL_i} \,.
\label{def: Woo}
\end{equation}
Since any permutation consists of a product of transpositions, we may assume $\( W_1^{\bullet \circ} , W_2^{\bullet \circ}, W_3^{\bullet \circ} \) = \( (ab), {\bf 1}, {\bf 1} \) \in S_{L_1} \otimes S_{L_2} \otimes S_{L_3}$ without loss of generality.
Let us represent the Wick contractions of $\{ U_i^\bullet \}$ by
\begin{multline}
\contraction{\langle \tr (}{\Phi}{^{\hA^{(1)}_a} \, \Phi^{\hA^{(1)}_b} \dots ) \, \tr (}{\Phi}
\contraction{\langle \tr (\Phi^{\hA^{(1)}_a} \, \Phi^{\hA^{(1)}_b} \dots ) \, \tr (}{\Phi}{^{\hA^{(2)}_c} \, \Phi^{\hA^{(2)}_d} \dots ) \, \tr (}{\Phi}
\contraction[2ex]{\langle \tr (\Phi^{\hA^{(1)}_a} \, }{\Phi}{^{\hA^{(1)}_b} \dots ) \, \tr (\Phi^{\hA^{(2)}_c} \, }{\Phi}
\contraction[2ex]{\langle \tr (\Phi^{\hA^{(1)}_a} \, \Phi^{\hA^{(1)}_b} \dots ) \, \tr (\Phi^{\hA^{(2)}_c} \, }{\Phi}{^{\hA^{(2)}_d} \dots ) \, \tr (\Phi^{\hA^{(3)}_e} \, }{\Phi}
\langle \tr (\Phi^{\hA^{(1)}_a} \, \Phi^{\hA^{(1)}_b} \dots ) \, \tr (\Phi^{\hA^{(2)}_c} \, \Phi^{\hA^{(2)}_d} \dots ) \, \tr (\Phi^{\hA^{(3)}_e} \, \Phi^{\hA^{(3)}_f} \dots ) \rangle 
\\
= \vev{ \Phi^{A^{(1)}_a} \Phi^{A^{(2)}_c} \Phi^{A^{(3)}_e} } 
\vev{ \Phi^{A^{(1)}_b} \Phi^{A^{(2)}_d} \Phi^{A^{(3)}_f} } \dots
\neq 0.
\label{tripleWick U-1}
\end{multline}
Then, the Wick contractions of $\{ U_i^\circ \}$ are written as
\begin{multline}
\contraction{\langle \tr (\Phi^{\hA^{(1)}_a} \, }{\Phi}{^{\hA^{(1)}_b} \dots ) \, \tr (}{\Phi}
\contraction{\langle \tr (\Phi^{\hA^{(1)}_a} \, \Phi^{\hA^{(1)}_b} \dots ) \, \tr (}{\Phi}{^{\hA^{(2)}_c} \, \Phi^{\hA^{(2)}_d} \dots ) \, \tr (}{\Phi}
\contraction[2ex]{\langle \tr (}{\Phi}{^{\hA^{(1)}_a} \, \Phi^{\hA^{(1)}_b} \dots ) \, \tr (\Phi^{\hA^{(2)}_c} \, }{\Phi}
\contraction[2ex]{\langle \tr (\Phi^{\hA^{(1)}_a} \, \Phi^{\hA^{(1)}_b} \dots ) \, \tr (\Phi^{\hA^{(2)}_c} \, }{\Phi}{^{\hA^{(2)}_d} \dots ) \, \tr (\Phi^{\hA^{(3)}_e} \, }{\Phi}
\langle \tr (\Phi^{\hA^{(1)}_a} \, \Phi^{\hA^{(1)}_b} \dots ) \, \tr (\Phi^{\hA^{(2)}_c} \, \Phi^{\hA^{(2)}_d} \dots ) \, \tr (\Phi^{\hA^{(3)}_e} \, \Phi^{\hA^{(3)}_f} \dots ) \rangle 
\\
= \vev{ \Phi^{A^{(1)}_b} \Phi^{A^{(2)}_c} \Phi^{A^{(3)}_e} } 
\vev{ \Phi^{A^{(1)}_a} \Phi^{A^{(2)}_d} \Phi^{A^{(3)}_f} } \dots
\neq 0.
\label{tripleWick U-2}
\end{multline}
Since both \eqref{tripleWick U-1} and \eqref{tripleWick U-2} are non-zero, and since $\Phi = (X,Y,Z)$ have orthogonal inner products, we should have $\Phi^{A^{(1)}_a}=\Phi^{A^{(1)}_b}$\,.
This implies that $W_i^{\bullet \circ} \in S_{l_i} \otimes S_{m_i} \otimes S_{n_i} \otimes {\bf 1}_{\olL_i}$\,, which is part of the flavor symmetry \eqref{flavor symmetry C123}.

\bigskip
The range of $\{ U_i \}$ in \eqref{partial Ui ranges} now becomes
\begin{equation}
\begin{aligned}
U_1 \ &\in \ S_{l_1} \otimes S_{m_1} \otimes S_{n_1} \otimes {\bf 1}_{\olL_1} \equiv \cS_1
\\
U_2 \ &\in \ S_{l_2} \otimes S_{m_2} \otimes S_{n_2} \otimes {\bf 1}_{\olL_2} \equiv \cS_2
\\
U_3 \ &\in \ S_{l_3} \otimes S_{m_3} \otimes S_{n_3} \otimes {\bf 1}_{\olL_3} \equiv \cS_3
\end{aligned}
\label{inequivalent Ui ranges}
\end{equation}
The sum over $(\cS_1 , \cS_2, \cS_3)$ counts each inequivalent Wick pairing more than once.
The multiplicity comes from the residual redundancy \eqref{residual redundancy},
\begin{equation}
\abs{S_{\olL_1} \otimes S_{\olL_2} \otimes S_{\olL_3}} = \olL_1 ! \, \olL_2 ! \, \olL_3 ! \,.
\end{equation}
The number of inequivalent Wick contractions is given by
\begin{equation}
\abs{\rm Wick} \equiv
\abs{ \frac{\cS_1 \otimes \cS_2 \otimes \cS_3}{S_{\olL_1} \otimes S_{\olL_2} \otimes S_{\olL_3} } }
= \prod_{i=1}^3 \frac{l_i ! \, m_i ! \, n_i !}{\olL_i \!}
\label{def:abs Wick}
\end{equation}

\subsubsection{The OPE coefficients simplified}

We collected all non-vanishing Wick contractions by restricting the sum $\{ U_i \}$ over the ranges \eqref{inequivalent Ui ranges}. The OPE coefficient \eqref{CXZm comp2} becomes
\begin{multline}
\tilde C_{123} = 
\( \prod_{i=1}^3 \frac{L_i!}{l_i! \, m_i! \, n_i! \, \olL_i ! } \) \ 
\sum_{\hat R \, \vdash \hL} 
\frac{\Dim_{N_c} (\hat R)}{d_{R_1} d_{R_2} d_{R_3}} \ \times
\\[1mm]
\sum_{ \{ T_i , \mu_i \} } \( \prod_{i=1}^3 
\proj^{\hat R \to \, {\rm sub}}_{\hat I_i \hat J_i} \) 
\sum_{U_1 \in \cS_1 } \ \sum_{U_2 \in \cS_2 } \ \sum_{U_3 \in \cS_3 } \ 
D^{\hat R}_{\hat J_1\hat I_2} (U_1 U_2^{-1}) \,
D^{\hat R}_{\hat J_2\hat I_3} (U_2 U_3^{-1}) \,
D^{\hat R}_{\hat J_3\hat I_1} ( U_3 U_1^{-1}) .
\label{CXZm comp3a}
\end{multline}
Recall that the projector is equal to the product of branching coefficients, $\proj = \cB \, \cB^T$ as in \eqref{def:proj sub}.
We can simplify the second line by using the identity of branching coefficients 
\eqref{perm matrix after branching2} 
\begin{equation}
\sum_{\hat J} D^{\hat R}_{\hat I \hat J} (u \circ v \circ w) \, B^{\hat R \to (q,r,s) \nu}_{\hat J \to (j,k,l)}
= \sum_{a,b,c} D^{q}_{aj} (u) \, D^{r}_{bk} (v) \, D^{s}_{cl} (w) \, 
B^{\hat R \to (q,r,s) \nu}_{\hat I \to (a,b,c)} \,.
\end{equation}
If we bring $U_k = u_k \otimes v_k \otimes w_k$ and $U_k^{-1} = u_k^{-1} \otimes v_k^{-1} \otimes w_k^{-1}$ across the double branching coefficients $\cB$ or $\cB^T$\,, they annihilate each other; see \eqref{CXZm comp quiver}.

Let us define a triple-projector product
\begin{equation}
\cI_{123}^{\hat R \to \, {\rm sub}} \equiv 
\proj_{\hat I_1 \hat I_2}^{\hat R \to \, {\rm sub}} \,
\tilde{\proj}_{\hat I_2 \hat I_3}^{\hat R \to \, {\rm sub}} \,
\dbtilde{\proj}_{\hat I_3 \hat I_1}^{\hat R \to \, {\rm sub}}
\label{def: cI123}
\end{equation}
where we used the symbols $\tilde \proj$ and $\dbtilde{\proj}$ to keep in mind that the branching coefficients come from different restrictions of $S_{\hL}$\,.
Then 
\begin{equation}
\begin{aligned}
\tilde C_{123} &= 
\( \prod_{i=1}^3 \frac{L_i!}{l_i! \, m_i! \, n_i! } \) \abs{\rm Wick} \,
\sum_{\hat R \, \vdash \hL} 
\frac{\Dim_{N_c} (\hat R)}{d_{R_1} d_{R_2} d_{R_3}} \,
\sum_{ \{ T_i , \mu_i \} } \cI_{123}^{\hat R \to \, {\rm sub}}
\\[1mm]
&= \( \prod_{i=1}^3 \frac{L_i!}{\olL_i!} \)
\sum_{\hat R \, \vdash \hL} 
\frac{\Dim_{N_c} (\hat R)}{d_{R_1} d_{R_2} d_{R_3}} \,
\sum_{ \{ T_i , \mu_i \} } \cI_{123}^{\hat R \to \, {\rm sub}}
\end{aligned}
\label{CXZm comp3}
\end{equation}
where we used \eqref{def:abs Wick}.

In the notation of the quiver calculus in Appendix \ref{app:quiver calc}, we can express the above calculation as
\begin{align}
\tilde C_{123} &\sim 
\sum_{\hat R \, \vdash \hL} \frac{\Dim_{N_c} (\hat R)}{d_{R_1} d_{R_2} d_{R_3}} \ 
\sum_{ \{ U_i \in \cS_i \} } \ 
\begin{tikzpicture}[node distance=10mm, baseline=(current bounding box.center)]
\begin{scope}[decoration={markings, mark=at position 0.7 with {\arrow{Straight Barb[]}}}]
\node (proj) {};
\node (u13) [midperm, shape border rotate=90, below of=proj, yshift=-10mm] {};
\node (u32) [midperm, shape border rotate=210, above right of=proj, xshift=14mm, yshift=8mm] {};
\node (u21) [midperm, shape border rotate=330, above left of=proj, xshift=-14mm, yshift=8mm] {};
\node (u13id) at (u13) {$U_1^{-1} U_3$};
\node (u32id) at (u32) {$U_2 U_3^{-1}$};
\node (u21id) at (u21) {$U_1 U_2^{-1}$};
\node (br1p) [branch, left of=u13, xshift=-10mm] {$\nu_{1+}$};
\node (br3m) [branch3, right of=u13, xshift=10mm] {$\nu_{3-}$};
\node (br3p) [branch3, below right of=u32, xshift=4mm, yshift=-8mm] {$\nu_{3+}$};
\node (br2m) [branch2, above left of=u32, xshift=-4mm, yshift=8mm] {$\nu_{2-}$};
\node (br2p) [branch2, above right of=u21, xshift=4mm, yshift=8mm] {$\nu_{2+}$};
\node (br1m) [branch, below left of=u21, xshift=-4mm, yshift=-8mm] {$\nu_{1-}$};
\draw [lineR] (u13.top side) -- (br1p);
\draw [lineR] (br3m) -- (u13.bottom side);
\draw [lineR] (u32.top side) -- (br3p);
\draw [lineR] (br2m) -- (u32.bottom side);
\draw [lineR] (u21.top side) -- (br2p);
\draw [lineR] (br1m) -- (u21.bottom side);
\draw [liner1] (br1p) .. controls (-1.9, -.6) .. (br1m);
\draw [liner2] (br1p) .. controls (-3.2, -1.2) .. (br1m);
\draw [liner1] (br3p) .. controls (1.9, -.6) .. (br3m);
\draw [liner2] (br3p) .. controls (3.2, -1.2) .. (br3m);
\draw [liner1] (br2p) .. controls (0,2.4) .. (br2m);
\draw [liner2] (br2p) .. controls (0,3.6) .. (br2m);
\draw [lines2] (br1p) -- (br1m);
\draw [lines2] (br2p) -- (br2m);
\draw [lines2] (br3p) -- (br3m);
\end{scope}
\end{tikzpicture}
\notag \\[2mm]
&\sim
\sum_{\hat R \, \vdash \hL} \frac{\Dim_{N_c} (\hat R)}{d_{R_1} d_{R_2} d_{R_3}} \ 
\abs{\rm Wick} \ \ 
\begin{tikzpicture}[node distance=10mm, baseline=(current bounding box.center)]
\begin{scope}[decoration={markings, mark=at position 0.7 with {\arrow{Straight Barb[]}}}]
\node (proj) {};
\node (u13) {};
\node (u32) {};
\node (u21) {};
\node (br1m) [branch, left of=u13, xshift=-8mm] {$\nu_{1-}$};
\node (br3p) [branch3, right of=u13, xshift=8mm] {$\nu_{3+}$};
\node (br3m) [branch3, below right of=u32, xshift=4mm, yshift=-8mm] {$\nu_{3-}$};
\node (br2p) [branch2, above left of=u32, xshift=-3mm, yshift=8mm] {$\nu_{2+}$};
\node (br2m) [branch2, above right of=u21, xshift=3mm, yshift=8mm] {$\nu_{2-}$};
\node (br1p) [branch, below left of=u21, xshift=-4mm, yshift=-8mm] {$\nu_{1+}$};
\draw [lineR] (br1m) -- (br2p);
\draw [lineR] (br2m) -- (br3p);
\draw [lineR] (br3m) -- (br1p);
\draw [liner1] (br1p) .. controls (-1, -.6) .. (br1m);
\draw [liner2] (br1p) .. controls (-1.9, -.9) .. (br1m);
\draw [liner1] (br3p) .. controls (1, -.6) .. (br3m);
\draw [liner2] (br3p) .. controls (1.9, -.9) .. (br3m);
\draw [liner1] (br2p) .. controls (0,1) .. (br2m);
\draw [liner2] (br2p) .. controls (0,2.1) .. (br2m);
\draw [lines2] (br1p) -- (br1m);
\draw [lines2] (br2p) -- (br2m);
\draw [lines2] (br3p) -- (br3m);
\end{scope}
\end{tikzpicture} 
\label{CXZm comp quiver}
\end{align}
From this diagram, we see that $\cI_{123}^{\hat R \to \, {\rm sub}}$ in \eqref{def: cI123} is also a triple product of the transformation matrices \eqref{def:intertwining mat}.

\subsection{Sum over the triple-projector products}\label{sec:sum projectors}

We compute the OPE coefficients by evaluating a sum over the triple-projector products,
\begin{equation}
\sum_{ \{ T_i , \mu_i \} } \cI_{123}^{\hat R \to \, {\rm sub}}
=
\sum_{ T_1 \, \vdash \olL_1 }
\sum_{ T_2 \, \vdash \olL_2 }
\sum_{ T_3 \, \vdash \olL_3 }
\sum_{\mu_1, \mu_2 , \mu_3}
\proj_{\hat I_1 \hat I_2}^{\hat R \to \, {\rm sub}} \,
\tilde{\proj}_{\hat I_2 \hat I_3}^{\hat R \to \, {\rm sub}} \,
\dbtilde{\proj}_{\hat I_3 \hat I_1}^{\hat R \to \, {\rm sub}}
\label{cI123 comp}
\end{equation}
where the projector is given by \eqref{def:proj sub}. 
The main idea is to decompose each projector further into a sum of sub-projectors, so that we can make use of the orthogonality of the sub-projectors on the fully-split space, $V_{FS}$.

Below we discuss the two cases $\tilde C_{\circ\circ\circ}$ in \eqref{def:tilde Cooo} and $\tilde C^{XYZ}_{\vec h}$ in \eqref{def:tilde CXYZ} separately.

\subsubsection{Case of $\tilde C_{\circ\circ\circ}$}\label{sec:tilde Cooo final}

Recall that $\tilde C_{\circ\circ\circ}$ is a linear combination of $C_{\circ\circ\circ}$ given in \eqref{def:Cooo}.
The Wick-contraction matrix of $C_{\circ\circ\circ}$ after a partial gauge-fixing \eqref{Wick mat part} is given by
\begin{equation}
\[ \hA^{(i)}_{U_i (p)} \] = 
\begin{bmatrix}
{\bf 1}_1 & \dots  & {\bf 1}_{\olL_1} & Z_{U_1(\olL_1+1)} & \dots & Z_{U_1(L_3)}  & 
Z_{U_1(L_3+1)} & \dots & Z_{U_1(\hL)} \\[2mm]
\tilde Z_{U_2(1)} & \dots  & \tilde Z_{U_2(\olL_1)} & {\bf 1}_{\olL_1+1} & \dots & {\bf 1}_{L_3} &
\tilde Z_{U_2(L_3+1)} & \dots & \tilde Z_{U_2(\hL)} \\[2mm]
\olZ_{U_3(1)} & \dots  & \olZ_{U_3(\olL_1)} & 
\olZ_{U_3(\olL_1+1)} & \dots & \olZ_{U_3(L_3)} & {\bf 1}_{L_3+1} & \dots & {\bf 1}_{\hL}
\end{bmatrix} 
\label{Wick mat part Cooo}
\end{equation}
which shows that $\cS_i = S_{L_i} \otimes S_{\olL_i}$ in place of \eqref{inequivalent Ui ranges}.
We represent \eqref{Wick mat part Cooo} as in the following figure,
\begin{equation}
\begin{tikzpicture}[node distance=10mm, baseline=(current bounding box.center)]
\node at (-3.5,2) {$\hat \cO_1$};
\node at (-3.5,0) {$\hat \cO_2$};
\node at (-3.5,-2) {$\hat \cO_3$};
\draw (-2.2,2.5) -- (-2.2,-3.5);
\draw (-1.2,2.5) -- (-1.2,-3.5);
\draw (0,2.5) -- (0,-3.5);
\draw (2.4,2.5) -- (2.4,-3.5);
\draw[thick] (-1.2,2.3) rectangle (0,1.7);
\filldraw[fill=white, thick] (-1.2,2.3) rectangle (2.4,1.7);
\filldraw[fill=black!35, thick] (-2.2,0.3) rectangle (-1.2,-0.3);
\filldraw[fill=black!35, thick] (0,0.3) rectangle (2.4,-0.3);
\filldraw[fill=black!80, thick] (-2.2,-1.7) rectangle (0,-2.3);
\filldraw[fill=white, thick] (-2.2,1.3) rectangle (2.4,0.7);
\filldraw[fill=white, thick] (-2.2,-0.7) rectangle (2.4,-1.3);
\filldraw[fill=white, thick] (-2.2,-2.7) rectangle (2.4,-3.3);
\node at (.6,2) {$Z$};
\node [white] at (-1.7,0) {$\tilde Z$};
\node [white] at (1.2,0) {$\tilde Z$};
\node [white] at (-1.1,-2) {$\olZ$};
\node at (-1.7,2) {${\bf 1}$};
\node at (-.6,0) {${\bf 1}$};
\node at (1.2,-2) {${\bf 1}$};
\node at (0,1) {\small $U_1 U_2^{-1}$};
\node at (0,-1) {\small $U_2 U_3^{-1}$};
\node at (0,-3) {\small $U_3 U_1^{-1}$};
\node at (-1.55,-3.85) {\small $1,2, \dots$};
\node at (1.85,-3.85) {\small $\dots,\hL$};
\end{tikzpicture}
\label{Wick structure Cooo}
\end{equation}

Let us choose the fully-split space as
\begin{equation}
V_{FS} = V_{\olL_1} \otimes V_{\olL_2} \otimes V_{\olL_3} 
\end{equation}
which induces the restriction $S_{\hL} \downarrow S_{FS}$\,, where
\begin{equation}
S_{FS} = S_{\olL_1} \otimes S_{\olL_2} \otimes S_{\olL_3} \,.
\end{equation}
On the space $V_{FS}$\,, the states decompose as
\begin{equation}
\ket{ \atop{\hat R}{\, \hat I} } 
= \ket{ \matop{R_i & T_i}{I_i & c_i}{\mu_i} }
(B^T)^{\hat R \to (R_i,T_i), \mu_i}_{\hat I \to (I_i,c_i)}
= \ket{\matop{Q_i & Q'_i & T_i}{b_i & b'_i & c_i}{\mu_i \, \rho_i}}
(B^T)^{\hat R \to (R_i,T_i), \mu_i}_{\hat I \to (I_i,c_i)} \, 
(B^T)^{R_i \to (Q_i, Q'_i), \rho_i}_{I_i \to (b_i,b'_i)} 
\label{double branched state}
\end{equation}
where we used \eqref{app:branching coeffs}.
We introduce the fully-split branching coefficients by
\begin{align}
\fB^{\hat R \to (R_i,T_i), \mu_i \to (Q_i,Q'_i,T_i), (\mu_i,\rho_i)}_{\hat I \longrightarrow (b_i,b'_i,c_i)}
= \sum_{I_i=1}^{d_{R_i}} B^{\hat R \to (R_i,T_i), \mu_i}_{\hat I \to (I_i,c_i)} \,
B^{R_i \to (Q_i, Q'_i), \rho_i}_{I_i \to (b_i,b'_i)}
\end{align}
and the corresponding sub-projector by
\begin{multline}
\fP_{\hat I \hat J} ^{\hat R \to (R_i,T_i), \mu_i \to (Q_i,Q'_i,T_i), (\mu_i,\rho_i)} 
\\
= \sum_{b,b',c} \fB^{\hat R \to (R_i,T_i), \mu_i \to (Q_i,Q'_i,T_i), (\mu_i,\rho_i)}_{\hat I \longrightarrow (b,b',c)}
(\fB^T)^{\hat R \to (R_i,T_i), \mu_i \to (Q_i,Q'_i,T_i), (\mu_i,\rho_i)}_{\hat J \longrightarrow (b,b',c)}
\end{multline}
We rewrite the original projectors in \eqref{def:proj sub} as a sum over sub-projectors on $V_{FS}$ as
\begin{equation}
\begin{aligned}
\proj_{\hat I \hat J}^{\hat R \to (R_1, T_1), \mu_1, \rho_1}
&= \sum_{Q_1,Q'_1,\rho_1}
\fP_{\hat I \hat J}^{\hat R \to (R_1,T_1), \mu_1 \to (Q_1,Q'_1,T_1), (\mu_1,\rho_1)}
\\
\tilde \proj_{\hat I \hat J}^{\hat R \to (R_2, T_2), \mu_2, \rho_2}
&= \sum_{Q_2,Q'_2,\rho_2}
\tilde \fP_{\hat I \hat J}^{\hat R \to (R_2,T_2), \mu_2 \to (Q_2,Q'_2,T_2), (\mu_2,\rho_2)} 
\\
\dbtilde{\proj}_{\hat I \hat J}^{\hat R \to (R_3, T_3), \mu_3, \rho_3}
&= \sum_{Q_3,Q'_3,\rho_3}
\dbtilde{\fP}_{\hat I \hat J}^{\hat R \to (R_3,T_3), \mu_3 \to (Q_3,Q'_3,T_3), (\mu_3,\rho_3)} \,.
\end{aligned}
\label{resolved proj Cooo}
\end{equation}
By construction, all sub-projectors follow from the same restriction
\begin{equation}
S_{\hL} \downarrow S_{FS}, \qquad
\hat R = \bigoplus_{Q,Q',T} \bigoplus_{\eta=1}^{g(Q,Q',T; \hat R)} \,
( Q \otimes Q' \otimes T )_\eta
\label{def:eta mult}
\end{equation}
and all sub-representations should be synchronized when evaluating $\cI_{123}^{\hat R \to \, {\rm sub}}$ in \eqref{cI123 comp}. The states can also be decomposed as
\begin{equation}
\ket{ \atop{\hat R}{\, \hat I} } 
= \ket{\matop{Q & Q' & T}{b & b' & c}{\eta}}
(B^T)^{\hat R \to (Q,Q',T), \eta}_{\hat I \to (b,b',c)} 
\label{double branched state2}
\end{equation}
in addition to \eqref{double branched state}. 
The consistency of the two decompositions suggests that the multiplicity labels can be rewritten as
\begin{equation}
\xi_i \equiv \{ \mu_i \,, \rho_i \}, \qquad
1 \le \xi_i \le g(Q_i,Q'_i;R_i) \, g(R_i,T_i;\hat R) .
\label{identify eta}
\end{equation}

In \eqref{resolved proj Cooo}, the representations $T_i$ come from the Fourier transform of identity fields ${\bf 1}$, and $Q_i, Q'_i$ come from the non-identity fields, $Z, \tilde Z, \olZ$\,. 
Since the OPE coefficient $C_{\circ\circ\circ}$ has the Wick-contraction structure given in \eqref{Wick structure Cooo}, we should identify the representations $\{ Q_i \,, Q'_i \,, T_i \}$ with those acting on the constituent of $V_{FS}$ as
\begin{equation}
\begin{aligned}
T_1 = Q'_2 = Q_3 \ &\in {\rm Hom} (V_{\olL_1})
\\[1mm]
Q_1 = T_2 = Q'_3 \ &\in {\rm Hom} (V_{\olL_2})
\\[1mm]
Q'_1 = Q_2 = T_3 \ &\in {\rm Hom} (V_{\olL_3}) .
\end{aligned}
\label{identify TQQ}
\end{equation}
We can show \eqref{identify TQQ} from another argument.
The triple-projector product is equal to the product of generalized Racah-Wigner tensors in Appendix \ref{app:genRW},
\begin{equation}
\trb{\, \hat R} \( 
\fP_{\hat I \hat J}^{\hat R \to \dots \to (Q_1,Q'_1,T_1), \xi_1 } \,
\tilde \fP_{\hat I \hat J}^{\hat R \to \dots \to (Q_2,Q'_2,T_2), \xi_2 } \,
\dbtilde{\fP}_{\hat I \hat J}^{\hat R \to \dots \to (Q_3,Q'_3,T_3), \xi_3 } 
\)
\\
= \tr ( U_{\hat R} \tilde U_{\hat R} \dbtilde{U}_{\hat R} )
\end{equation}
which we conjecture as \eqref{trUUU conjecture},
\begin{align}
\sum_{\xi_1 , \xi_2 , \xi_3}
\tr (U_{\hat R} \tilde U_{\hat R} \dbtilde{U}_{\hat R} )
&= \delta^{T_1 Q'_2} \, \delta^{Q'_2 Q_3} \,
\delta^{Q_1 T_2} \, \delta^{T_2 Q'_3} \,
\delta^{Q'_1 Q_2} \, \delta^{Q_2 T_3} \,
\( \prod_{i=1}^3 d_{Q_i} \) \, \cG_{123}
\label{def:cG123 Cooo} \\[1mm]
\cG_{123} &= \frac{g(Q_1,Q_2;R_{1}) \, g(R_{1}, Q_3; \hat R) \, 
g(Q_2,Q_3;R_{2}) \, g(R_{2}, Q_1; \hat R) \, 
g(Q_3,Q_1;R_{3}) \, g(R_{3}, Q_2; \hat R)}{g (Q_1, Q_2, Q_3; \hat R)^2} \,.
\notag
\end{align}

The three-point function \eqref{CXZm comp3} becomes
\begin{equation}
\tilde C_{\circ\circ\circ} = \( \prod_{i=1}^3 \frac{L_i!}{\olL_i!} \)
\sum_{\hat R \, \vdash \hL} 
\frac{\Dim_{N_c} (\hat R)}{d_{R_1} d_{R_2} d_{R_3}} \ 
\sum_{Q_1 \vdash \olL_2 } 
\sum_{Q_2 \vdash \olL_3 }
\sum_{Q_3 \vdash \olL_1 } 
\( \prod_{i=1}^3 d_{Q_i} \) \, \cG_{123} \,.
\label{Cooo comp I123}
\end{equation}
Here, the Littlewood-Richardson coefficients in $\cG_{123}$ put constraints on the sum over $\{ Q_i \}$.
In other words, we should find all $\{ Q_i \} = \{ Q_i^\star \}$ such that
\begin{equation}
R_1 = Q_1^\star \otimes Q_2^\star \,, \quad
R_2 = Q_2^\star \otimes Q_3^\star \,, \quad
R_3 = Q_3^\star \otimes Q_1^\star \,, \quad
\hat R = Q_1^\star \otimes Q_2^\star \otimes Q_3^\star 
\label{def:Q123 star}
\end{equation}
The conditions \eqref{def:Q123 star} can be summarized as
\begin{equation}
\begin{tikzpicture}[node distance=10mm, baseline=(current bounding box.center)]
\node at (-3.5,1) {$\widetilde{\cO}_1$};
\node at (-3.5,0) {$\widetilde{\cO}_2$};
\node at (-3.5,-1) {$\widetilde{\cO}_3$};
\draw (-2.2,1.6) -- (-2.2,-1.6);
\draw (-1.2,1.6) -- (-1.2,-1.6);
\draw (0,1.6) -- (0,-1.6);
\draw (2.4,1.6) -- (2.4,-1.6);
\draw[thick] (-1.2,1.3) rectangle (0,0.7);
\filldraw[fill=white, thick] (-1.2,1.3) rectangle (2.4,0.7);
\filldraw[fill=black!35, thick] (-2.2,0.3) rectangle (-1.2,-0.3);
\filldraw[fill=black!35, thick] (0,0.3) rectangle (2.4,-0.3);
\filldraw[fill=black!80, thick] (-2.2,-0.7) rectangle (0,-1.3);
\node at (.6,1) {$R_1$};
\node [white] at (1.8,0) {$R_2$};
\node [white] at (-1.1,-1) {$R_3$};
\node at (-1.7,1) {$Q_3^\star$};
\node at (-.6,0) {$Q_1^\star$};
\node at (1.2,-1) {$Q_2^\star$};
\draw[thick] (-2.2,-2) rectangle (2.4,-2.6);
\node at (0.1,-2.3) {$\hat R$};
\end{tikzpicture}
\label{Rep structure Cooo}
\end{equation}

\paragraph{Extremal case.}

As a check, consider the situation $L_1+L_2=L_3=\hL$. From \eqref{Rep structure Cooo}, this corresponds to
\begin{equation}
Q_2 = \emptyset, \qquad
R_1 = Q_1 \,, \qquad 
R_2 = Q_3 \,, \qquad
\hat R = R_3 \,.
\end{equation}
We get
\begin{equation}
\cG_{123} = \frac{g(R_{1}, Q_3; \hat R) \, \, g(R_{2}, Q_1; \hat R) \, g(Q_3,Q_1;R_{3}) }{g (Q_1, Q_3; \hat R)^2}
= g(R_1, R_2 ; R_3)
\end{equation}
and therefore
\begin{equation}
\tilde C_{\circ\circ\circ} = L_3! \, 
\frac{\Dim_{N_c} (R_3)}{ d_{R_3}} \, g(R_1, R_2 ; R_3) .
\label{extremal limit Cooo}
\end{equation}
This result agrees with the literature \cite{Corley:2001zk} including the normalization of the two-point function given in \eqref{BPS hw normalization}.

\subsubsection{Case of $\tilde C^{XYZ}_{\vec h}$}\label{sec:tilde CXYZ final}

Our discussion is quite parallel to Section \ref{sec:tilde Cooo final}.
Recall that $\tilde C^{XYZ}_{\vec h}$ is a linear combination of $C^{XYZ}_{\vec h}$ given in \eqref{def:CXYZ vec}.
We represent the Wick-contraction matrix by
\begin{equation}
\begin{tikzpicture}[node distance=10mm, baseline=(current bounding box.center)]
\tikzmath{\xm=-4; \xp=4.8; \yp=2.5; \ym=-3.5;}
\node at (-5.5,2) {$\hat \cO_1$};
\node at (-5.5,0) {$\hat \cO_2$};
\node at (-5.5,-2) {$\hat \cO_3$};
\draw (\xm,\yp) -- (\xm,\ym);
\draw (-2.2,\yp) -- (-2.2,\ym);
\draw (-1,\yp) -- (-1,\ym);
\draw (0,\yp) -- (0,\ym);
\draw (1.6,\yp) -- (1.6,\ym);
\draw (3.4,\yp) -- (3.4,\ym);
\draw (\xp,\yp) -- (\xp,\ym);
\filldraw[fill=black!80, thick] (-4,2.3) rectangle (-2.2,1.7);
\filldraw[fill=black!50, thick] (-1,2.3) rectangle (0,1.7);
\filldraw[fill=white, thick] (1.6,2.3) rectangle (\xp,1.7);
\filldraw[fill=black!80, thick] (-2.2,0.3) rectangle (-1,-0.3);
\filldraw[fill=black!50, thick] (-1,0.3) rectangle (1.6,-0.3);
\draw[thick] (1.6,0.3) rectangle (3.4,-0.3);
\filldraw[fill=black!80, thick] (-4,-1.7) rectangle (-1,-2.3);
\filldraw[fill=black!50, thick] (0,-1.7) rectangle (1.6,-2.3);
\draw[thick] (3.4,-1.7) rectangle (\xp,-2.3);
\filldraw[fill=white, thick] (\xm,1.3) rectangle (\xp,0.7);
\filldraw[fill=white, thick] (\xm,-0.7) rectangle (\xp,-1.3);
\filldraw[fill=white, thick] (\xm,-2.7) rectangle (\xp,-3.3);
\node [white] at (-3.1,2) {$\olX^{\ell_{31}-h_2}$};
\node [white] at (-1.6,0) {$\olX^{h_1}$};
\node [white] at (-2.5,-2) {$X^{\ell_{31}-h_2+h_1}$};
\node [white] at (-0.5,2) {$\olY^{h_3}$};
\node [white] at (0.3,0) {$Y^{\ell_{23}-h_1+h_3}$};
\node [white] at (0.8,-2) {$\olY^{\ell_{23}-h_1}$};
\node at (3.2,2) {$Z^{\ell_{12}-h_3+h_2}$};
\node at (2.5,0) {$\olZ^{\ell_{12}-h_3}$};
\node at (4.1,-2) {$\olZ^{h_2}$};
\node at (-1.6,2) {${\bf 1}$};
\node at (0.8,2) {${\bf 1}$};
\node at (-3.1,0) {${\bf 1}$};
\node at (4.1,0) {${\bf 1}$};
\node at (-0.5,-2) {${\bf 1}$};
\node at (2.5,-2) {${\bf 1}$};
\node at (0.4,1) {\small $U_1 U_2^{-1}$};
\node at (0.4,-1) {\small $U_2 U_3^{-1}$};
\node at (0.4,-3) {\small $U_3 U_1^{-1}$};
\node at (-3.3,-3.85) {\small $1,2, \dots$};
\node at (4.2,-3.85) {\small $\dots,\hL$};
\end{tikzpicture}
\label{Wick mat structure CXYZ}
\end{equation}
where $h_i$ are constrained by \eqref{ranges oh h},
\begin{equation}
0 \le h_1 \le \ell_{23} = \olL_1 \,, \qquad
0 \le h_2 \le \ell_{31} = \olL_2 \,, \qquad
0 \le h_3 \le \ell_{12} = \olL_3 \,.
\label{ranges oh h2}
\end{equation}

We choose the fully-split space as
\begin{equation}
V_{FS} = V_{\ell_{31}-h_2} \otimes V_{h_1} \otimes V_{h_3} \otimes V_{\ell_{23}-h_1} 
\otimes V_{\ell_{12}-h_3} \otimes V_{h_2}
\end{equation}
and decompose the original projectors \eqref{def:proj sub}.
From \eqref{Wick mat structure CXYZ}, one finds that the new branch coefficients are needed for
\begin{equation}
\begin{aligned}
S_{\ell_{12}-h_3 + h_2} \downarrow ( S_{\ell_{12}-h_3} \otimes S_{h_2} ) 
\qquad {\rm and } \qquad
S_{\ell_{23}} \downarrow ( S_{h_1} \otimes S_{\ell_{23}-h_1} ) 
&\qquad {\rm for} \ \ \cO_1 
\\
S_{\ell_{23}-h_1 + h_3} \downarrow ( S_{\ell_{23}-h_1} \otimes S_{h_3} ) 
\qquad {\rm and } \qquad
S_{\ell_{31}} \downarrow ( S_{\ell_{31}-h_2} \otimes S_{h_2} ) 
&\qquad {\rm for} \ \ \cO_2 
\\
S_{\ell_{31}-h_2 + h_1} \downarrow ( S_{\ell_{31}-h_2} \otimes S_{h_1} ) 
\qquad {\rm and } \qquad
S_{\ell_{12}} \downarrow ( S_{h_3} \otimes S_{\ell_{12}-h_3} ) 
&\qquad {\rm for} \ \ \cO_3 \,.
\end{aligned}
\end{equation}
For example, we rewrite the states for $\cO_1$ on the space $V_{FS}$ as
\begin{equation}
\begin{aligned}
\ket{ \atop{\hat R}{\, \hat I} } 
&= \ket{ \matop{R_1 & T_1}{I_1 & c_1}{\mu_1} }
(B^T)^{\hat R \to (R_1,T_1), \mu_1}_{\hat I \to (I_1,c_1)}
\\
&= \ket{ \matop{q_1  & r_1 & s_1 & T_1}{j_1 & k_1 & l_1 & c_1}{\mu_1  \, \nu_{1 \mp}} }
(B^T)^{\hat R \to (R_1,T_1), \mu_1}_{\hat I \to (I_1,c_1)}
(B^T)^{R_1 \to (q_1,r_1,s_1), \nu_{1 \mp}}_{I_1 \to (j_1,k_1,l_1)}
\\
&= \ket{ \matop{q_1 & r_1 & s'_1 & s''_1 & t'_1 & t''_1}{j_1 & k_1 & l'_1 & l''_1 & c'_1 & c''_1}{\mu_1  \, \nu_{1 \mp} \, \rho_1 \, \zeta_1 } } \ \times
\\
&\hspace{25mm}
\brT^{\hat R \to (R_1,T_1), \mu_1}_{\hat I \to (I_1,c_1)}
\brT^{R_1 \to (q_1,r_1,s_1), \nu_{1 \mp}}_{I_1 \to (j_1,k_1,l_1)}
\brT^{s_1 \to (s'_1,s''_1), \rho_1 }_{I_1 \to (l'_1,l''_1)}
\brT^{T_1 \to (t'_1,t''_1), \zeta_1 }_{c_1 \to (c'_1,c''_1)}
\end{aligned}
\end{equation}
and introduce the fully-split branching coefficients by
\begin{align}
\fB^{\hat R \to \dots \to (q_1 , r_1, s'_1 ,  s''_1 , t'_1 , t''_1), \mu_1, \nu_{1 \mp} , \rho_1 , \zeta_1}_{\hat I \longrightarrow (j_1 , k_1 , l'_1 , l''_1 , c'_1 , c''_1)}
= 
B^{\hat R \to (R_1,T_1), \mu_1}_{\hat I \to (I_1,c_1)}
B^{R_1 \to (q_1,r_1,s_1), \nu_{1 \mp}}_{I_1 \to (j_1,k_1,l_1)}
B^{s_1 \to (s'_1,s''_1), \rho_1 }_{I_1 \to (l'_1,l''_1)}
B^{T_1 \to (t'_1,t''_1), \zeta_1 }_{c_1 \to (c'_1,c''_1)} \,.
\end{align}
The original projector \eqref{def:proj sub} becomes a sum over the sub-projectors $\fP = \fB \, \fB^T$, 
\begin{equation}
\proj^{\hat R \to \bs{RT}_{1-,1+}}_{\hat I_1 \hat J_1} 
=
\sum_{s'_1,s''_1,t'_1,t''_1,\rho_1,\zeta_1}
\fP_{\hat I \hat J}^{\hat R \to \dots \to (q_1 , r_1, s'_1 ,  s''_1 , t'_1 , t''_1), \mu_1, \nu_{1 \mp} , \rho_1 , \zeta_1}
\end{equation}
and similarly
\begin{equation}
\begin{aligned}
\tilde{\proj}^{\hat R \to \bs{RT}_{2-,2+}}_{\hat I_2 \hat J_2} 
&=
\sum_{r'_2,r''_2,t'_2,t''_2,\rho_2,\zeta_2}
\fP_{\hat I \hat J}^{\hat R \to \dots \to (q_2 , r'_2, r''_2 ,  s_2 , t'_2 , t''_2), \mu_2, \nu_{2 \mp} , \rho_2 , \zeta_2}
\\[1mm]
\dbtilde{\proj}^{\hat R \to \bs{RT}_{3-,3+}}_{\hat I_3 \hat J_3}
&=
\sum_{q'_3,q''_3,t'_3,t''_3,\rho_3,\zeta_3}
\fP_{\hat I \hat J}^{\hat R \to \dots \to (q'_3 , q''_3, r_3 , s_3 , t'_3 , t''_3), \mu_3, \nu_{3 \mp} , \rho_3 , \zeta_3} \,.
\end{aligned}
\end{equation}
When summing over $\{ t'_i, t''_i \}$ we can forget the constraint $t'_i \otimes t''_i \simeq T_i$\,, because the OPE coefficient \eqref{CXZm comp3a} contains sums over $\{ T_i \}$.

All sub-projectors come from the irreducible decompositions of $\hat R$ under the restriction $S_{\hL} \downarrow S_{FS}$\,,
\begin{equation}
\hat R = \bigoplus_{q',q'',r',r'',s',s''} \bigoplus_{\eta=1}^{g(q',q'',r',r'',s',s''; \hat R)} \,
( q' \otimes q'' \otimes r' \otimes r'' \otimes s' \otimes s'' )_\eta
\label{def:eta mult}
\end{equation}
Since the OPE coefficient $C^{XYZ}_{\vec h}$ has the Wick contraction structure of \eqref{Wick mat structure CXYZ}, we should identify the representations as
\begin{alignat}{9}
q_1 = t'_2 = q'_3 \ &\in {\rm Hom} (V_{\ell_{31}-h_2}),
&\qquad
t'_1 = q_2 = q''_3 \ &\in {\rm Hom} (V_{h_1})
\notag \\[1mm]
r_1 = r'_2 = t'_3 \ &\in {\rm Hom} (V_{h_3}),
&\qquad
t''_1 = r''_2 = r_3 \ &\in {\rm Hom} (V_{\ell_{23}-h_1} )
\label{identify qqrrss} \\[1mm]
s'_1 = s_2 = t''_3 \ &\in {\rm Hom} (V_{\ell_{12}-h_3}) ,
&\qquad
s''_1 = t''_2 = s_3 \ &\in {\rm Hom} (V_{h_2})
\notag
\end{alignat}
and replace the multiplicity labels by 
\begin{equation}
\xi_{i \mp} = \pare{ \mu_i, \nu_{i \mp} , \rho_i , \xi_i }.
\label{identify etamp}
\end{equation}
Again, the trace over the product of sub-projectors is given by the generalized Racah-Wigner tensors \eqref{def:genRW2},
\begin{multline}
\trb{\, \hat R} \( 
\fP_{\hat I_1 \hat I_2}^{\hat R \to \dots \to (q_1 , r_1, s'_1 ,  s''_1 , t'_1 , t''_1), \xi_{1-}, \xi_{1+} } \,
\fP_{\hat I_2 \hat I_3}^{\hat R \to \dots \to (q_2 , r'_2, r''_2 ,  s_2 , t'_2 , t''_2), \xi_{2-}, \xi_{2+} } \,
\fP_{\hat I_3 \hat I_1}^{\hat R \to \dots \to (q'_3 , q''_3, r_3 , s_3 , t'_3 , t''_3), \xi_{3-}, \xi_{3+} }
\)
\\
= \tr ( W_{\hat R} \tilde W_{\hat R} \dbtilde{W}_{\hat R} ).
\end{multline}
From the identity of the projectors \eqref{restricted projector relation2}, this becomes
\begin{equation}
\begin{aligned}
\tr (W_{\hat R} \, \tilde W_{\hat R}  \, \dbtilde{W}_{\hat R} ) &= 
\Bigl( \cD_{123} \, d_{q_1} \, d_{q_2} \, d_{r_1} \, d_{r_3} \, d_{s_2} \, d_{s_3} \Bigr)
\delta^{\xi_{1-} \, \xi_{2+} } \,
\delta^{\xi_{2-} \, \xi_{3+} } \,
\delta^{\xi_{3-} \, \xi_{1+} } 
\\[1mm]
\cD_{123} &= \delta^{q_1 t'_2} \, \delta^{q_1 q'_3} \,
\delta^{t'_1 q_2} \, \delta^{q_2 q''_3} \,
\delta^{r_1 r'_2} \, \delta^{r_1 t'_3} \,
\delta^{t''_1 r_3} \, \delta^{''_2 r_3} \,
\delta^{s'_1 s_2} \, \delta^{s_2 t''_3} \, 
\delta^{s''_1 s_3} \, \delta^{t''_2 s_3} \,.
\end{aligned}
\end{equation}
We need to sum over the representations and multiplicity labels.
We conjecture that the result is given by \eqref{trWWW conjecture},
\begin{equation}
\begin{gathered}
\sum_{\xi_\mp, \xi'_\mp, \xi''_\mp} \tr (W_{\hat R} \, \tilde W_{\hat R}  \, \dbtilde{W}_{\hat R} ) 
=
\Bigl( \cD_{123} \, d_{q_1} \, d_{q_2} \, d_{r_1} \, d_{r_3} \, d_{s_2} \, d_{s_3} \Bigr)
\oldelta^{\, \nu_{1-} \, \nu_{2+} } \,
\oldelta^{\, \nu_{2-} \, \nu_{3+} } \,
\oldelta^{\, \nu_{3-} \, \nu_{1+} } \, \cG_{123}
\\
\cG'_{123} = \frac{
\abs{\cM_{R_1, s_1,\nu_{1-}}} \abs{\cM_{R_1, s_1,\nu_{1+}}} 
\abs{\cM_{R_2, r_2,\nu_{2-}}}\abs{\cM_{R_2, r_2,\nu_{2+}}} 
\abs{\cM_{R_3,q_3,\nu_{3-}}}\abs{\cM_{R_3, q_3,\nu_{3+}}} }{\abs{\cM_{\rm tot}}^3 } 
\end{gathered}
\label{def:cG'123}
\end{equation}
where $\cM_{R, r,\nu}$ is the slice of the total multiplicity space constrained by $(R,r,\nu)$.

The three-point function \eqref{CXZm comp3} becomes
\begin{equation}
\tilde C^{XYZ}_{\vec h} = \( \prod_{i=1}^3 \frac{L_i!}{\olL_i!} \)
\sum_{\hat R \, \vdash \hL} 
\frac{\Dim_{N_c} (\hat R)}{d_{R_1} d_{R_2} d_{R_3}} 
\( d_{q_1} \, d_{q_2} \, d_{r_1} \, d_{r_3} \, d_{s_2} \, d_{s_3} \)
\oldelta^{\, \nu_{1-} \, \nu_{2+} } \,
\oldelta^{\, \nu_{2-} \, \nu_{3+} } \,
\oldelta^{\, \nu_{3-} \, \nu_{1+} } \, 
\cG'_{123} \,.
\label{CXYZ comp I123}
\end{equation}
Here $\{q_i, r_i , s_i \}$ must be consistent with $\bsR_i$ in \eqref{extended rep basis}. 
This condition is implicitly included in the definition of $\oldelta$ in \eqref{def:oldelta}.
In other words, the OPE coefficients are non-zero only if $\(q_1, q_2, r_1, r_3, s_2, s_3 \)$ satisfy
\begin{gather}
q_1 \otimes q_2 = q_3 ,\quad
r_1 \otimes r_3 = r_2 ,\quad
s_2 \otimes s_3 = s_1 ,\qquad
q_1 \otimes q_2 \otimes r_1 \otimes r_3 \otimes s_2 \otimes s_3 
= \hat R
\label{def:R123 mult} \\[1mm]
( R_1 )_{\nu_{1 \mp}} = q_1 \otimes r_1 \otimes \( s_2 \otimes s_3 \), \quad
( R_2 )_{\nu_{2\mp}} = q_2 \otimes \( r_1 \otimes r_3 \) \otimes s_2, \quad
( R_3 )_{\nu_{3\mp}} = \( q_1 \otimes q_2 \) \otimes r_3 \otimes s_3
\notag
\end{gather}
which can be represented by
\begin{equation}
\begin{tikzpicture}[node distance=10mm, baseline=(current bounding box.center)]
\tikzmath{\xm=-4; \xp=4.8; \yp=1.7; \ym=-1.7;}
\node at (-5.5,1) {$\widetilde{\cO}_1$};
\node at (-5.5,0) {$\widetilde{\cO}_2$};
\node at (-5.5,-1) {$\widetilde{\cO}_3$};
\draw (\xm,\yp) -- (\xm,\ym);
\draw (-2.2,\yp) -- (-2.2,\ym);
\draw (-1,\yp) -- (-1,\ym);
\draw (0,\yp) -- (0,\ym);
\draw (1.6,\yp) -- (1.6,\ym);
\draw (3.4,\yp) -- (3.4,\ym);
\draw (\xp,\yp) -- (\xp,\ym);
\filldraw[fill=black!80, thick] (-4,1.3) rectangle (-2.2,0.7);
\filldraw[fill=black!50, thick] (-1,1.3) rectangle (0,0.7);
\filldraw[fill=white, thick] (1.6,1.3) rectangle (\xp,0.7);
\filldraw[fill=black!80, thick] (-2.2,0.3) rectangle (-1,-0.3);
\filldraw[fill=black!50, thick] (-1,0.3) rectangle (1.6,-0.3);
\draw[thick] (1.6,0.3) rectangle (3.4,-0.3);
\filldraw[fill=black!80, thick] (-4,-0.7) rectangle (-1,-1.3);
\filldraw[fill=black!50, thick] (0,-0.7) rectangle (1.6,-1.3);
\draw[thick] (3.4,-0.7) rectangle (\xp,-1.3);
\node [white] at (-3.1,1) {$q_1$};
\node [white] at (-1.6,0) {$q_2$};
\node [white] at (-2.5,-1) {$q_3$};
\node [white] at (-0.5,1) {$r_1$};
\node [white] at (0.3,0) {$r_2$};
\node [white] at (0.8,-1) {$r_3$};
\node at (3.2,1) {$s_1$};
\node at (2.5,0) {$s_2$};
\node at (4.1,-1) {$s_3$};
\node at (-1.6,1) {$q_2$};
\node at (0.8,1) {$r_3$};
\node at (-3.1,0) {$q_1$};
\node at (4.1,0) {$s_3$};
\node at (-0.5,-1) {$r_1$};
\node at (2.5,-1) {$s_2$};
\draw[thick] (\xm,-2) rectangle (\xp,-2.6);
\node at (0.4,-2.3) {$\hat R$};
\end{tikzpicture}
\label{Wick rep structure CXYZ}
\end{equation}

We find some difference from the case of $\tilde C_{\circ\circ\circ}$ in \eqref{Cooo comp I123}.
First, we do not have a sum over $\(q_1^\star, q_2^\star, r_1^\star, r_3^\star, s_2^\star, s_3^\star \)$. 
This is because $\tilde C^{XYZ}_{\vec h}$ has the same structure of the Wick contractions as the extremal correlators for each flavor $X, Y, Z$.\footnote{Recall that $\vev{\olZ \olZ} = 0$ whereas any of $\vev{ Z, \tilde Z}, \vev{\tilde Z \olZ}, \vev{\olZ Z} $ are non-zero.}
Thus, the first line of \eqref{def:R123 mult} is trivial.
Second, there is no sum over $\{ \nu_{i\mp} \}$, because $\{ \nu_{i\mp} \}$ are part of the operator data $\bsR_i = \pare{ R_i , (q_i, r_i, s_i), \nu_{i-}, \nu_{i+} }$. 
We should pick up the right combination of multiplicities consistent with $\bsR_i$\,.

\paragraph{Extremal case.}

Consider the situation where the operators consist of $Z$ or $\olZ$ only. This means
\begin{equation}
\begin{gathered}
0 = h_1 = \ell_{31} - h_2 = h_3 \,, \quad
\ell_{23} = 0, \qquad
V_{FS} = V_{\ell_{12}} \otimes V_{\ell_{31}}
\\
q_i = r_i = \emptyset, \quad
R_i = s_i \,, \qquad
\hat R = R_1 \,.
\end{gathered}
\end{equation}
In particular, we do not need to specify $\nu_{i \mp}$\,.

The quantity $\cG'_{123}$ becomes
\begin{equation}
\cG'_{123} = \frac{\abs{\cM_{R_1}}^2 \abs{\cM_{R_2}}^2 \abs{\cM_{R_3}}^2 }{\abs{\cM_{\rm tot}}^3 } 
= g(R_2, R_3 ; R_1)
\end{equation}
where we used
\begin{equation}
\abs{\cM_{R_1}} = 1, \qquad
\abs{\cM_{R_2}} = \abs{\cM_{R_3}} = \abs{\cM_{\rm tot}} = g(R_2, R_3 ; R_1).
\end{equation}
The three-point function \eqref{CXYZ comp I123} becomes
\begin{equation}
\tilde C^{XYZ}_{\vec h} = L_1 ! \,
\frac{\Dim_{N_c} (R_1)}{d_{R_1} } \, g(R_2, R_3 ; R_1) 
\end{equation}
which agrees with \eqref{extremal limit Cooo} after relabeling.

In Appendix \ref{app:resLR} we consider the restricted Littlewood-Richardson coefficients, which are related to the extremal three-point functions of different type.

\section{Background independence at large $N_c$}\label{sec:bg indep}

We study the tree-level three-point functions in the representation basis, and check the background independence conjectured in \cite{deMelloKoch:2018ert}.
Our proof is based on the conjectured relations for the generalized Racah-Wigner tensor in Appendix \ref{app:genRW}.

\subsection{The LLM operators}\label{sec:LLM ops}

Let us review the argument on the large-$N_c$ background independence \cite{deMelloKoch:2018ert}.
They mapped the $\cN=4$ SYM operators with the $\cO(N_c^0)$ canonical dimensions to those with the $\cO(N_c^2)$ canonical dimensions by attaching a large number of background boxes.
We call the latter LLM operators, because they correspond to stringy excitations on the LLM geometry.
Recall that the LLM geometries are the half-BPS solutions of IIB supergravity.
This implies that the addition of $\cO(N_c^2)$ boxes should consist of a single holomorphic scalar like $\sim Z^{N_c^2}$.

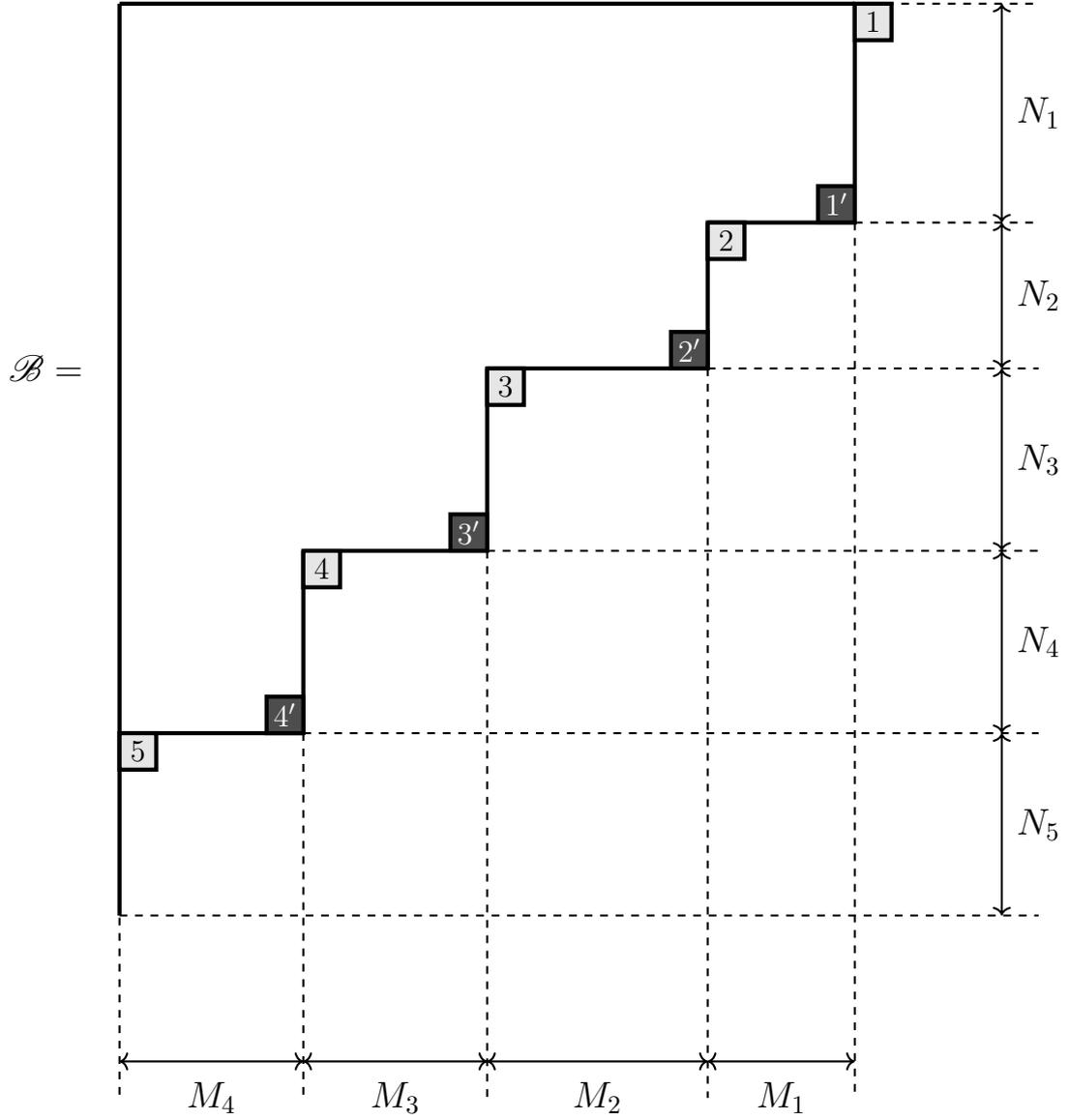
\begin{figure}[ht]
\begin{center}
\begin{tikzpicture}[x=5mm, y=5mm, baseline=(current bounding box.center)]
\tikzmath{\xp=25; \ym=-30; \xpp=\xp-1; \ymm=\ym+1;}
\draw[ultra thick] (0,0) -- (0,-25);
\draw[ultra thick] (0,-20) -- (5,-20);
\draw[ultra thick] (5,-20) -- (5,-15);
\draw[ultra thick] (5,-15) -- (10,-15);
\draw[ultra thick] (10,-15) -- (10,-10);
\draw[ultra thick] (10,-10) -- (16,-10);
\draw[ultra thick] (16,-10) -- (16,-6);
\draw[ultra thick] (16,-6) -- (20,-6);
\draw[ultra thick] (20,-6) -- (20,0);
\draw[ultra thick] (20,0) -- (0,0);
\draw[thick, dashed] (20,0) -- (\xp,0);
\draw[thick, dashed] (20,-6) -- (\xp,-6);
\draw[thick, dashed] (16,-10) -- (\xp,-10);
\draw[thick, dashed] (10,-15) -- (\xp,-15);
\draw[thick, dashed] (5,-20) -- (\xp,-20);
\draw[thick, dashed] (0,-25) -- (\xp,-25);
\draw[thick, dashed] (0,-20) -- (0,\ym);
\draw[thick, dashed] (5,-20) -- (5,\ym);
\draw[thick, dashed] (10,-15) -- (10,\ym);
\draw[thick, dashed] (16,-10) -- (16,\ym);
\draw[thick, dashed] (20,-6) -- (20,\ym);
\draw [thick, <->] (\xpp,0) -- (\xpp,-6);
\draw [thick, <->] (\xpp,-6) -- (\xpp,-10);
\draw [thick, <->] (\xpp,-10) -- (\xpp,-15);
\draw [thick, <->] (\xpp,-15) -- (\xpp,-20);
\draw [thick, <->] (\xpp,-20) -- (\xpp,-25);
\draw [thick, <->] (0,\ymm) -- (5,\ymm);
\draw [thick, <->] (5,\ymm) -- (10,\ymm);
\draw [thick, <->] (10,\ymm) -- (16,\ymm);
\draw [thick, <->] (16,\ymm) -- (20,\ymm);
\node at (\xp,-3) {\large $N_1$};
\node at (\xp,-8) {\large $N_2$};
\node at (\xp,-12.5) {\large $N_3$};
\node at (\xp,-17.5) {\large $N_4$};
\node at (\xp,-22.5) {\large $N_5$};
\node at (18,\ym) {\large $M_1$};
\node at (13,\ym) {\large $M_2$};
\node at (7.5,\ym) {\large $M_3$};
\node at (2.5,\ym) {\large $M_4$};
\draw[ultra thick, fill=black!10] (20,0) rectangle (21,-1);
\node at (20.5,-0.5) {$1$};
\draw[ultra thick, fill=black!70] (19,-5) rectangle (20,-6);
\node[white] at (19.5,-5.5) {$1'$};
\draw[ultra thick, fill=black!10] (16,-6) rectangle (17,-7);
\node at (16.5,-6.5) {$2$};
\draw[ultra thick, fill=black!70] (15,-9) rectangle (16,-10);
\node[white] at (15.5,-9.5) {$2'$};
\draw[ultra thick, fill=black!10] (10,-10) rectangle (11,-11);
\node at (10.5,-10.5) {$3$};
\draw[ultra thick, fill=black!70] (9,-14) rectangle (10,-15);
\node[white] at (9.5,-14.5) {$3'$};
\draw[ultra thick, fill=black!10] (5,-15) rectangle (6,-16);
\node at (5.5,-15.5) {$4$};
\draw[ultra thick, fill=black!70] (4,-19) rectangle (5,-20);
\node[white] at (4.5,-19.5) {$4'$};
\draw[ultra thick, fill=black!10] (0,-20) rectangle (1,-21);
\node at (0.5,-20.5) {$5$};
\node at (-2,-10) {\large $\scrB =$};
\end{tikzpicture}
\end{center}
\caption{The general background Young diagram $\scrB$ having a staircase shape, which corresponds to the LLM geometry of concentric shapes by AdS/CFT. All $M_i$ and $N_i$ are $\cO(N_c)$, and $\sum_i N_i =N_c$\,. The gray and black boxes represent localized string excitations. To define the operation $\blp$ we should choose one gray box.}
\label{fig:LLM tableau}
\end{figure}

For simplicity, we consider the operator mixing in the $\alg{su}(2)$ sector, at one-loop in $\lambda$ at any $N_c$\,.
We expand the dilatation eigenstates in terms of the restricted Schur basis as
\begin{equation}
\fD_1 \, \cO_\Delta = \Delta_1 \, \cO_\Delta \,, \qquad
\cO_\Delta = \sum_{R,r,s,\nu_\mp} c_{R,(r,s),\nu_-,\nu_+} \, \cO^{R,(r,s),\nu_-,\nu_+} \,.
\end{equation}
We denote the action of the one-loop dilatation on the restricted Schur basis by
\begin{equation}
\fD_1 \, \cO^{R,(r,s),\nu_-,\nu_+}
= \sum_{T,t,u,\mu_-,\mu_+} \, N^{R,(r,s),\nu_-,\nu_+}_{T,(t,u),\mu_-,\mu_+} \, 
\cO^{T,(t,u),\mu_-,\mu_+} 
\end{equation}
and define the LLM operator by
\begin{equation}
\cO_\Delta \ \to \ \cO_\Delta^{\rm LLM}
= \sum_{R,(r,s),\nu_\mp} c_{R,(r,s),\nu_-,\nu_+} \, \cO^{+R,(+r,s),\nu_-,\nu_+} \,.
\end{equation}
The operation $r \to (\blp r)$ can be exemplified as
\begin{equation}
r = 
\ytableausetup{centertableaux,boxsize=3mm}
\ytableaushort {\none,\none} * {3,1}
\quad \to \quad
(\blp r) =
\ytableausetup{centertableaux,boxsize=3mm}
\ytableaushort {\none,\none,\none,\none,\none,\none,\none,\none,\none}
* {10,10,10,10,10,8,6,5,5} * [*(xgray)]{10,10,10,10,10,5,5,5,5} 
\end{equation}
Here there are $\cO(1)$ white boxes, and $\cO(N_c^2)$ gray boxes in total.
Each edge of the gray block has the length of $\cO(N_c)$.
The general form of the background Young diagram $\scrB$ is shown in Figure \ref{fig:LLM tableau}.

We specify a corner of the background Young diagram $\scrB$, and consider a set of all Young diagrams attached to that corner. This set of states has many interesting properties.
First, from the Littlewood-Richardson rule, we find
\begin{equation}
g(r,s;R) \simeq g(\blp r,s; \blp R), \qquad (N_c \gg 1). 
\label{LLM LR identity}
\end{equation}
This allows us to use the same multiplicity labels $\nu_\mp$ before and after the $\blp$ operation.
Note that the tensor product $(\blp r) \otimes s$ contains representations in which boxes are attached to multiple corners of $\scrB$. However, the overlap between such states and $(\blp r)$ is suppressed by $1/N_c$\,.
Second, the hook length of $(\blp r)$ factorizes as \cite{deMelloKoch:2018ert}
\begin{equation}
\frac{\hook_{\blp r}}{\hook_r \, \hook_{\scrB}} \simeq ( \eta_\scrB )^{\abs{r}} \qquad
(N_c \gg 1)
\end{equation}
where $\eta_{\scrB}$ is the factor which depends only on $\scrB$,
\begin{equation}
\eta_\scrB \equiv \prod_{k=1}^C \frac{L(k,C)}{L(k,C)-N_k} \, \prod_{l=C+1}^D \frac{L(C+1,l)}{L(C+1,l)-M_l} \,,
\qquad
L(a,b) = \sum_{k=a}^b \( M_k + N_k \)
\end{equation}
assuming that the small diagram $r$ is put at the $C$-th corner of $\scrB$ in Figure \ref{fig:LLM tableau}.
It follows that
\begin{equation}
\frac{(\abs{\scrB} + \abs{r})!}{\abs{\scrB}!} \simeq \abs{\scrB}^{\abs{r}} , \qquad
\frac{d_{\blp r}}{d_r \, d_{\scrB}} \simeq \frac{1}{\abs{r}!} \( \frac{\abs{\scrB}}{\eta_\scrB} \)^{\abs{r}} \qquad
(N_c \gg 1).
\end{equation}
Since position of the $C$-th corner is $(i,j) = (1 + \sum_{l=C+1}^D M_l , 1 + \sum_{k=1}^C N_k )$, from \eqref{def:DimNR wtNR} we get
\begin{equation}
\frac{\Dim_{N_c} (\blp R)}{\Dim_{N_c} (\scrB)} \simeq \Dim_{N'_c} (R), \qquad
N'_c = N_c + \sum_{l=C+1}^D M_l - \sum_{k=1}^C N_k \,.
\label{def:N'_c}
\end{equation}

\clearpage

In \cite{deMelloKoch:2018ert} they found that the operator mixing coefficients satisfy the identity
\begin{equation}
N^{+R,(+r,s),\nu_-,\nu_+}_{+T,(+t,u),\mu_-,\mu_+}
\simeq
N^{R,(r,s),\nu_-,\nu_+}_{T,(t,u),\mu_-,\mu_+} \qquad
(N_c \gg 1)
\end{equation}
showing that
\begin{equation}
\fD_1 \, \cO_\Delta^{\rm LLM} \simeq \Delta_1 \, \cO_\Delta^{\rm LLM} \qquad
(N_c \gg 1) .
\end{equation}

\subsection{Tree-level OPE coefficients}

We revisit two types of OPE coefficients in Section \ref{sec:3pt}.
We will show that the OPE coefficients of non-extremal three-point functions in $\cN=4$ SYM are essentially same as those of the LLM operators, after redefinition of $N_c$\,.

\subsubsection{Adding a background tableau to $\tilde C_{\circ\circ\circ}$}

Recall that $\tilde C_{\circ\circ\circ}$ is given by \eqref{Cooo comp I123},
\begin{equation}
\begin{aligned}
\tilde C_{\circ\circ\circ} &= 
\Vev{
\hat \cO_1^{R_1 (\olL_1)} [Z,{\bf 1}] \,
\hat \cO_2^{R_2 (\olL_2)} [\tilde Z, {\bf 1}] \,
\hat \cO_3^{R_3 (\olL_3)} [\olZ, {\bf 1}] }
\\[1mm]
&= \( \prod_{i=1}^3 \frac{L_i!}{\olL_i!} \)
\sum_{\hat R \, \vdash \hL} 
\frac{\Dim_{N_c} (\hat R)}{d_{R_1} d_{R_2} d_{R_3}} \ 
\sum_{Q_1 \vdash \olL_2 } 
\sum_{Q_2 \vdash \olL_3 }
\sum_{Q_3 \vdash \olL_1 } 
\( \prod_{i=1}^3 d_{Q_i} \) \, \cG_{123} \,.
\end{aligned}
\end{equation}

We obtain the OPE coefficients of the LLM operators by the substitution $Q_1 \to (\blp Q_1)$, while leaving $Q_2 \,, Q_3$ as before. From \eqref{def:Q123 star} it follows that
\begin{equation}
\begin{gathered}
(\blp R_1) = ( \blp Q_1) \otimes Q_2 \,, \qquad
R_2 = Q_2 \otimes Q_3 \,, \qquad
(\blp R_3 )= Q_3 \otimes (\blp Q_1) 
\\[1mm]
(\blp \hat R) = (\blp Q_1 )\otimes Q_2 \otimes Q_3 
\end{gathered}
\end{equation}
and thus
\begin{align}
\tilde C_{\circ\circ\circ}^{\rm LLM} &\equiv
\Vev{
\hat \cO_1^{\blp R_1 (\olL_1)} [Z,{\bf 1}] \,
\hat \cO_2^{R_2 (\olL_2)} [\tilde Z, {\bf 1}] \,
\hat \cO_3^{\blp R_3 (\olL_3)} [\olZ, {\bf 1}] } 
\\
&= \frac{(\blp L_1)! L_2! (\blp L_3)!}{\olL_1! (\blp \olL_2)! \olL_3!} \,
\sum_{(+\hat R) \, \vdash (+\hL)} 
\frac{\Dim_{N_c} (\blp \hat R)}{d_{\blp R_1} d_{R_2} d_{\blp R_3}} \ 
\sum_{(\blp Q_1) \, \vdash (\blp \olL_2) } 
\sum_{Q_2 \, \vdash \olL_3 }
\sum_{Q_3 \, \vdash \olL_1 } 
\( d_{\blp Q_1} d_{Q_2} d_{Q_3} \) \, \cG_{123}^{\rm LLM} \,.
\notag
\end{align}
By using the identities in Section \ref{sec:LLM ops}, we find
\begin{equation}
\tilde C_{\circ\circ\circ}^{\rm LLM} \simeq
( \eta_{\scrB} )^{\hL} \,
\wt_{N_c} (\scrB) \,
\frac{L_1! L_2! L_3!}{\olL_1! \olL_2! \olL_3!}
\sum_{\hat R \, \vdash \hL} 
\frac{\Dim_{N'_c} (\hat R)}{d_{R_1} d_{R_2} d_{R_3}} \ 
\sum_{Q_1 \vdash \olL_2 } 
\sum_{Q_2 \vdash \olL_3 }
\sum_{Q_3 \vdash \olL_1 } 
\( d_{Q_1} d_{Q_2} d_{Q_3} \) \, \cG_{123} \,.
\end{equation}
If we remove the $\scrB$\,-dependent prefactor $( \eta_{\scrB} )^{\hL} \, \wt_{N_c} (\scrB)$, the OPE coefficient $\tilde C_{\circ\circ\circ}^{\rm LLM}$ agrees with $\tilde C_{\circ\circ\circ}$ up to the redefinition of $N_c \to N'_c$ in \eqref{def:N'_c}.

\subsubsection{Adding a background tableau to $\tilde C^{XYZ}_{\vec h}$}

Recall that $\tilde C^{XYZ}_{\vec h}$ is given by \eqref{CXYZ comp I123},
\begin{equation}
\begin{aligned}
\tilde C^{XYZ}_{\vec h} &= 
\Vev{
\hat \cO_1^{\bsR_1 (\olL_1)} [\olX,\olY,Z,{\bf 1}] \,
\hat \cO_2^{\bsR_2 (\olL_2)} [\olX,Y,\olZ, {\bf 1}] \,
\hat \cO_3^{\bsR_3 (\olL_3)} [X,Y,\olZ, {\bf 1}] }
\\[1mm]
&=
\( \prod_{i=1}^3 \frac{L_i!}{\olL_i!} \)
\sum_{\hat R \, \vdash \hL} 
\frac{\Dim_{N_c} (\hat R)}{d_{R_1} d_{R_2} d_{R_3}} 
\( d_{q_1} \, d_{q_2} \, d_{r_1} \, d_{r_3} \, d_{s_2} \, d_{s_3} \)
\oldelta^{\, \nu_{1-} \, \nu_{2+} } \,
\oldelta^{\, \nu_{2-} \, \nu_{3+} } \,
\oldelta^{\, \nu_{3-} \, \nu_{1+} } \, 
\cG'_{123} 
\end{aligned}
\end{equation}
where $\bsR_i$ is defined in \eqref{extended rep basis} as
\begin{equation}
\bsR_i = \pare{ R_i , (q_i, r_i, s_i), \nu_{i-}, \nu_{i+} }, \qquad \( R_i \vdash L_i  \).
\end{equation}

We obtain the OPE coefficients in the LLM background by the substitution $(s_1 \,, s_2 \,, s_3) \to (\blp s_1 \,, \blp s_2, s_3)$\,, while $q_i, r_i$ are the same as before. 
From \eqref{def:R123 mult} we find
\begin{equation}
\begin{gathered}
q_1 \otimes q_2 = q_3 ,\quad
r_1 \otimes r_3 = r_2 ,\quad
(\blp s_2) \otimes s_3 = (\blp s_1) ,\qquad
q_1 \otimes q_2 \otimes r_1 \otimes r_3 \otimes (\blp s_2) \otimes s_3 
= \hat R
\label{def:R123 mult LLM} \\[1mm]
( \blp R_1 )_{\nu_{1 \mp}} = q_1 \otimes r_1 \otimes \Big( ( \blp s_2) \otimes s_3 \Big)
\\
( \blp R_2 )_{\nu_{2\mp}} = q_2 \otimes \Big( r_1 \otimes r_3 \Big) \otimes ( \blp s_2 )
\\
( R_3 )_{\nu_{3\mp}} = \Big( q_1 \otimes q_2 \Big) \otimes r_3 \otimes s_3 \,.
\end{gathered}
\end{equation}
It follows that
\begin{multline}
(\tilde C^{XYZ}_{\vec h})^{\rm LLM} =
\( \frac{(\blp L_1)! (\blp L_2)! L_3!}{\olL_1! \olL_2! (\blp \olL_3)!} \)
\sum_{\hat R \, \vdash \hL} 
\frac{\Dim_{N_c} (\blp \hat R)}{d_{\blp R_1} d_{\blp R_2} d_{R_3}} 
\( d_{q_1} \, d_{q_2} \, d_{r_1} \, d_{r_3} \, d_{\blp s_2} \, d_{s_3} \) \ \times
\\
\oldelta^{\, \nu_{1-} \, \nu_{2+} } \,
\oldelta^{\, \nu_{2-} \, \nu_{3+} } \,
\oldelta^{\, \nu_{3-} \, \nu_{1+} } \, 
\cG^{' \, {\rm LLM}}_{123} \,.
\end{multline}
At large $N_c$, we can simplify this results following our discussion in Section \ref{sec:LLM ops} as
\begin{multline}
(\tilde C^{XYZ}_{\vec h})^{\rm LLM} =
\frac{\olL_3!}{(\olL_3-|r_1|)! } \,
\( \frac{\eta_\scrB}{\abs{\scrB}} \)^{|r_1|} \,
\eta_\scrB^{\hL} \, \wt_{N_c} (\scrB) \ \times
\\ 
\frac{L_1! L_2! L_3!}{\olL_1! \olL_2! \olL_3!}
\sum_{\hat R \, \vdash \hL} 
\frac{\Dim_{N'_c} (\hat R)}{d_{R_1} d_{R_2} d_{R_3}} 
\( d_{q_1} \, d_{q_2} \, d_{r_1} \, d_{r_3} \, d_{s_2} \, d_{s_3} \) \,
\oldelta^{\, \nu_{1-} \, \nu_{2+} } \,
\oldelta^{\, \nu_{2-} \, \nu_{3+} } \,
\oldelta^{\, \nu_{3-} \, \nu_{1+} } \, 
\cG'_{123} \,.
\end{multline}
The first line is a numerical prefactor, and the second line agrees with $(\tilde C^{XYZ}_{\vec h})$ by the redefinition of $N_c \to N'_c$ in \eqref{def:N'_c}.

\section{Conclusion and Outlook}

In this paper, we have studied general non-extremal three-point functions of scalar multi-trace operators at tree level valid for any values of $N_c$ in gauge theory including $\cN=4$ SYM, by using the representation theory of symmetric groups.

We made full use of various new mathematical techniques.
The quiver calculus of \cite{Pasukonis:2013ts} gives a collection of diagrammatic method which simplifies various objects in the representation theory.
The generalized Racah-Wigner tensor is introduced as an extension of the $6j$ symbols.
We conjectured formulae about the invariant products of the generalized Racah-Wigner tensors, written in terms of the Littlewood-Richardson coefficients.

With these formulae, we provide strong evidence on the large $N_c$ background independence, a correspondence between small ($\cO(N_c^0)$) and huge ($\cO(N_c^2)$) operators of $\cN=4$ SYM.
The background independence has been checked for two-point functions as well as extremal three-point functions. Our argument demonstrates that it extends to non-extremal three-point functions.
These results will clarify the properties of stringy excitations on the LLM backgrounds, particularly how they differ from the usual strings on \AdSxS.

\bigskip
Let us comment on some important future directions.

The first direction is to find a connection with the integrability results of the planar $\cN=4$ SYM.
Clearly, the operators in the representation basis are not the eigenstates of the dilatation operator of $\cN=4$ SYM. One should think of the representation basis as a tool for the finite $N_c$ computation.
The two-point functions of single-trace operators in the $\alg{su}(2)$ sector have been computed in this way \cite{Bhattacharyya:2008xy,Mattioli:2016eyp}, generalizing the old results of the complex matrix model \cite{Ginibre65,MehtaBook}.
A particularly interesting question is to determine the so-called octagon frame, namely the tree-level part of the ``simplest'' four-point functions of $\cN=4$ SYM in the large charge limit \cite{Bargheer:2019kxb}.
The finite group methods developed in this paper can be used for the exact finite\,-$N_c$ computation, because it is a generalization of the character expansion methods familiar in the matrix models \cite{Kostov:1996bs,Kostov:1997bn,Kristjansen:2002bb}. 

The second direction is to refine our computation.
The conjectured formula for the invariant products of generalized Racah-Wigner tensor should be proven.
The computation of the $n$-point functions in the representation basis is also important.
It is interesting to ask whether one can bootstrap four-point functions out of two- and three-point data.

The third direction is to investigate a possible relation between quiver calculus and knot theory.
The $6j$ symbol of the unitary group has been extensively studied in the context of knot theory and integrable systems \cite{KR90}.
Since the $6j$ symbols of symmetrical groups are related to those of unitary groups, the quiver calculus could give a new insight into the study of knot polynomials.
For example, some non-trivial conjectures about the $6j$ symbols have been made \cite{Itoyama:2012re,Nawata:2013ooa,Morozov:2018fnb}, though most of them discuss the multiplicity-free cases only.
Since the new invariants $\cG_{123}$ and $\cG'_{123}$ discussed in this paper are closely related to the multiplicity structure, studying similar quantity in the case of unitary groups is a fascinating problem.

Finally, we hope to find a clear understanding of the AdS/CFT correspondence of the operators with huge anomalous dimensions, including giant gravitons \cite{McGreevy:2000cw,Balasubramanian:2001nh} and the fluctuation in the LLM geometry \cite{Koch:2016jnm,deMelloKoch:2018tlb,deMelloKoch:2018ert}.
Some correlation functions have been studied such as three giants \cite{Takayanagi:2002nv,Hirano:2018xmh,deMelloKoch:2019dda}, two giants and one single-trace \cite{Bissi:2011dc,Caputa:2012yj,Lin:2012ey,Kristjansen:2015gpa,Jiang:2019zig,Jiang:2019xdz,Chen:2019gsb,Kim:2019gcq}.

\subsubsection*{Acknowledgments}

RS thanks Robert de Mello Koch and Sanjaye Ramgoolam for their comments on the manuscript, and is obliged to Korea Institute for Advanced Study where this research has been initiated.

\appendix
\section{Survey of finite-group representation theory}\label{app:notation}

We explain our notation and formulae used in the main text, while providing a brief survey of the representation theory of finite groups.
Our notation is similar to the one used in \cite{Kimura:2016bzo}.
For more details on finite groups, see textbooks like \cite{GoldschmidtBook,StanleyBook2}.

\subsection{Basic notation}\label{app:basic notation}

The symmetric group permuting $L$ elements is denoted by $S_L$\,.
We denote the conjugacy class of $S_L$ by
\begin{equation}
\cc_\alpha = \frac{1}{|S_L|} \sum_{\gamma \in S_L} \gamma \alpha \gamma^{-1} ,
\label{def:conjugacy}
\end{equation}
The $\delta$-function over $S_L$ (or $\bb{C}[S_L]$) is defined by
\begin{equation}
\delta ( \beta ) = 
\begin{cases}
1 &\qquad (\beta = {\bf 1} \in S_L) \\
0 &\qquad ({\rm otherwise}).
\end{cases}
\label{def:delta fn}
\end{equation}
A permutation cycle is denoted by $(12 \dots L) \in \bb{Z}_L$\,.
Any element of $S_L$ consists of permutation cycles.
The number of length-$k$ cycles in $\sigma \in S_L$ is denoted by $\cyc_k (\sigma)$. The number of cycles in $\sigma$ is
\begin{equation}
C(\sigma) = \sum_k \cyc_k (\sigma) 
\end{equation}
so that $C({\bf id}) = C( (1)(2) \dots (L) ) = L$.

A partition of $L$, or equivalently a Young diagram with $L$ boxes, is denoted by $R \vdash L$.
Define
\begin{alignat}{9}
d_R &= \frac{L!}{\hook_R} \,, &\qquad
\hook_R &= \prod_{(i,j) \in R} \Big( \text{hook length at }(i,j) \Big)
\label{def:dR hookR} \\
\Dim_{N} (R) & 
= \frac{d_R}{L!} \, \wt_{N} (R) \,, &\qquad
\wt_{N} (R) &= \prod_{(i,j) \in R} (N+i-j)
\label{def:DimNR wtNR} 
\end{alignat}
where $d_R$ is the dimension of $R$ as the representation of $S_L$\,, and $\Dim_{N} (R)$ is the dimension of $R$ as the representation of $U(N)$.\footnote{$\wt_{N} (R)$ is also denoted by $f_R$ in the literature, e.g. \cite{deMelloKoch:2007rqf}.}
For example, $\hook_R$ and $\wt_{N} (R)$ of the Young tableau $R = {\tiny \yng(4,2)}$ are given by
\begin{equation}
\begin{aligned}
\small \young(5421,21) \qquad &\Rightarrow \quad
\hook_{\tiny \yng(4,2)} = 5 \times 4 \times 2 \times 2 \times 1 \times 1
\\[3mm]
\ytableausetup{mathmode, boxsize=2.3em, aligntableaux=center}
\begin{ytableau}
\text{\scriptsize $N$} & \text{\scriptsize $N+1$} & \text{\scriptsize $N+2$} & \text{\scriptsize $N+3$} \\
\text{\scriptsize $N-1$} & \text{\scriptsize $N$} 
\end{ytableau} \ 
\qquad &\Rightarrow \quad
\wt_N \( \tiny \yng(4,2) \) = \(N-1\) N^2 \(N+1\)\(N+2\)\(N+3\) .
\end{aligned}
\label{example:hook-weight}
\end{equation}

\bigskip
We assume that all representations are real and orthogonal.\footnote{The orthogonal form of the Young-Yamanouchi basis satisfies these conditions.}
Denote the $I$-th component of the irreducible representation $R$ of $S_L$ by $\ket{\atopfrac{R}{I}}$, with $I=1,2,\dots, d_R$\,. Introduce the dual basis by
\begin{equation}
\Vev{ \atop{R}{I} \, \Big| \, \atop{S}{J} } = \delta^{RS} \, \delta_{IJ} \,.
\end{equation}
Let $D^R_{IJ} (\sigma)$ be the representation matrix of $\sigma \in S_{m+n}$ of the representation $R \vdash L$, 
\begin{equation}
D^R_{IJ} (\sigma) = \Vev{ \atop{R}{I} \, \Big| \, \sigma \, \Big| \, \atop{R}{J} } 
= D^R_{JI} (\sigma^{-1}) .
\label{real unitary rep}
\end{equation}
The character of the representation $R$ for the group element $\sigma$ is denoted by\footnote{Often we sum over the repeated indices of matrices. The symbol $\sum$ is written explicitly in Appendix \ref{app:notation}.}
\begin{equation}
\chi^R (\sigma) = \sum_{I=1}^{d_R} D^R_{II} (\sigma) .
\label{def:irrep char}
\end{equation}

By restricting $\sigma \in S_L = S_{m+n}$ to $S_m \otimes S_n$\,, we obtain the irreducible decomposition\footnote{The restriction to a subgroup is also called subduction in the literature.}
\begin{equation}
R = \bigoplus_{\substack{r \vdash m\\ s \vdash n}} g(r,s;R) \( r \otimes s \) 
= \bigoplus_{\substack{r \vdash m\\ s \vdash n}} \bigoplus_{\nu=1}^{g(r,s;R)} \( r \otimes s \)_\nu
\label{LR coeff by decomposition}
\end{equation}
where $g(r,s;R)$ is the Littlewood-Richardson coefficient. It counts the number of $r \otimes s$ appearing in the irreducible decomposition of $R$.
The subscript $\nu$ is called the multiplicity label.
With an appropriate change of basis,\footnote{This appropriate basis is called the split basis.} we can transform the representation matrix into a block-diagonal form,
\begin{equation}
D^R_{IJ} (\sigma) = B
\begin{pmatrix}
D^{r^{(1)} \otimes s^{(1)}}_{i_1 j_1} (\sigma) & & & \\
& D^{r^{(2)} \otimes s^{(2)}}_{i_2 j_2} (\sigma) & & \\
& & D^{r^{(3)} \otimes s^{(3)}}_{i_3 j_3} (\sigma) & & \\
& & & \ddots \ 
\end{pmatrix} B^{T} \qquad
(\sigma \in S_m \otimes S_n)
\label{block diagonal DR}
\end{equation}
such that it matches \eqref{LR coeff by decomposition}.
By definition of the irreducible decomposition, there are no off-block-diagonal elements including the multiplicity labels. For general $\sigma \in S_{m+n}$\,, the matrix \eqref{block diagonal DR} has off-block-diagonal elements.\footnote{The restricted Schur basis should have off-block-diagonal elements with respect to the multiplicity labels, which can be checked by counting the dimensions \cite{Mattioli:2016eyp}.}

Let $\ket{\atopfrac{r, s}{i,j} \, \nu }$ be an orthonormal basis of $r \otimes s$ at the $\nu$-th multiplicity, satisfying
\begin{equation}
\Vev{ \matop{r_1 & s_1}{i_1 & j_1}{\nu_1} \, \Big| \, \matop{r_2 & s_2}{i_2 & j_2}{\nu_2} }
= \delta^{r_1r_2} \, \delta^{s_1s_2} \, \delta^{\nu_1 \nu_2} \, \delta_{i_1 i_2} \, \delta_{j_1 j_2} 
\end{equation}
for $\nu_k=1,2, \dots , g(r_k,s_k;R)$.
The rotation matrix is called the branching coefficients, defined by
\begin{equation}
B^{R \to (r, s), \nu}_{I \to (i, j)} = \Vev{ \atop{R}{I} \, \Big| \, \matop{r&s}{i&j}{\nu} } ,\qquad
\brT^{R \to (r, s), \nu}_{I \to (i, j)} = \Vev{ \matop{r&s}{i&j}{\nu}  \, \Big| \, \atop{R}{I} } \,.
\label{app:branching coeffs}
\end{equation}

\subsection{Branching coefficients}\label{app:branching}

We find from \eqref{block diagonal DR} that the branching coefficients satisfy the completeness relations
\begin{align}
\sum_{r,s, \nu} \sum_{i,j} B^{R \to (r, s), \nu}_{I \to (i, j)} \, \brT^{R \to (r, s), \nu}_{J \to (i, j)} 
&= \delta_{I,J} 
\label{product branching1} \\
\sum_{I} \brT^{R \to (r_1, r_2), \nu}_{I \to (i_1, i_2)} \, B^{R \to (s_1, s_2), \mu}_{I \to (j_1, j_2)} 
&= \delta^{r_1,s_1} \, \delta^{r_2,s_2} \, \delta^{\nu\mu} \, \delta_{i_1,j_1} \, \delta_{i_2,j_2} \,.
\label{product branching2} 
\end{align}
In \eqref{product branching2}, we assume that two product representations $r_1 \otimes r_2$ and $s_1 \otimes s_2$ descend from the same restriction $S_{m+n} \downarrow (S_m \otimes S_n)$. 
If they descend from different restrictions, then the two branching coefficients $B$ and $\tilde B$ are unrelated, and we obtain another orthogonal matrix
\begin{equation}
\sum_{I} \brT^{R \to (r_1, r_2), \nu}_{I \to (i_1, i_2)} \, \tilde B^{R \to (s_1, s_2), \mu}_{I \to (j_1, j_2)} 
= \Vev{ \, \matop{r_1 & r_2}{i_1 & i_2}{\nu} \ \Big| \ \matop{s_1 & s_2}{j_1 & j_2}{\mu} \, }.
\label{def:intertwining mat}
\end{equation}
For example, given two irreducible decompositions
\begin{alignat}{9}
S_6 &\downarrow (S_4 \otimes S_2),
&\qquad {\tiny \yng(5,1)} \ &= \ {\tiny \yng(4)} \otimes {\tiny \yng(2)} \ \ 
\oplus \ \ {\tiny \yng(4)} \otimes {\tiny \yng(1,1)} \ \ 
\oplus \ \ {\tiny \yng(3,1)} \otimes {\tiny \yng(2)}
\notag \\[1mm]
S_6 &\downarrow (S_3 \otimes S_3),
&\qquad {\tiny \yng(5,1)} \ &= \ {\tiny \yng(3)} \otimes {\tiny \yng(3)} \ \ 
\oplus \ \ {\tiny \yng(3)} \otimes {\tiny \yng(2,1)} \ \ 
\oplus \ \ {\tiny \yng(2,1)} \otimes {\tiny \yng(3)}
\end{alignat}
any pairs $r_1 \otimes r_2$ and $s_1 \otimes s_2$ from different restrictions can have non-vanishing overlap, e.g.
\begin{equation}
\Vev{ \, \atop{ {\tiny \yng(4)} \otimes {\tiny \yng(2)} }{i_1, i_2} \ \Big| \  
\atop{{\tiny \yng(2,1)} \otimes {\tiny \yng(3)}}{j_1, j_2} \, } \neq 0 .
\label{overlap example}
\end{equation}
Sometimes we take the coordinates explicitly in order to distinguish $S_{m+n} \downarrow (S_m \otimes S_n)$ and $S_{m+n} \downarrow (S_n \otimes S_m)$. For example, the following two restrictions
\begin{equation}
\begin{aligned}
S_{m+n} \downarrow (S_m \otimes S_n) \ &\sim \ 
{\rm Permute} \( \{ 1,2, \dots, m \} \) \times {\rm Permute} \( \{ m+1, \dots m+n \} \)
\\
S_{m+n} \downarrow (S_n \otimes S_m) \ &\sim \ 
{\rm Permute} \( \{ 1,2, \dots, n \} \) \times {\rm Permute} \( \{ n+1, \dots n+m \} \)
\end{aligned}
\end{equation}
define different branching coefficients, $B^{R \to (r_1, r_2), \nu}_{I \to (i_1, i_2)}$ and $\tilde B^{R \to (s_1, s_2), \mu}_{I \to (j_1, j_2)}$\,.

\bigskip
From \eqref{block diagonal DR}, we obtain the following identities for the matrix elements of $\gamma = \gamma_1 \circ \gamma_2 \in S_m \otimes S_n$
\begin{equation}
D^R_{IJ} (\gamma_1 \circ \gamma_2)
= \sum_{r_1,r_2,\nu} \sum_{i,j,k,l} 
D^{r_1}_{ik} (\gamma_1) \, D^{r_2}_{jl} (\gamma_2) \, 
B^{R \to (r_1,r_2) \nu}_{I \to (i,j)} \, \brT^{R \to (r_1,r_2) \nu}_{J \to (k,l)} 
\label{perm matrix after branching}
\end{equation}
By multiplying $B^{R \to (r_1,r_2) \nu}_{J \to (k',l')}$ to \eqref{perm matrix after branching} and summing over $J$, we find
\begin{equation}
\sum_J D^R_{IJ} (\gamma_1 \circ \gamma_2) \, B^{R \to (r_1,r_2) \nu}_{J \to (k,l)}
= \sum_{i,j} D^{r_1}_{ik} (\gamma_1) \, D^{r_2}_{jl} (\gamma_2) \, B^{R \to (r_1,r_2) \nu}_{I \to (i,j)} \,.
\label{perm matrix after branching2} 
\end{equation}
Again, by multiplying $\brT^{R \to (r_1,r_2) \mu}_{I \to (i',j')}$ to \eqref{perm matrix after branching2} and summing over $J$, we find
\begin{equation}
\sum_{I,J} D_{IJ}^R (\gamma_1 \circ \gamma_2) \, 
\brT^{R \to (r_1, r_2) \mu}_{I \to (i,j)} \, B^{R \to (r_1, r_2) \nu}_{J \to (k,l)} 
= \delta^{\mu \nu} \, D^{r_1}_{ik} (\gamma_1) \, D^{r_2}_{jl} (\gamma_2) .
\label{perm matrix after branching3}
\end{equation}
In the RHS, the matrix elements of $\gamma_1 \circ \gamma_2$ in the split basis are independent of the multiplicity labels $\mu, \nu$. This can be understood also from the construction of the Young-Yamanouchi basis.

\bigskip
The branching coefficients \eqref{app:branching coeffs} for general restriction $S_L \downarrow (S_{m_1} \otimes S_{m_2} \otimes \dots \otimes S_{m_\ell})$ are given by
\begin{equation}
B^{R \to (r_1 \,, r_2 \,, \dots \,, r_\ell), \nu}_{I \to (i_1 \,, i_2 \,, \dots \,, i_\ell)} = 
\Vev{ \atop{R}{I} \, \Big| \, \matop{r_1 & r_2 & \dots & r_\ell}{i_1 & i_2 & \dots & i_\ell}{\nu} } ,
\quad
\brT^{R \to (r_1 \,, r_2 \,, \dots \,, r_\ell), \nu}_{I \to (i_1 \,, i_2 \,, \dots \,, i_\ell)} = 
\Vev{ \matop{r_1 & r_2 & \dots & r_\ell}{i_1 & i_2 & \dots & i_\ell}{\nu} \, \Big| \, \atop{R}{I} }
\label{app:gen branching coeffs}
\end{equation}
for $\nu = 1, 2, \dots, g(r_1 \,, r_2 \,, \dots \,, r_\ell;R)$.
The generalized split basis can be defined by the branching coefficients as in \eqref{block diagonal DR}.
The formula \eqref{perm matrix after branching} is generalized as
\begin{multline}
D^R_{IJ} (\gamma_1 \circ \gamma_2 \circ \dots \circ \gamma_\ell)
\\
= \sum_{r_1,r_2,\nu} \sum_{i,j,k,l} 
D^{r_1}_{i_1 k_1} (\gamma_1) \, D^{r_2}_{i_2 k_2} (\gamma_2) \, \dots \, D^{r_\ell}_{i_\ell k_\ell} (\gamma_\ell) \, 
B^{R \to (r_1 \,, r_2 \,, \dots \,, r_\ell), \nu}_{I \to (i_1 \,, i_2 \,, \dots \,, i_\ell)} \,
\brT^{R \to (r_1 \,, r_2 \,, \dots \,, r_\ell), \nu}_{J \to (k_1 \,, k_2 \,, \dots \,, k_\ell)} \,.
\label{gen:perm matrix after branching}
\end{multline}
for $\gamma = \gamma_1 \circ \gamma_2 \circ \dots \circ \gamma_\ell \in (S_{m_1} \otimes S_{m_2} \otimes \dots \otimes S_{m_\ell})$.


\subsection{Restricted Schur basis}\label{app:RSB}

Consider the restriction $S_M \downarrow (S_{m_1} \otimes S_{m_2} \otimes S_{m_3})$ with $M=m_1+m_2+m_3$\,, which corresponds to the multi-trace operators with three complex scalars in \eqref{def:perm multi-trace}.

Define the restricted Schur characters by using the branching coefficients \cite{Pasukonis:2013ts},
\begin{equation}
\chi^{R,(r_1,r_2,r_3),\nu_+,\nu_-} (\sigma) 
\equiv \sum_{I,J} \sum_{i,j,k} B^{R \to (r_1,r_2,r_3) \nu_+}_{I \to (i,j,k)} \, 
\brT^{R \to (r_1,r_2,r_3), \nu_-}_{J \to (i,j,k)} \,
D^R_{IJ} (\sigma) , \qquad (\sigma \in S_M).
\label{app:restricted character}
\end{equation}
Define the operator in the restricted Schur basis by
\begin{equation}
\cO^{R,(r_1,r_2,r_3),\nu_+,\nu_-} [X,Y,Z] = \frac{1}{m_1! \, m_2! \, m_3!} \sum_{\alpha \in S_{M}} \chi^{R,(r_1,r_2,r_3),\nu_+,\nu_-} (\alpha) \,
\tr_{\! M} \( \alpha \, X^{\otimes m_1} \, Y^{\otimes m_2} \, Z^{\otimes m_3} \) .
\label{def:RSB operator}
\end{equation}
The inverse transformation from the restricted Schur basis to the permutation basis is
\begin{multline}
\tr_{\! M} \( \alpha \, X^{\otimes m_1} \, Y^{\otimes m_2} \, Z^{\otimes m_3} \) 
\\[1mm]
= \frac{m_1! \, m_2! \, m_3!}{M!}
\sum_{R, r_1, r_2, r_3, \mu_+, \mu_-} 
\frac{d_R}{d_{r_1} d_{r_2} d_{r_3}} \,
\chi^{R,(r_1,r_2,r_3),\mu_+,\mu_-} (\alpha) \,
\cO^{R,(r_1,r_2,r_3),\mu_+,\mu_-}
\end{multline}
which can be checked by the row orthogonality of the restricted characters \eqref{restricted row orthogonality},
\begin{equation}
\frac{1}{M!} \sum_{\sigma \in S_{M}} \chi^{R,(r_1,r_2,r_3),\nu_+,\nu_-} (\sigma)
\chi^{S,(s_1,s_2, s_3),\mu_+,\mu_-} (\sigma)
= \frac{d_{r_1} d_{r_2} d_{r_3}}{d_R} \, \delta^{RS} \delta^{r_1 s_1} \delta^{r_2 s_2} \delta^{r_3 s_3} \delta^{\nu_+ \mu_+} \delta^{\nu_- \mu_-} \,.
\end{equation}
As discussed in Section \ref{sec:diag 2pt}, the tree-level two-point function is
\begin{multline}
\Vev{ \cO^{R,(r_1,r_2,r_3)(\nu_+,\nu_-)} [X,Y,Z] (x) \,
\cO^{S,(s_1,s_2,s_3)(\mu_+,\mu_-)} [\olX,\olY,\olZ] (0)
} 
\\[1mm]
= \frac{\wt_N (R)}{\abs{x}^{2M}} \, \frac{\hook_R}{\hook_{r_1} \hook_{r_2} \hook_{r_3}} \, \delta^{RS} \, \delta^{r_1 s_1} \, \delta^{r_2 s_2} \, \delta^{r_3 s_3} \,
\delta^{\nu_+ \mu_+} \, \delta^{\nu_- \mu_-} \,.
\end{multline}

\subsection{Formulae}\label{app:formulae}

The formulae for the irreducible characters and the restricted characters will be summarized below.
For simplicity, we mostly consider the restriction $S_{m+n} \downarrow (S_m \otimes S_n)$.
Generalization to $S_M \downarrow (\otimes_k S_{m_k})$ is straightforward.

\paragraph{Character Orthogonality.}

Let $R, S$ be the irreducible representations of $S_L$\,.
The representation matrices satisfy the grand orthogonality relation 
\begin{align}
\sum_{\sigma\in S_{L}}D^R_{ij}(\sigma)D^S_{kl}(\sigma^{-1})=\frac{L!}{d_R} \, \delta_{il}\delta_{jk} \,.
\label{orthogonality_representation}
\end{align}
By taking the trace, we obtain the row (or first) orthogonality relation of irreducible characters,
\begin{align}
\sum_{\sigma\in S_{L}} \chi^R(\sigma) \chi^S (\sigma^{-1}) = L! \, \delta^{RS} \,.
\label{row orthogonality characters}
\end{align}
The irreducible characters also satisfy the column (or second) orthogonality relation,
\begin{equation}
\sum_{R \, \vdash L} \chi^R (\sigma) \chi^R (\tau) = \sum_{\gamma \in S_L} \delta (\sigma \gamma \tau \gamma^{-1} )
= \begin{cases}
|\cc_\sigma| &\quad (\cc_\sigma = \cc_\tau) \\
0 &\quad ({\rm otherwise}) 
\end{cases}
\label{column orthogonality characters}
\end{equation}
where $|\cc_\sigma|$ is the number of elements in a given conjugacy class \eqref{def:conjugacy}.
This relation follows from the fact that any class function can be expanded by irreducible characters
\begin{equation}
f (\sigma) = f (\gamma \sigma \gamma^{-1}), \quad (\forall \gamma \in S_L)
\qquad \Leftrightarrow \qquad
f (\sigma) = \sum_{R \, \vdash L} {\tilde f}_R \, \chi^R (\sigma).
\label{class fn expand}
\end{equation}
As a corollary, the $\delta$-function can be written as
\begin{equation}
\delta ( \beta ) = \frac{1}{L!} \sum_{R \, \vdash L} d_R \, \chi^R (\beta).
\label{delta fn2}
\end{equation}

\paragraph{Multiplicity label.}

There are several ways to understand Littlewood-Richardson coefficients.

The first way is by restriction $S_{m+n} \downarrow (S_m \otimes S_n)$ as in \eqref{LR coeff by decomposition}
\begin{equation}
R = \bigoplus_{\substack{r \vdash m\\ s \vdash n}} g(r,s;R) \( r \otimes s \) .
\label{LR coeff by decomposition2}
\end{equation}
The second way is by induction,
\begin{equation}
r \otimes s = \bigoplus_R g(r,s;R) \, R 
\label{LR coeff by induction}
\end{equation}
Frobenius reciprocity guarantees the consistency between \eqref{LR coeff by induction} and \eqref{LR coeff by decomposition2}.
Finally, the Littlewood-Richardson coefficient can be computed by
\begin{equation}
g(r, s ; R) = \frac{1}{| S_m \otimes S_n|} \sum_{\alpha \in S_m} \sum_{\beta \in S_n} 
\chi^{r} (\alpha) \, \chi^{s} (\beta) \, \chi^R (\alpha \circ \beta) 
\label{def:Littlewood Richardson}
\end{equation}
where $\alpha \circ \beta \in S_m \otimes S_n \subset S_{m+n}$\,.

The generalized Littlewood-Richardson coefficient for $\otimes_{k=1}^l S_{m_k}$ is given by
\begin{equation}
g(r_1 , r_2, \dots, r_l ; R) = \frac{1}{| \otimes_{k=1}^l S_{m_k} |} 
\sum_{ \{ \sigma_k \in S_{m_k} \} }
\( \prod_{k=1}^l \chi^{r_k} (\sigma_k) \) \chi^R (\sigma_1 \circ \sigma_2 \circ \dots \circ \sigma_l) .
\label{def:gen LR}
\end{equation}
They satisfy a recursion relation
\begin{equation}
\sum_{R \, \vdash M} g(r_1 , r_2, \dots, r_l ; R) \, g(R , r_{l+1} ; S) 
= g(r_1 , r_2, \dots, r_{l+1} ; S), \qquad
\( M = \sum_{k=1}^l m_k \)
\label{gen LR recursion}
\end{equation}
which can be shown from \eqref{column orthogonality characters}.
The equation \eqref{gen LR recursion} implies an important identity for multiple branching coefficients
\begin{gather}
B^{S \to (r_1, r_2, \dots, r_{l+1}), \eta}_{I \to (a_1, a_2 , \dots, a_{l+1})}
= \sum_{R} \sum_{A=1}^{d_R} 
B^{S \to (R, r_{l+1}), \mu}_{I \to (A,a_{l+1})} 
B^{R \to (r_1, r_2, \dots, r_l), \rho}_{A \to (a_1, a_2, \dots, a_l)}
\label{multiple branching identity} \\
\eta = 1,2, \dots, g(r_1,r_2, \dots , r_{l+1}; S), \quad
\mu = 1,2, \dots, g(R, r_{l+1} ; S ), \quad
\rho = 1,2, \dots, g(r_1, r_2, \dots, r_l; R).
\notag
\end{gather}

\paragraph{Schur-Weyl duality.}

The quantity $N^{C(\sigma)}$ is a class function.
We obtain its irreducible decomposition \eqref{class fn expand} by using the Schur-Weyl duality \cite{Corley:2001zk} as
\begin{equation}
N^{C(\sigma)} = \sum_{R \, \vdash L} \Dim_N (R) \, \chi^R (\sigma) .
\label{decompose Nc(sigma)}
\end{equation}
Note that $\Dim_N (R)=0$ if the height of the Young diagram $R$ is larger than $N$, as can be seen from \eqref{def:DimNR wtNR}.
By applying the grand orthogonality relation \eqref{orthogonality_representation}, we find
\begin{equation}
\sum_{\sigma \in S_L} D^S_{IJ} (\sigma) N^{C(\sigma)}
= \delta_{IJ} \, \Dim_N (S) \, \hook_S 
= \delta_{IJ} \, \wt_N (S) .
\end{equation}
By multiplying the branching coefficients as in \eqref{def:restricted projector}, we obtain another formula \cite{deMelloKoch:2007rqf}
\begin{align}
\sum_{\sigma \in S_{m+n}} \chi^{R,(r,s),\nu_+,\nu_-} (\sigma) \, N^{C(\sigma)} 
= \delta^{\nu_+ \nu_-} \, d_r d_s \wt_N (R) .
\end{align}

\paragraph{Restricted projector.}

We define the restricted projector
\begin{equation}
\proj^{R,(r_1,r_2),\nu_+,\nu_-} = \frac{d_R}{(m+n)!} \sum_{\sigma \in S_{m+n}} \chi^{R,(r_1,r_2),\nu_+,\nu_-} (\sigma) \, \sigma \ \in \ \bb{C} [S_{m+n}]
\label{def:restricted projector}
\end{equation}
so that \cite{Mattioli:2016eyp}
\begin{align}
\chi^{R,(r_1,r_2),\nu_+,\nu_-} (\sigma) &= \chi^R \( \proj^{R,(r_1,r_2),\nu_+,\nu_-} \sigma \)
\label{restricted projector relation1}
\\[1mm]
\proj^{R,(r_1,r_2),\nu_+,\nu_-} \, \proj^{S,(s_1,s_2),\mu_+,\mu_-} 
&=
\delta^{RS} \, \delta^{r_1s_1} \, \delta^{r_2s_2} \, \delta^{\nu_- \mu_+} \, \proj^{R,(r_1,r_2),\nu_+,\mu_-} \,.
\label{restricted projector relation2}
\end{align}
By comparing \eqref{restricted projector relation1} and \eqref{app:restricted character}, one finds
\begin{equation}
\proj_{IJ}^{R,(r_1,r_2),\nu_+,\nu_-} \equiv
D^R_{IJ} \( \proj^{R,(r_1,r_2),\nu_+,\nu_-} \) = 
\sum_{i,j} B^{R \to (r_1, r_2) \nu_+}_{I \to (i,j)} \, 
\brT^{R \to (r_1, r_2), \nu_-}_{J \to (i,j)} \,.
\label{def:restricted projector R}
\end{equation}
It follows that
\begin{equation}
\chi^R \( \proj^{R,(r_1,r_2),\nu_+,\nu_-} \) = 
\sum_{I} \sum_{i,j} B^{R \to (r_1,r_2) \nu_+}_{I \to (i,j)} \, 
\brT^{R \to (r_1, r_2), \nu_-}_{I \to (i,j)} 
= \delta^{\nu_+ \nu_-} \, d_{r_1} \, d_{r_2} \,.
\label{restricted projector trace}
\end{equation}
The restricted projector is useful for fixing the normalization.
These formulae as well as the following identities can be proven by using the quiver calculus in Appendix \ref{app:quiver calc}.

\paragraph{Restricted Character Orthogonality.}

The restricted characters \eqref{app:restricted character} satisfy the identities
\begin{alignat}{9}
\chi^{R,(r,s),\nu_+,\nu_-} (\sigma) &= \chi^{R,(r,s),\nu_-,\nu_+} (\sigma^{-1}) & &
\label{restricted inverse}\\[1mm]
\chi^{R,(r,s),\nu_+,\nu_-} (\gamma \sigma \gamma^{-1}) &= \chi^{R,(r,s),\nu_+,\nu_-} (\sigma)
&\qquad &(\forall \gamma \in S_m \otimes S_n) 
\label{restricted conjugacy}\\[1mm]
\chi^{R,(r,s),\nu_+,\nu_-} (\sigma_1 \circ \sigma_2) &= \delta^{\nu_+ \nu_-} \, 
\chi^{r} (\sigma_1) \, \chi^{s} (\sigma_2)
&\qquad &(\forall \sigma_1 \circ \sigma_2 \in S_m \otimes S_n) 
\label{restricted trivial}
\end{alignat}
where the last relation is consistent with \eqref{perm matrix after branching3}.
The row and column orthogonality relations \eqref{column orthogonality characters} are generalized as
\begin{align}
\frac{1}{(m+n)!} \sum_{\sigma \in S_{m+n}} \chi^{R,(r_1,r_2),\nu_+,\nu_-} (\sigma)
\chi^{S,(s_1,s_2),\mu_+,\mu_-} (\sigma)
&= \frac{d_{r_1} d_{r_2}}{d_R} \, \delta^{RS} \delta^{r_1 s_1} \delta^{r_2 s_2} \delta^{\nu_+ \mu_+} \delta^{\nu_- \mu_-}
\label{restricted row orthogonality} \\
\sum_{R,r_1,r_2,\nu_+,\nu_-} \frac{d_R}{d_{r_1} d_{r_2}} \, \chi^{R,(r_1,r_2),\nu_+,\nu_-} (\sigma)
\chi^{R,(r_1,r_2),\nu_+,\nu_-} (\tau)
&= \frac{(m+n)!}{m! n!}
\sum_{\gamma \in S_m \otimes S_n} \delta (\gamma \sigma \gamma^{-1} \tau^{-1} ) .
\label{restricted column orthogonality}
\end{align}

One can generalize the grand orthogonality relation \eqref{orthogonality_representation} with the branching coefficients in two ways.
First, let $R$ and $S$ be the irreducible representations of $S_{m+n}$\,.
A sum over $S_{m+n}$ gives
\begin{multline}
\frac{1}{(m+n)!} \sum_{\sigma \in S_{m+n}} D_{IJ}^R (\sigma) \, 
B^\dagger{}^{R \to (r_1, r_2) \nu_+}_{I \to (i,j)} \, B^{R \to (r_1, r_2) \nu_-}_{J \to (k,l)}
D_{MN}^S (\sigma) \, 
B^\dagger{}^{S \to (s_1, s_2) \mu_+}_{M \to (m,n)} \, B^{S \to (s_1, s_2) \mu_-}_{N \to (p,q)}
\\
= \frac{\delta^{RS}}{d_R} \,
\delta^{\nu_+\mu_+} \, \delta^{\nu_-\mu_-} \,
\delta^{r_1,s_1} \, \delta^{r_2, s_2} \, 
\delta_{i,m} \, \delta_{j,n} \, \delta_{k,p} \, \delta_{l,q} 
\label{restricted grand orthogonality}
\end{multline}
which reduces to \eqref{restricted row orthogonality} by taking the trace over $r_1 \otimes r_2 = s_1 \otimes s_2$.
Second, let $(r_1, r_2)$ and $(s_1, s_2)$ be the irreducible representations of $S_m \otimes S_n$\,.
A sum over $S_m \otimes S_n$ gives
\begin{multline}
\frac{1}{m! \, n!} \sum_{\sigma \in S_m \otimes S_n} D_{IJ}^R (\sigma) \, 
B^\dagger{}^{R \to (r_1, r_2) \nu_+}_{I \to (i,j)} \, B^{R \to (r_1, r_2) \nu_-}_{J \to (k,l)}
D_{MN}^S (\sigma) \, 
B^\dagger{}^{S \to (s_1, s_2) \mu_+}_{M \to (m,n)} \, B^{S \to (s_1, s_2) \mu_-}_{N \to (p,q)}
\\
= \frac{\delta^{r_1 s_1} \delta^{r_2 s_2}}{d_{r_1} d_{r_2}} \,
\delta^{\nu_+\nu_-} \, \delta^{\mu_+\mu_-} \,
\delta_{i,m} \, \delta_{j,n} \, \delta_{k,p} \, \delta_{l,q} 
\label{restricted grand orthogonality2}
\end{multline}
where we used \eqref{perm matrix after branching3}

\section{Quiver calculus}\label{app:quiver calc}

Let us introduce a graphical notation of various representation-theoretical objects following \cite{Pasukonis:2013ts}.
We denote the indices of $R \vdash L=(m+n)$ by a double line, and those of $r_1 \vdash m$ or $r_2 \vdash n$ by a single line. We use different lines to distinguish two set of representations $\{ R, (r_1, r_2) \}$ and $\{ S, (s_1, s_2) \}$.

The matrix representation of a permutation group element is represented by
\begin{equation}
\begin{tikzpicture}[node distance=10mm, baseline=(current bounding box.center)]
\begin{scope}[decoration={markings, mark=at position 0.7 with {\arrow{Straight Barb[]}}}]
 \node (sig1) [perm] {$\sigma$};
 \node (eq) [left of=sig1,xshift=-4mm] {$D^R_{IJ} (\sigma) \ =\ $};
 \node (id1) [above of=sig1, yshift=3mm] {$I$};
 \node (id2) [below of=sig1, yshift=-3mm] {$J$};                   
 \draw [lineR] (id1) -- (sig1);
 \draw [lineR] (sig1) -- (id2);
 \node (eq2) [right of=sig1] {$=\ $};
 \node (sig2) [perm, right of=eq2] {$\sigma$};
 \node (id3) [above of=sig2, yshift=3mm] {$J$};
 \node (id4) [below of=sig2, yshift=-3mm] {$I$};                   
 \draw [lineR] (sig2) -- (id3);
 \draw [lineR] (id4) -- (sig2);
 \node (eq3) [right of=sig2] {$=\ $};
 \node (sig3) [perm2, right of=eq3] {$\sigma^{-1}$};
 \node (id5) [above of=sig3, yshift=3mm] {$I$};
 \node (id6) [below of=sig3, yshift=-3mm] {$J$};                   
 \draw [lineR] (sig3) -- (id5);
 \draw [lineR] (id6) -- (sig3);
\end{scope}
\end{tikzpicture}
\end{equation}
by using \eqref{real unitary rep}. Note that the matrix transposition is represented as flipping all the arrow directions.
The composition of permutations is
\begin{equation}
\begin{tikzpicture}[node distance=10mm, baseline=(current bounding box.center)]
\begin{scope}[decoration={markings, mark=at position 0.7 with {\arrow{Straight Barb[]}}}]
 \node (sig1) [perm] {$\sigma \, \tau$};
 \node (eq) [left of=sig1,xshift=-30mm] {$\ds D^R_{IJ} (\sigma \tau) \ = \ \sum_{K=1}^{d_R} D^R_{IK} (\sigma) D^R_{KJ} (\tau) \ =\ $};
 \node (id1) [above of=sig1, yshift=3mm] {$I$};
 \node (id2) [below of=sig1, yshift=-3mm] {$J$};                   
 \draw [lineR] (id1) -- (sig1);
 \draw [lineR] (sig1) -- (id2);
 \node (eq2) [right of=sig1] {$=\ $};
 \node (sig2) [perm, above right of=eq2, yshift=1mm] {$\sigma$};
 \node (sig3) [perm, below of=sig2, yshift=-1mm] {$\tau$};
 \node (id3) [above of=sig2, yshift=2mm] {$I$};
 \node (id4) [below of=sig3, yshift=-2mm] {$J$};                   
 \draw [lineR] (sig2) -- (sig3);
 \draw [lineR] (id3) -- (sig2);
 \draw [lineR] (sig3) -- (id4);
\end{scope}
\end{tikzpicture}
\end{equation}
The grand orthogonality relation \eqref{orthogonality_representation} is
\begin{equation}
\frac{1}{L!} \, \sum_{\sigma \in S_L} = \ \ 
\begin{tikzpicture}[node distance=10mm, baseline=(current bounding box.center)]
\begin{scope}[decoration={markings, mark=at position 0.7 with {\arrow{Straight Barb[]}}}]
 \node (sig1) [perm] {$\sigma$};
 \node (id1) [above of=sig1, yshift=2mm] {$I$};
 \node (id2) [below of=sig1, yshift=-2mm] {$J$};
 \draw [lineR] (id1) -- (sig1);
 \draw [lineR] (sig1) -- (id2);
 \node (sig2) [perm, right of=sig1, xshift=2mm] {$\sigma^{-1}$};
 \node (id3) [above of=sig2, yshift=2mm] {$K$};
 \node (id4) [below of=sig2, yshift=-2mm] {$L$};
 \draw [lineS] (id3) -- (sig2);
 \draw [lineS] (sig2) -- (id4);
 \node (eq) [right of=sig2, xshift=4mm] {$\ds \ \ = \ \frac{\delta^{RS}}{d_R}\ $};
 \node (id5) [above right of=eq, xshift=3mm, yshift=4mm] {$I$};
 \node (id6) [below right of=eq, xshift=3mm, yshift=-4mm] {$J$};
 \node (id7) [right of=id5] {$K$};
 \node (id8) [right of=id6] {$L$};
 \draw [thick, double distance=1pt, -{Straight Barb[]}] (id5) -- (id8);  
 \draw [thick, double distance=1pt, -{Straight Barb[]}] (id7) -- (id6);
\end{scope}
\end{tikzpicture}
= \ \frac{\delta^{RS}}{d_R} \, \delta_{IL} \, \delta_{JK}
\label{grand orthogonality quiver1}
\end{equation}
or equivalently
\begin{equation}
\begin{tikzpicture}[node distance=10mm, baseline=(current bounding box.center)]
\begin{scope}[decoration={markings, mark=at position 0.7 with {\arrow{Straight Barb[]}}}]
 \node (sig10) [perm] {$\sigma$};
 \node (sum0) [left of=sig10, xshift=-3mm] {\raisebox{-6mm}{$\ds \frac{1}{L!} \, \sum_{\sigma \in S_L}$}};
 \node (id10) [above of=sig10, yshift=2mm] {$I$};
 \node (id20) [below of=sig10, yshift=-2mm] {$J$};
 \draw [lineR] (id10) -- (sig10);
 \draw [lineR] (sig10) -- (id20);
 \node (sig20) [perm, right of=sig10, xshift=2mm] {$\sigma$};
 \node (id30) [above of=sig20, yshift=2mm] {$K$};
 \node (id40) [below of=sig20, yshift=-2mm] {$L$};
 \draw [lineS] (id30) -- (sig20);
 \draw [lineS] (sig20) -- (id40);
 \node (sig1) [perm, right of=sig20, xshift=21mm] {$\sigma$};
 \node (sum) [left of=sig1, xshift=-4mm] {\raisebox{-6mm}{$\ds =\frac{1}{L!} \, \sum_{\sigma \in S_L}$}};
 \node (id1) [above of=sig1, yshift=2mm] {$I$};
 \node (id2) [below of=sig1, yshift=-2mm] {$J$};
 \draw [lineR] (id1) -- (sig1);
 \draw [lineR] (sig1) -- (id2);
 \node (sig2) [perm2, right of=sig1, xshift=2mm] {$\sigma^{-1}$};
 \node (id3) [above of=sig2, yshift=2mm] {$K$};
 \node (id4) [below of=sig2, yshift=-2mm] {$L$};
 \draw [lineS] (sig2) -- (id3);
 \draw [lineS] (id4) --  (sig2);
 \node (eq) [right of=sig2, xshift=4mm] {$\ds \ \ = \ \frac{\delta^{RS}}{d_R}\ $};
 \node (eq) [right of=sig2, xshift=4mm] {$\ds \ \ = \ \frac{\delta^{RS}}{d_R}\ $};
 \node (id5) [above right of=eq, xshift=3mm, yshift=4mm] {$I$};
 \node (id6) [below right of=eq, xshift=3mm, yshift=-4mm] {$J$};
 \node (id7) [right of=id5] {$K$};
 \node (id8) [right of=id6] {$L$};
 \draw [thick, double distance=1pt, -{Straight Barb[]}] (id5) .. controls (8.4,.2) .. (id7);  
 \draw [thick, double distance=1pt, -{Straight Barb[]}] (id8) .. controls (8.4,-.2) .. (id6);
\end{scope}
\end{tikzpicture}
= \ \frac{\delta^{RS}}{d_R} \, \delta_{IK} \, \delta_{JL} \,.
\label{grand orthogonality quiver2}
\end{equation}
The branching coefficients \eqref{app:branching coeffs} are represented as
\begin{equation}
B^{R \to (r_1, r_2) \nu}_{I \to (i,j)} =
\begin{tikzpicture}[node distance=10mm, baseline=(current bounding box.center)]
\begin{scope}[decoration={markings, mark=at position 0.7 with {\arrow{Straight Barb[]}}}]
 \node (br) [branch] {$\nu$};
 \node (id1) [above of=br, yshift=3mm] {$I$};
 \node (id2) [below left of=br, yshift=-4mm] {$i$};
 \node (id3) [below right of=br, yshift=-4mm] {$j$};
 \draw [lineR] (id1) -- (br);
 \draw [liner1] (br) -- (id2);
 \draw [liner2] (br) -- (id3);
\end{scope}
\end{tikzpicture}
\hspace{18mm}
\brT^{R \to (r_1, r_2) \nu}_{I \to (i,j)} =
\begin{tikzpicture}[node distance=10mm, baseline=(current bounding box.center)]
\begin{scope}[decoration={markings, mark=at position 0.7 with {\arrow{Straight Barb[]}}}]
 \node (br2) [branch] {$\nu$};
 \node (id4) [above left of=br2, yshift=4mm] {$i$};
 \node (id5) [above right of=br2, yshift=4mm] {$j$};
 \node (id6) [below of=br2, yshift=-3mm] {$I$};
 \draw [lineR] (br2) -- (id6);
 \draw [liner1] (id4) -- (br2);
 \draw [liner2] (id5) -- (br2);
\end{scope}
\end{tikzpicture}
\end{equation}
We use double lines for the  indices of $S_{m+n}$\,, wavy lines for $S_m$ and straight lines for $S_n$\,.
The completeness relations of the branching coefficients \eqref{product branching1}, \eqref{product branching2} are
\begin{equation}
\begin{tikzpicture}[node distance=10mm, baseline=(current bounding box.center)]
\begin{scope}[decoration={markings, mark=at position 0.7 with {\arrow{Straight Barb[]}}}]
 \node (sum) {\raisebox{-6mm}{$\ds \sum_{r_1, r_2, \nu}$}};
 \node (br1) [branch, above right of=sum, xshift=6mm, yshift=-1mm] {$\nu$};
 \node (br2) [branch, below right of=sum, xshift=6mm, yshift=1mm] {$\nu$};
 \draw [liner1] (br1.south west) -- (br2.north west);
 \draw [liner2] (br1.south east) -- (br2.north east);
 \node (id1) [above of=br1,yshift=2mm] {$I$};
 \node (id2) [below of=br2,yshift=-2mm] {$J$};
 \draw [lineR] (id1) -- (br1);
 \draw [lineR] (br2) -- (id2);
\end{scope}
\end{tikzpicture}
\ \ = \ \ 
\begin{tikzpicture}[node distance=10mm, baseline=(current bounding box.center)]
\begin{scope}[decoration={markings, mark=at position 0.7 with {\arrow{Straight Barb[]}}}]
\node (id1) {$I$};
\node (id2) [below of=id1,yshift=-20mm] {$J$};
\draw [lineR] (id1) -- (id2);
\end{scope}
\end{tikzpicture}
\hspace{18mm}
\begin{tikzpicture}[node distance=10mm, baseline=(current bounding box.center)]
\begin{scope}[decoration={markings, mark=at position 0.7 with {\arrow{Straight Barb[]}}}]
 \node (br1) [branch] {$\nu$};
 \node (br2) [branch, below of=br1, yshift=-4mm] {$\mu$};
 \draw [lineR] (br1) -- (br2);
 \node (id1) [above left of=br1, yshift=4mm] {$i$};
 \node (id2) [above right of=br1, yshift=4mm] {$j$};
 \node (id3) [below left of=br2, yshift=-4mm] {$k$};
 \node (id4) [below right of=br2, yshift=-4mm] {$l$};
 \draw [liner1] (id1) -- (br1);
 \draw [liner2] (id2) -- (br1);
 \draw [lines1] (br2) -- (id3);
 \draw [lines2] (br2) -- (id4);
\end{scope}
\end{tikzpicture}
= \ \delta^{\nu\mu} \, \delta^{r_1 s_1} \, \delta^{r_2 s_2} \ \ 
\begin{tikzpicture}[node distance=10mm, baseline=(current bounding box.center)]
\begin{scope}[decoration={markings, mark=at position 0.7 with {\arrow{Straight Barb[]}}}]
\node (id1) {$i$};
\node (id2) [right of=id1, xshift=-5mm] {$j$};
\node (id3) [below of=id1, yshift=-20mm] {$k$};
\node (id4) [below of=id2, yshift=-20mm] {$l$};
\draw [liner1] (id1) -- (id3);
\draw [liner2] (id2) -- (id4);
\end{scope}
\end{tikzpicture}
\end{equation}
where we assumed that $r_1 \otimes r_2$ and $s_1 \otimes s_2$ follow from the same restriction of $R$.
If the two product representations descend from different restrictions, we get the orthogonal matrix \eqref{def:intertwining mat}
\begin{equation}
\begin{tikzpicture}[node distance=10mm, baseline=(current bounding box.center)]
\begin{scope}[decoration={markings, mark=at position 0.7 with {\arrow{Straight Barb[]}}}]
 \node (br1) [branch] {$\nu$};
 \node (br2) [branch2, below of=br1, yshift=-4mm] {$\mu$};
 \draw [lineR] (br1) -- (br2);
 \node (id1) [above left of=br1, yshift=4mm] {$i$};
 \node (id2) [above right of=br1, yshift=4mm] {$j$};
 \node (id3) [below left of=br2, yshift=-4mm] {$k$};
 \node (id4) [below right of=br2, yshift=-4mm] {$l$};
 \draw [liner1] (id1) -- (br1);
 \draw [liner2] (id2) -- (br1);
 \draw [lines1] (br2) -- (id3);
 \draw [lines2] (br2) -- (id4);
\end{scope}
\end{tikzpicture}
= \ 
\begin{tikzpicture}[node distance=10mm, baseline=(current bounding box.center)]
 \node (ct) [branch] {$\ds \frac{\nu}{\mu}$};
 \node (id5) [above left of=ct, yshift=8mm] {$i$};
 \node (id6) [above right of=ct, yshift=8mm] {$j$};
 \node (id7) [below left of=ct, yshift=-8mm] {$k$};
 \node (id8) [below right of=ct, yshift=-8mm] {$l$};
 \draw [liner1] (id5) -- (ct);
 \draw [liner2] (id6) -- (ct);
 \draw [lines1] (ct) -- (id7);
 \draw [lines2] (ct) -- (id8);
\end{tikzpicture}
\end{equation}
The relation \eqref{perm matrix after branching2} is expressed as
\begin{equation}
\begin{tikzpicture}[node distance=10mm, baseline=(current bounding box.center)]
\begin{scope}[decoration={markings, mark=at position 0.7 with {\arrow{Straight Barb[]}}}]
 \node (br) [branch] {$\nu$};
 \node (id1) [above of=br, yshift=4mm] {$I$};
 \node (gam1) [perm, below left of=br, yshift=-4mm] {$\gamma_1$};
 \node (gam2) [perm, below right of=br, yshift=-4mm] {$\gamma_2$};
 \node (id2) [below of=gam1, yshift=-2mm] {$i$};
 \node (id3) [below of=gam2, yshift=-2mm] {$j$};
 \draw [lineR] (id1) -- (br);
 \draw [liner1] (br) -- (gam1);
 \draw [liner2] (br) -- (gam2);
 \draw [liner1] (gam1) -- (id2);
 \draw [liner2] (gam2) -- (id3);
 \node (eq2) [below right of=br, xshift=12mm] {$\ =\ $};
 \node (br2) [branch, right of=eq2, xshift=6mm, yshift=-3mm] {$\nu$};
 \node (gam3) [perm, above of=br2, yshift=4mm] {$\gamma_1 \circ \gamma_2$};
 \node (id4) [below left of=br2, yshift=-4mm] {$i$};
 \node (id5) [below right of=br2, yshift=-4mm] {$j$};
 \node (id6) [above of=gam3, yshift=4mm] {$I$};
 \draw [lineR] (id6) -- (gam3);
 \draw [lineR] (gam3) -- (br2);
 \draw [liner1] (br2) -- (id4);
 \draw [liner2] (br2) -- (id5);
\end{scope}
\end{tikzpicture}
\end{equation}
The identity for multiple branching coefficients \eqref{multiple branching identity} is
\begin{equation}
\begin{tikzpicture}[node distance=10mm, baseline=(current bounding box.center)]
\begin{scope}[decoration={markings, mark=at position 0.7 with {\arrow{Straight Barb[]}}}]
\node (id0) {$I$};
\node (br) [branch,below of=id0, yshift=-5mm] {$\eta$};
\node (id4) [below of=br, yshift=-15mm, xshift=-2mm] {$a_l$};
\node (id5) [right of=id4, xshift=4mm] {$a_{l+1}$};
\node (id3) [left of=id4] {$\dots$};
\node (id2) [left of=id3] {$a_2$};
\node (id1) [left of=id2] {$a_1$};
\draw[lineR] (id0) -- (br);
\draw [liner1] (br) .. controls (-1.5,-2) .. (id1);
\draw [liner2] (br) .. controls (-1.2,-2.5) .. (id2);
\draw [lines1] (br) .. controls (-.2,-3) .. (id4);
\draw [lines2] (br) .. controls (.5,-2) .. (id5);
\end{scope}
\end{tikzpicture}
= \ \sum_{S} 
\begin{tikzpicture}[node distance=10mm, baseline=(current bounding box.center)]
\begin{scope}[decoration={markings, mark=at position 0.7 with {\arrow{Straight Barb[]}}}]
\node (id0) {$I$};
\node (br) [branch, below of=id0, yshift=-5mm] {$\mu$};
\node (br2) [branch, below left of=br, yshift=-5mm] {$\rho$};
\node (id4) [below of=br, yshift=-20mm, xshift=-2mm] {$a_l$};
\node (id5) [right of=id4, xshift=4mm] {$a_{l+1}$};
\node (id3) [left of=id4] {$\dots$};
\node (id2) [left of=id3] {$a_2$};
\node (id1) [left of=id2] {$a_1$};
\node (idS) [below left of=br, xshift=-1mm, yshift=3mm] {$S$};
\draw[lineR] (id0) -- (br);
\draw[lineS] (br) -- (br2);
\draw [liner1] (br2) .. controls (-2,-3) .. (id1);
\draw [liner2] (br2) .. controls (-1.7,-3.5) .. (id2);
\draw [lines1] (br2) .. controls (-.5,-3.5) .. (id4);
\draw [lines2] (br) .. controls (.8,-2) .. (id5);
\end{scope}
\end{tikzpicture}
\end{equation}

The character and the restricted characters are
\begin{equation}
\begin{tikzpicture}[node distance=10mm, baseline=(current bounding box.center)]
\begin{scope}[decoration={markings, mark=at position 0.7 with {\arrow{Straight Barb[]}}}]
 \node (sig1) [perm] {$\sigma$};
 \node (chi1) [left of=sig1,xshift=-20mm] {$\chi^{R} (\sigma) \, = \, \chi^{R} (\sigma^{-1}) \, = \ $};
 \draw [lineR] (sig1.south) .. controls (-1,-3) and (-1,3) .. (sig1.north);
 \node (sig2) [perm, right of=sig1, xshift=55mm] {$\sigma$};
 \node (chi) [left of=sig2,xshift=-20mm] {$\chi^{R (r_1,r_2) (\nu_+,\nu_-)} (\sigma) \ =\ $};
 \node (br1) [branch, above of=sig2, yshift=4mm] {$\nu_+$};
 \node (br2) [branch, below of=sig2, yshift=-4mm] {$\nu_-$}; 
 \draw [lineR] (br1) -- (sig2);
 \draw [lineR] (sig2) -- (br2);
 \draw [liner1] (br2.west) .. controls (5.5,-0.5) and (5.5,0.5) .. (br1.west);
 \draw [liner2] (br2.east) .. controls (7.5,-0.5) and (7.5,0.5) .. (br1.east);
 \node (sig3) [perm, right of=sig2, xshift=16mm] {$\sigma^{-1}$};
 \node (chi3) [left of=sig3, xshift=-3mm] {$\ = \ \ $};
 \node (br3) [branch, above of=sig3, yshift=4mm] {$\nu_+$};
 \node (br4) [branch, below of=sig3, yshift=-4mm] {$\nu_-$}; 
 \draw [lineR] (sig3) -- (br3);
 \draw [lineR] (br4) -- (sig3);
 \draw [liner1] (br3.west) .. controls (8.1,0.5) and (8.1,-0.5) .. (br4.west);
 \draw [liner2] (br3.east) .. controls (10.1,0.5) and (10.1,-0.5) .. (br4.east);
\end{scope}
\end{tikzpicture}
\end{equation}
We can show the row orthogonality of the restricted character as
\begin{equation}
\begin{aligned}
\begin{tikzpicture}[node distance=10mm, baseline=(current bounding box.center)]
\begin{scope}[decoration={markings, mark=at position 0.7 with {\arrow{Straight Barb[]}}}]
 \node (sig) [perm] {$\sigma$};
 \node (sum) [left of=sig, xshift=-8mm] {\raisebox{-6mm}{$\ds \frac{1}{L!} \, \sum_{\sigma \in S_L}$}};
 \node (br1) [branch, above of=sig, yshift=4mm] {$\nu_+$};
 \node (br2) [branch, below of=sig, yshift=-4mm] {$\nu_-$}; 
 \draw [lineR] (br1) -- (sig);
 \draw [lineR] (sig) -- (br2);
 \draw [liner1] (br2.west) .. controls (-.8,-0.5) and (-.8,0.5) .. (br1.west);
 \draw [liner2] (br2.east) .. controls (.8,-0.5) and (.8,0.5) .. (br1.east);
 \node (sig2) [perm, right of=sig, xshift=10mm] {$\sigma^{-1}$};
 \node (br3) [branch, above of=sig2, yshift=4mm] {$\mu_+$};
 \node (br4) [branch, below of=sig2, yshift=-4mm] {$\mu_-$}; 
 \draw [lineR] (sig2) -- (br3);
 \draw [lineR] (br4) -- (sig2);
 \draw [lines1] (br3.west) .. controls (1.2,0.5) and (1.2,-0.5) .. (br4.west);
 \draw [lines2] (br3.east) .. controls (2.8,0.5) and (2.8,-0.5) .. (br4.east);
\end{scope}
\end{tikzpicture}
\ &= \ \frac{\delta^{RS}}{d_R} \ \ 
\begin{tikzpicture}[node distance=10mm, baseline=(current bounding box.center)]
\begin{scope}[decoration={markings, mark=at position 0.6 with {\arrow{Straight Barb[]}}}]
 \node (ct) [circle] {};
 \node (br5) [branch, above left of=ct, yshift=4mm] {$\nu_+$};
 \node (br6) [branch, below left of=ct, yshift=-4mm] {$\nu_-$};
 \node (br7) [branch, right of=br5, xshift=13mm] {$\mu_+$};
 \node (br8) [branch, right of=br6, xshift=13mm] {$\mu_-$};
 \draw [lineR] (br5.south) .. controls (0.5,0.2) .. (br7.south);
 \draw [lineR] (br8.north) .. controls (0.5,-0.2) .. (br6.north);
 \draw [liner1] (br6.west) .. controls (-1.4,-0.5) and (-1.4,0.5) .. (br5.west);
 \draw [liner2] (br6.east) .. controls (0.2,-0.5) and (0.2,0.5) .. (br5.east);
 \draw [lines1] (br7.west) .. controls (.8,0.5) and (.8,-0.5) .. (br8.west);
 \draw [lines2] (br7.east) .. controls (2.4,0.5) and (2.4,-0.5) .. (br8.east);
\end{scope}
\end{tikzpicture}
\\
&= \frac{d_{r_1} d_{r_2}}{d_R} \, \delta^{RS} \, \delta^{\nu_+ \mu_+} \, \delta^{\nu_- \mu_-} \, \delta^{r_1 s_1} \, \delta^{r_2 s_2} .
\end{aligned}
\end{equation}
To show the column orthogonality, we insert the resolution of identity on the irreducible representation $R$ by \eqref{orthogonality_representation},
\begin{equation}
\delta_{il} \, \delta_{jk} = \frac{d_R}{L!} \, \sum_{\gamma\in S_{L}}D^R_{ij}(\gamma)D^R_{kl}(\gamma^{-1}),
\qquad (i,j,k,l=1, 2, \dots, d_R).
\end{equation}
We obtain
\begin{equation}
\begin{tikzpicture}[node distance=10mm, baseline=(current bounding box.center)]
\begin{scope}[decoration={markings, mark=at position 0.7 with {\arrow{Straight Barb[]}}}]
 \node (sig1) [perm] {$\sigma$};
 \node (sum) [left of=sig1, xshift=-6mm] {\raisebox{-6mm}{$\ds \sum_{R \, \vdash L}$}};
 \draw [lineR] (sig1.north) .. controls (-1,3) and (-1,-3) .. (sig1.south);
 \node (sig2) [perm, right of=sig1, xshift=5mm] {$\tau$};
 \draw [lineR] (sig2.north) .. controls (.7,3) and (.7,-3) .. (sig2.south);
 \node (eq) [right of=sig2, xshift=10mm] {\raisebox{-6mm}{$\ds \ = \ \sum_{R \, \vdash L} \, \frac{d_R}{L!} \,\sum_{\gamma \in S_L} $}};
 \node (sig3) [perm, right of=eq, xshift=10mm] {$\sigma$};
 \node (gam1) [perm, shape border rotate=90, above right of=sig3, xshift=5mm, yshift=3mm] {$\gamma$};
 \node (gam2) [perm, shape border rotate=270, below right of=sig3, xshift=5mm, yshift=-3mm] {$\gamma^{-1}$}; 
 \node (sig4) [perm, right of=sig3, xshift=16mm] {$\tau^{-1}$};
 \draw [lineR] (sig3) -- (gam1.west);
 \draw [lineR] (gam2.west) -- (sig3);
 \draw [lineR] (gam1.east) -- (sig4);
 \draw [lineR] (sig4) -- (gam2.east);
\end{scope}
\end{tikzpicture}
\ \ = \sum_{\gamma \in S_L} \delta (\sigma \gamma \tau^{-1} \gamma^{-1})
\end{equation}
where we used \eqref{delta fn2}. Note that
\begin{equation}
\sum_{\gamma \in S_L} \delta (\sigma \gamma \tau^{-1} \gamma^{-1})
=
\sum_{\omega \in S_L} \delta (\sigma \omega \tau \omega^{-1}), \qquad
(\omega \tau = \gamma \in S_L).
\label{flipping identity}
\end{equation}
Similarly, we can derive the column orthogonality for the restricted characters \eqref{restricted column orthogonality}. By using
\begin{equation}
\begin{aligned}
\delta_{il} \, \delta_{jk} = \frac{d_{r_1}}{m!} \, \sum_{\gamma\in S_m}D^{r_1}_{ij}(\gamma_1)D^{r_1}_{kl}(\gamma_1^{-1}),
\qquad (i,j,k,l=1, 2, \dots, d_{r_1})
\\
\delta_{mq} \, \delta_{np} = \frac{d_{r_2}}{n!} \, \sum_{\gamma\in S_n}D^{r_2}_{mn}(\gamma_2)D^{r_2}_{pq}(\gamma_2^{-1}),
\qquad (i,j,k,l=1, 2, \dots, d_{r_2})
\end{aligned}
\end{equation}
we find
\begin{equation}
\begin{aligned}
\begin{tikzpicture}[node distance=10mm, baseline=(current bounding box.center)]
\begin{scope}[decoration={markings, mark=at position 0.7 with {\arrow{Straight Barb[]}}}]
 \node (sig) [perm] {$\sigma$};
 \node (sum) [left of=sig, xshift=-15mm] {\raisebox{-6mm}{$\ds \sum_{R,r_1,r_2, \nu_+ , \nu_-} \frac{d_R}{d_{r_1} d_{r_2}}$}};
 \node (br1) [branch, above of=sig, yshift=4mm] {$\nu_+$};
 \node (br2) [branch, below of=sig, yshift=-4mm] {$\nu_-$}; 
 \draw [lineR] (br1) -- (sig);
 \draw [lineR] (sig) -- (br2);
 \draw [liner1] (br2.west) .. controls (-.8,-0.5) and (-.8,0.5) .. (br1.west);
 \draw [liner2] (br2.east) .. controls (.8,-0.5) and (.8,0.5) .. (br1.east);
 \node (sig2) [perm, right of=sig, xshift=10mm] {$\tau$};
 \node (br3) [branch, above of=sig2, yshift=4mm] {$\nu_+$};
 \node (br4) [branch, below of=sig2, yshift=-4mm] {$\nu_-$}; 
 \draw [lineR] (br3) -- (sig2);
 \draw [lineR] (sig2) -- (br4);
 \draw [liner1] (br4.west) .. controls (1.2,-0.5) and (1.2,0.5) .. (br3.west);
 \draw [liner2] (br4.east) .. controls (2.8,-0.5) and (2.8,0.5) .. (br3.east);
\end{scope}
\end{tikzpicture}
&= \sum_{R,r_1, r_2, \nu_+ , \nu_-} \frac{d_R}{m! n!} \, \sum_{\substack{\gamma_1 \in S_m\\ \gamma_2 \in S_n}} \ \ 
\begin{tikzpicture}[node distance=10mm, baseline=(current bounding box.center)]
\begin{scope}[decoration={markings, mark=at position 0.7 with {\arrow{Straight Barb[]}}}]
 \node (sig) [perm] {$\sigma$};
 \node (br1) [branch, above of=sig, yshift=4mm] {$\nu_+$};
 \node (br2) [branch, below of=sig, yshift=-4mm] {$\nu_-$}; 
 \node (gam1a) [perm, above of=br1] {$\gamma_1$};
 \node (gam1b) [perm, right of=gam1a, xshift=6mm] {$\gamma_2$};
 \node (gam2a) [perm, below of=br2] {$\gamma_1^{-1}$}; 
 \node (gam2b) [perm, right of=gam2a, xshift=6mm] {$\gamma_2^{-1}$}; 
 \node (br3) [branch, below of=gam1b] {$\nu_+$};
 \node (br4) [branch, above of=gam2b] {$\nu_-$}; 
 \node (sig2) [perm, above of=br4, yshift=4mm] {$\tau^{-1}$};
 \draw [lineR] (br1) -- (sig);
 \draw [lineR] (sig) -- (br2);
 \draw [lineR] (sig2) -- (br3);
 \draw [lineR] (br4) -- (sig2);
 \draw [liner1] (gam1a.north west) .. controls (-1,2) and (-1.6,2) .. (br1.west);
 \draw [liner2] (gam1b.south west) -- (br1.east);
 \draw [liner1] (br3.west) -- (gam1a.south east);
 \draw [liner2] (br3.east) .. controls (3.2,2) and (2.4,2) .. (gam1b.north east);
 \draw [liner1] (br2.west) .. controls (-1.6,-2) and (-1,-2) .. (gam2a.bottom left corner);
 \draw [liner2] (br2.east) -- (gam2b.north west);
 \draw [liner1] (gam2a.north east) -- (br4.west);
 \draw [liner2] (gam2b.bottom right corner)  .. controls (2.4,-2) and (3.2,-2) .. (br4.east);
\end{scope}
\end{tikzpicture}
\\[2mm]
&= \sum_{R,r_1,r_2, \nu_+ , \nu_-} \frac{d_R}{m! n!} \, \sum_{\gamma \in S_m \otimes S_n} \ \ 
\begin{tikzpicture}[node distance=10mm, baseline=(current bounding box.center)]
\begin{scope}[decoration={markings, mark=at position 0.7 with {\arrow{Straight Barb[]}}}]
 \node (sig) [perm] {$\sigma$};
 \node (gam1) [perm, above of=sig, yshift=2mm] {$\gamma$};
 \node (gam2) [perm, below of=sig, yshift=-2mm] {$\gamma^{-1}$}; 
 \node (br1) [branch, above right of=gam1, xshift=4mm, yshift=2mm] {$\nu_+$};
 \node (br2) [branch, below right of=gam2, xshift=4mm, yshift=-2mm] {$\nu_-$}; 
 \node (br3) [branch, below right of=br1, xshift=4mm] {$\nu_+$};
 \node (br4) [branch, above right of=br2, xshift=4mm] {$\nu_-$}; 
 \node (sig2) [perm, above of=br4, yshift=4mm] {$\tau^{-1}$};
 \draw [lineR] (gam1) -- (sig);
 \draw [lineR] (sig) -- (gam2);
 \draw [lineR] (sig2) -- (br3);
 \draw [lineR] (br4) -- (sig2);
 \draw [lineR] (br1) -- (gam1);
 \draw [lineR] (gam2) -- (br2);
 \draw [liner1] (br3.west) -- (br1.south);
 \draw [liner2] (br3.east) .. controls (3,2) .. (br1.north east);
 \draw [liner1] (br2.north) -- (br4.west);
 \draw [liner2] (br2.south east) .. controls (3,-2) .. (br4.east);
\end{scope}
\end{tikzpicture}
\\[1mm]
&= \sum_{R \, \vdash L} \, \frac{d_R}{m! n!} \ \ 
\begin{tikzpicture}[node distance=10mm, baseline=(current bounding box.center)]
\begin{scope}[decoration={markings, mark=at position 0.7 with {\arrow{Straight Barb[]}}}]
 \node (sig) [perm] {$\sigma$};
 \node (gam1) [perm, shape border rotate=270, above right of=sig, xshift=3mm, yshift=3mm] {$\gamma$};
 \node (gam2) [perm, shape border rotate=90, below right of=sig, xshift=3mm, yshift=-3mm] {$\gamma^{-1}$}; 
 \node (sig2) [perm, right of=sig, xshift=12mm] {$\tau^{-1}$};
 \draw [lineR] (gam1) -- (sig);
 \draw [lineR] (sig) -- (gam2);
 \draw [lineR] (sig2) -- (gam1);
 \draw [lineR] (gam2) -- (sig2);
\end{scope}
\end{tikzpicture}
\\[2mm]
&= \frac{(m+n)!}{m! n!} \, \sum_{\gamma \in S_m \otimes S_n}
\delta (\sigma \gamma^{-1} \tau^{-1} \gamma) .
\end{aligned}
\end{equation}
In the last line, we cannot use \eqref{flipping identity}, because $\gamma \in S_m \otimes S_n \subsetneq S_{m+n}$.

We can show the restricted grand orthogonality \eqref{restricted grand orthogonality} by
\begin{equation}
\begin{aligned}
\begin{tikzpicture}[node distance=10mm, baseline=(current bounding box.center)]
\begin{scope}[decoration={markings, mark=at position 0.7 with {\arrow{Straight Barb[]}}}]
 \node (sig) [perm] {$\sigma$};
 \node (sum) [left of=sig, xshift=-8mm] {\raisebox{-6mm}{$\ds \frac{1}{L!} \, \sum_{\sigma \in S_L}$}};
 \node (br1) [branch, above of=sig, yshift=3mm] {$\nu_+$};
 \node (br2) [branch, below of=sig, yshift=-3mm] {$\nu_-$}; 
 \node (id1) [above left of=br1, yshift=3mm] {$i$};
 \node (id2) [above right of=br1, yshift=3mm] {$j$};
 \node (id3) [below left of=br2, yshift=-3mm] {$k$};
 \node (id4) [below right of=br2, yshift=-3mm] {$l$};
 \draw [lineR] (br1) -- (sig);
 \draw [lineR] (sig) -- (br2);
 \draw [liner1] (id1) -- (br1.north west);
 \draw [liner2] (id2) -- (br1.north east);
 \draw [liner1] (br2.south west) -- (id3);
 \draw [liner2] (br2.south east) -- (id4);
 \node (sig2) [perm, right of=sig, xshift=12mm] {$\sigma$};
 \node (br3) [branch, above of=sig2, yshift=3mm] {$\mu_+$};
 \node (br4) [branch, below of=sig2, yshift=-3mm] {$\mu_-$}; 
 \node (id5) [above left of=br3, yshift=3mm] {$m$};
 \node (id6) [above right of=br3, yshift=3mm] {$n$};
 \node (id7) [below left of=br4, yshift=-3mm] {$p$};
 \node (id8) [below right of=br4, yshift=-3mm] {$q$};
 \draw [lineS] (br3) -- (sig2);
 \draw [lineS] (sig2) -- (br4);
 \draw [lines1] (id5) -- (br3.north west);
 \draw [lines2] (id6) -- (br3.north east);
 \draw [lines1] (br4.south west) -- (id7);
 \draw [lines2] (br4.south east) -- (id8);
\end{scope}
\end{tikzpicture}
&= \frac{\delta^{RS}}{d_R} \,
\begin{tikzpicture}[node distance=10mm, baseline=(current bounding box.center)]
\begin{scope}[decoration={markings, mark=at position 0.7 with {\arrow{Straight Barb[]}}}]
 \node (sig) {};
 \node (br1) [branch, above of=sig, yshift=3mm] {$\nu_+$};
 \node (br2) [branch, below of=sig,, yshift=-3mm] {$\nu_-$}; 
 \node (id1) [above left of=br1, yshift=3mm] {$i$};
 \node (id2) [above right of=br1, yshift=3mm] {$j$};
 \node (id3) [below left of=br2, yshift=-3mm] {$k$};
 \node (id4) [below right of=br2, yshift=-3mm] {$l$};
 \draw [liner1] (id1) -- (br1.north west);
 \draw [liner2] (id2) -- (br1.north east);
 \draw [liner1] (br2.south west) -- (id3);
 \draw [liner2] (br2.south east) -- (id4);
 \node (sig2) [right of=sig, xshift=12mm] {};
 \node (br3) [branch, above of=sig2, yshift=3mm] {$\mu_+$};
 \node (br4) [branch, below of=sig2, yshift=-3mm] {$\mu_-$}; 
 \node (id5) [above left of=br3, yshift=3mm] {$m$};
 \node (id6) [above right of=br3, yshift=3mm] {$n$};
 \node (id7) [below left of=br4, yshift=-3mm] {$p$};
 \node (id8) [below right of=br4, yshift=-3mm] {$q$};
 \draw [lineR] (br1.south) .. controls (1,.2) .. (br3.south);
 \draw [lineR] (br2.north) .. controls (1,-.2) .. (br4.north);
 \draw [lines1] (br3.north west) -- (id5);
 \draw [lines2] (br3.north east) -- (id6);
 \draw [lines1] (id7) -- (br4.south west);
 \draw [lines2] (id8) -- (br4.south east);
\end{scope}
\end{tikzpicture}
\\[1mm]
&= \frac{\delta^{RS}}{d_R} \,
\delta^{\nu_+\mu_+} \, \delta^{\nu_-\mu_-} \,
\delta^{r_1,s_1} \, \delta^{r_2, s_2} \, 
\delta^{i,m} \, \delta^{j,n} \, \delta^{k,p} \, \delta^{l,q} \,.
\end{aligned}
\end{equation}

\paragraph{Restricted projector.}

The restricted projector \eqref{def:restricted projector} can be represented as
\begin{equation}
\proj^{R,(r_1,r_2),\nu_+,\nu_-} = \frac{d_R}{(m+n)!} \sum_{\sigma \in S_{m+n}} \sigma \cdot \ 
\begin{tikzpicture}[node distance=10mm, baseline=(current bounding box.center)]
\begin{scope}[decoration={markings, mark=at position 0.7 with {\arrow{Straight Barb[]}}}]
 \node (sig2) [perm] {$\sigma$};
 \node (br1) [branch, above of=sig2, yshift=4mm] {$\nu_+$};
 \node (br2) [branch, below of=sig2, yshift=-4mm] {$\nu_-$}; 
 \draw [lineR] (br1) -- (sig2);
 \draw [lineR] (sig2) -- (br2);
 \draw [liner1] (br2.west) .. controls (-1,-0.5) and (-1,0.5) .. (br1.west);
 \draw [liner2] (br2.east) .. controls (1,-0.5) and (1,0.5) .. (br1.east);
\end{scope}
\end{tikzpicture}
\end{equation}
which is an element of $\bb{C}[S_{m+n}]$ and not a number.
Its matrix elements are given by the branching coefficients \eqref{def:restricted projector R}, which can be shown by
\begin{equation}
\begin{aligned}
\proj_{IJ}^{R,(r_1,r_2),\nu_+,\nu_-} &= \frac{d_R}{(m+n)!} \sum_{\sigma \in S_{m+n}} \ \ 
\begin{tikzpicture}[node distance=10mm, baseline=(current bounding box.center)]
\begin{scope}[decoration={markings, mark=at position 0.7 with {\arrow{Straight Barb[]}}}]
 \node (sig1) [perm] {$\sigma$};
 \node (id1) [above of=sig1, yshift=2mm] {$I$};
 \node (id2) [below of=sig1, yshift=-2mm] {$J$};
 \draw [lineR] (id1) -- (sig1);
 \draw [lineR] (sig1) -- (id2); 
 \node (sig2) [perm, right of=sig1, xshift=7mm] {$\sigma$};
 \node (br1) [branch, above of=sig2, yshift=4mm] {$\nu_+$};
 \node (br2) [branch, below of=sig2, yshift=-4mm] {$\nu_-$}; 
 \draw [lineR] (br1) -- (sig2);
 \draw [lineR] (sig2) -- (br2);
 \draw [liner1] (br2.west) .. controls (.7,-0.5) and (.7,0.5) .. (br1.west);
 \draw [liner2] (br2.east) .. controls (2.7,-0.5) and (2.7,0.5) .. (br1.east);
\end{scope}
\end{tikzpicture}
\\
&= \frac{d_R}{(m+n)!} \sum_{\sigma \in S_{m+n}} \ \ 
\begin{tikzpicture}[node distance=10mm, baseline=(current bounding box.center)]
\begin{scope}[decoration={markings, mark=at position 0.7 with {\arrow{Straight Barb[]}}}]
 \node (sig1) [perm] {$\sigma$};
 \node (id1) [above of=sig1, yshift=2mm] {$I$};
 \node (id2) [below of=sig1, yshift=-2mm] {$J$};
 \draw [lineR] (id1) -- (sig1);
 \draw [lineR] (sig1) -- (id2); 
 \node (sig2) [perm, right of=sig1, xshift=7mm] {$\sigma^{-1}$};
 \node (br1) [branch, above of=sig2, yshift=4mm] {$\nu_+$};
 \node (br2) [branch, below of=sig2, yshift=-4mm] {$\nu_-$}; 
 \draw [lineR] (br2) -- (sig2);
 \draw [lineR] (sig2) -- (br1);
 \draw [liner1] (br1.west) .. controls (.7,0.5) and (.7,-0.5) .. (br2.west);
 \draw [liner2] (br1.east) .. controls (2.7,0.5) and (2.7,-0.5) .. (br2.east);
\end{scope}
\end{tikzpicture}
\ \ = \ \
\begin{tikzpicture}[node distance=10mm, baseline=(current bounding box.center)]
\begin{scope}[decoration={markings, mark=at position 0.8 with {\arrow{Straight Barb[]}}}]
 \node (br1) [branch] {$\nu_+$};
 \node (br2) [branch, below of=br1, yshift=-4mm] {$\nu_-$}; 
 \draw [liner1] (br1.south west) -- (br2.north west);
 \draw [liner2] (br1.south east) -- (br2.north east);
 \node (id1) [above of=br1,yshift=2mm] {$I$};
 \node (id2) [below of=br2,yshift=-2mm] {$J$};
 \draw [lineR] (id1) -- (br1);
 \draw [lineR] (br2) -- (id2);
\end{scope}
\end{tikzpicture}
\end{aligned}
\end{equation}
The identity \eqref{restricted projector relation2} follows from the calculation
\begin{equation}
\begin{aligned}
&\frac{d_R d_S}{(m+n)!^2} 
\sum_{\sigma, \tau \in S_{m+n}} \sigma \, \tau \cdot 
\begin{tikzpicture}[node distance=10mm, baseline=(current bounding box.center)]
\begin{scope}[decoration={markings, mark=at position 0.7 with {\arrow{Straight Barb[]}}}]
 \node (sig1) [perm] {$\sigma$};
 \node (br1) [branch, above of=sig1, yshift=4mm] {$\nu_+$};
 \node (br2) [branch, below of=sig1, yshift=-4mm] {$\nu_-$}; 
 \draw [lineR] (br1) -- (sig1);
 \draw [lineR] (sig1) -- (br2);
 \draw [liner1] (br2.west) .. controls (-1,-0.5) and (-1,0.5) .. (br1.west);
 \draw [liner2] (br2.east) .. controls (1,-0.5) and (1,0.5) .. (br1.east);
 \node (sig2) [perm, right of=sig1, xshift=10mm] {$\tau$};
 \node (br3) [branch, above of=sig2, yshift=4mm] {$\mu_+$};
 \node (br4) [branch, below of=sig2, yshift=-4mm] {$\mu_-$}; 
 \draw [lineS] (br3) -- (sig2);
 \draw [lineS] (sig2) -- (br4);
 \draw [lines1] (br4.west) .. controls (1.1,-0.5) and (1.1,0.5) .. (br3.west);
 \draw [lines2] (br4.east) .. controls (3.1,-0.5) and (3.1,0.5) .. (br3.east);
 \end{scope}
\end{tikzpicture}
\\
&= 
\frac{d_R d_S}{(m+n)!^2} \sum_{\sigma, \rho \, \in S_{m+n}} \rho \cdot  
\begin{tikzpicture}[node distance=10mm, baseline=(current bounding box.center)]
\begin{scope}[decoration={markings, mark=at position 0.7 with {\arrow{Straight Barb[]}}}]
 \node (sig1) [perm] {$\sigma$};
 \node (br1) [branch, above of=sig1, yshift=4mm] {$\nu_+$};
 \node (br2) [branch, below of=sig1, yshift=-4mm] {$\nu_-$}; 
 \draw [liner1] (br2.west) .. controls (-1,-0.5) and (-1,0.5) .. (br1.west);
 \draw [liner2] (br2.east) .. controls (1,-0.5) and (1,0.5) .. (br1.east);
 \node (sig2) [perm, above right of=sig1, xshift=13mm, yshift=-1mm] {$\sigma^{-1}$};
 \node (sig3) [perm, below right of=sig1, xshift=13mm, yshift=1mm] {$\rho$};
 \node (br3) [branch, above of=sig2, yshift=4mm] {$\mu_+$};
 \node (br4) [branch, below of=sig3, yshift=-4mm] {$\mu_-$}; 
 \draw [lineR] (br1) -- (sig1);
 \draw [lineR] (sig1) -- (br2);
 \draw [lineR] (br3) -- (sig2);
 \draw [lineR] (sig2) -- (sig3);
 \draw [lineR] (sig3) -- (br4);
  \draw [lines1] (br4.west) .. controls (1.1,-0.5) and (1.1,0.5) .. (br3.west);
 \draw [lines2] (br4.east) .. controls (3.1,-0.5) and (3.1,0.5) .. (br3.east);
 \end{scope}
\end{tikzpicture}
\\
&= \frac{\delta^{RS} \, d_R}{(m+n)!} \sum_{\rho \in S_{m+n}} \rho \cdot  \ \ 
\begin{tikzpicture}[node distance=10mm, baseline=(current bounding box.center)]
\begin{scope}[decoration={markings, mark=at position 0.7 with {\arrow{Straight Barb[]}}}]
 \node (sig1) {};
 \node (br1) [branch, above of=sig1, yshift=4mm] {$\nu_+$};
 \node (br2) [branch, below of=sig1, yshift=-4mm] {$\nu_-$}; 
 \draw [liner1] (br2.west) .. controls (-1,-0.5) and (-1,0.5) .. (br1.west);
 \draw [liner2] (br2.south) .. controls (-1,-2) and (-1.3,-0.5) .. (-1.3,0) .. controls (-1.3,0.5) and (-1,2) .. (br1.north);
 \node (sig2) [perm, right of=sig1, xshift=8mm] {$\rho$};
 \node (br3) [branch, above of=sig2, xshift=2mm, yshift=4mm] {$\mu_+$};
 \node (br4) [branch, below of=sig2, xshift=2mm, yshift=-4mm] {$\mu_-$}; 
 \draw [lineR] (br1) -- (sig2);
 \draw [lineR] (br3) -- (br2);
 \draw [lineR] (sig2) -- (br4);
 \draw [lines1] (br4.south) .. controls (3,-2) and (3.3,-0.5) .. (3.3,0) .. controls (3.3,0.5) and (3,2) .. (br3.north);
 \draw [lines2] (br4.east) .. controls (3.1,-0.5) and (3.1,0.5) .. (br3.east);
 \end{scope}
\end{tikzpicture}
\\
\ &= \delta^{RS} \, \delta^{r_1 s_1} \, \delta^{r_2 s_2} \, \delta^{\nu_- \mu_+} \, \frac{d_R}{(m+n)!} \sum_{\sigma \in S_{m+n}} \rho \cdot  \ \ 
\begin{tikzpicture}[node distance=10mm, baseline=(current bounding box.center)]
\begin{scope}[decoration={markings, mark=at position 0.7 with {\arrow{Straight Barb[]}}}]
 \node (sig1) [perm] {$\rho$};
 \node (br1) [branch, above of=sig1, yshift=4mm] {$\nu_+$};
 \node (br2) [branch, below of=sig1, yshift=-4mm] {$\mu_-$}; 
 \draw [lineR] (br1) -- (sig1);
 \draw [lineR] (sig1) -- (br2);
 \draw [liner1] (br2.west) .. controls (-1,-0.5) and (-1,0.5) .. (br1.west);
 \draw [liner2] (br2.east) .. controls (1,-0.5) and (1,0.5) .. (br1.east);
 \end{scope}
\end{tikzpicture}
\end{aligned}
\end{equation}

\section{Generalized Racah-Wigner tensor}\label{app:genRW}

The associativity of triple tensor-product representations gives rise to the $6j$ symbols, which is also called Wigner's $6j$ invariants \cite{WignerBook}, Racah $W$-coefficients \cite{Racah42} or recoupling coefficients \cite{Kramer68a},
\begin{equation}
\Biggl\{ \, \begin{matrix}
j_1 &j_2 & j_{1+2} \\
j_3 & J & j_{2+3}
\end{matrix} \, \Biggr\} : \, 
{\rm Hom} \Bigl( (j_1 \otimes j_2) \otimes j_3 , J \Bigr)
\ \to \ 
{\rm Hom} \Bigl( j_1 \otimes ( j_2 \otimes j_3 ) , J \Bigr).
\label{app:naive 6j}
\end{equation}
The problem of computing $6j$ symbol is called the Racah-Wigner calculus.

We construct a slightly general object from the branching coefficients. 
The generalized $6j$ symbol is covariant under the action of symmetric groups, and contains four multiplicity labels.

\subsection{Case of $\tilde C_{\circ\circ\circ}$}\label{app:Cooo}

Consider two ways of the double restriction
\begin{equation}
S_L \downarrow (S_{L_1+L_2} \otimes S_{L_3}) \downarrow (S_{L_1} \otimes S_{L_2} \otimes S_{L_3}) \,,
\qquad
S_L \downarrow (S_{L_1} \otimes S_{L_2+L_3}) \downarrow (S_{L_1} \otimes S_{L_2} \otimes S_{L_3} )
\end{equation}
with $L= L_1+L_2+L_3$\,, which corresponds to the calculation of $\tilde C_{\circ\circ\circ}$ in Section \ref{sec:tilde Cooo final}. They induce the irreducible decompositions
\begin{equation}
\begin{aligned}
\hat R &= \bigoplus_{R_{12}, q_3} g(R_{12}, q_3; \hat R) \, R_{12} \otimes q_3 
= \bigoplus_{q_1, q_2, q_3} g(q_1, q_2; R_{12}) g(R_{12}, q_3; \hat R) \, q_1 \otimes q_2 \otimes q_3 
\\[1mm]
\hat R &= \bigoplus_{R_{23}, q'_1} g(R_{23}, q'_1; \hat R) \, q'_1 \otimes R_{23} 
= \bigoplus_{q'_1, q'_2, q'_3} g(q'_2, q'_3; R_{23}) g(R_{23}, q'_1; \hat R) \, q'_1 \otimes q'_2 \otimes q'_3 \,.
\end{aligned}
\end{equation}
The corresponding branching coefficients are
\begin{alignat}{9}
\ket{ \atop{\hat R}{\, \hat I} } 
&= \ket{ \matop{R_{12} & q_3}{I & c}{\mu } } \, \brT^{\hat R \to (R_{12}, q_3), \mu}_{\hat I \to (I, c)} 
& &= \ket{ \matop{q_1 & q_2 & q_3}{a & b & c}{\mu \, \rho} } \, 
\brT^{\hat R \to (R_{12}, q_3), \mu}_{\hat I \to (I, c)} \, 
\brT^{R_{12} \to (q_1, q_2), \rho}_{I \to (a,b)} 
\notag\\[1mm]
&= \ket{ \matop{q'_1 & R_{23}}{a' & I'}{\mu' } } \, 
\tbrT^{\hat R \to (q'_1 , R_{23}), \mu'}_{\hat I \to (a', I')} 
& &= \ket{ \matop{q'_1 & q'_2 & q'_3}{a' & b' & c'}{\mu' \, \rho'} } \, 
\tbrT^{\hat R \to (q'_1 , R_{23}), \mu'}_{\hat I \to (a', I')} \, 
\tbrT^{R_{23} \to (q'_2, q'_3), \rho'}_{I' \to (b',c')} \,.
\label{partial branching}
\end{alignat}
The multiplicity labels $(\mu,\rho)$ and $(\mu',\rho')$ run over the spaces
\begin{equation}
\begin{aligned}
\xi \equiv (\mu,\rho) \in \cM_{12} \,, &\qquad
\abs{\cM_{12}} = g(q_1,q_2;R_{12}) \, g(R_{12}, q_3; \hat R)
\\[1mm]
\xi' \equiv (\mu',\rho') \in \cM_{23} \,, &\qquad
\abs{\cM_{23}} = g(q_2,q_3;R_{23}) \, g(R_{23}, q_1; \hat R) 
\end{aligned}
\label{ranges xis}
\end{equation}
which are subsets of the total multiplicity space induced by the irreducible decomposition
\begin{equation}
\begin{gathered}
\hat R = \bigoplus_{q_1, q_2, q_3} \ \bigoplus_{\eta \, \in \cM_{1,2,3}}
\( q_1 \otimes q_2 \otimes q_3 \)_\eta ,\qquad
\ket{ \atop{\hat R}{\, \hat I} } 
= \sum_{q_1, q_2, q_3, \eta}
\ket{ \matop{q_1 & q_2 & q_3}{a & b & c}{\eta} } \, 
\brT^{\hat R \to (q_1, q_2, q_3), \eta}_{\hat I \to (a, b, c)}
\\[1mm]
\eta \, \in \cM_{\rm tot} \,, \qquad
\abs{\cM_{\rm tot}} = g(q_1,q_2,q_3;\hat R).
\end{gathered}
\label{entire branching}
\end{equation}
From the identity \eqref{multiple branching identity}, we obtain the following relation between the branching coefficients in \eqref{partial branching} and \eqref{entire branching},
\begin{equation}
\begin{aligned}
\Vev{ \matop{\tilde q_1 & \tilde q_2 & \tilde q_3}{\tilde a & \tilde b & \tilde c}{\tilde \eta} \, \Big| \,
\matop{q_1 & q_2 & q_3}{a & b & c}{\mu \, \rho} }
&= \sum_{\tilde R_{12}}
\Vev{ \matop{\tilde q_1 & \tilde q_2 & \tilde q_3}{\tilde a & \tilde b & \tilde c}{\tilde \mu \, \tilde \rho} \, \Big| \,
\matop{q_1 & q_2 & q_3}{a & b & c}{\mu \, \rho} }
\\
&= \delta^{\tilde q_1 q_1} \, \delta^{\tilde q_2 q_2} \, \delta^{\tilde q_3 q_3} \, \delta^{\tilde \mu \mu} \, \delta^{\tilde \rho \rho} \,
\delta_{\tilde a a} \, \delta_{\tilde b b} \, \delta_{\tilde c c} 
\end{aligned}
\label{partial-entire overlap}
\end{equation}
where the RHS depends on $R_{12}$ through the multiplicity space of $(\mu, \rho)$ in \eqref{ranges xis}.

We define the orthogonal matrix \eqref{def:intertwining mat} between the two states by
\begin{align}
&U_{\hat R} \begin{pmatrix}[cccc|cc]
q_1 & q_2 & q_3 & R_{12} & \mu & \rho \\
q'_1 & q'_2 & q'_3 & R_{23} & \mu' & \rho'
\end{pmatrix}_{abc, a'b'c'} 
\equiv
\Vev{ \matop{q_1 & q_2 & q_3}{a & b & c}{\mu \, \rho} \, \Big| \,
\matop{q'_1 & q'_2 & q'_3}{a' & b' & c'}{\mu' \, \rho'} }
\label{def:URvev} \\
&\hspace{10mm}
= \sum_{\hat I=1}^{d_{\hat R}} \sum_{I=1}^{d_{R_{12}}} \sum_{I'=1}^{d_{R_{23}}}
\brT^{\hat R \to (R_{12}, q_3), \mu}_{\hat I \to (I, c)} \, 
\brT^{R_{12} \to (q_1, q_2), \rho}_{I \to (a,b)} \,
\tilde B^{\hat R \to (q'_1 , R_{23}), \mu'}_{\hat I \to (a', I')} \, 
\tilde B^{R_{23} \to (q'_2, q'_3), \rho'}_{I' \to (b',c')} 
\label{def:covariant 6j}
\end{align}
and call it the {\it generalized Racah-Wigner tensor}. 
Our notation is slightly redundant because the generalized Racah-Wigner tensor is proportional to $\prod_{i=1}^3 \delta^{q_i q'_i}$\,, which follows from \eqref{def:URvev}.
The usual $6j$ symbol for a symmetric group is given by
\begin{equation}
\tr (U_{\hat R}) \equiv
\sum_{a,b,c} U_{\hat R} \begin{pmatrix}[cccc|cc]
q_1 & q_2 & q_3 & R_{12} & \mu & \rho \\
q_1 & q_2 & q_3 & R_{23} & \mu' & \rho'
\end{pmatrix}_{abc, abc} \,.
\label{def:usual 6j}
\end{equation}
The generalized Racah-Wigner tensor can be depicted as
\begin{equation}
U_{\hat R} \begin{pmatrix}[cccc|cc]
q_1 & q_2 & q_3 & R_{12} & \mu & \rho \\
q'_1 & q'_2 & q'_3 & R_{23} & \mu' & \rho'
\end{pmatrix}_{abc, a'b'c'} = \ \ 
\begin{tikzpicture}[node distance=10mm, baseline=(current bounding box.center)]
\begin{scope}[decoration={markings, mark=at position 0.8 with {\arrow{Straight Barb[]}}}]
\node (id1) {$a'$};
\node (id2) [below of=id1, yshift=-3mm] {$b'$};
\node (id3) [below of=id2, yshift=-3mm] {$c'$};
\node (br23) [branch2, above right of=id3, xshift=8mm, yshift=-1mm] {$\rho'$};
\node (br1) [branch2, above right of=br23, xshift=8mm] {$\mu'$};
\node (br3) [branch, right of=br1, xshift=8mm] {$\mu\phantom{{}'}\!$};
\node (br12) [branch, above right of=br3, xshift=8mm, yshift=-1mm] {$\rho\phantom{{}'}\!$};
\node (id4) [above right of=br12, xshift=8mm] {$a$};
\node (id5) [below of=id4, yshift=-3mm] {$b$};
\node (id6) [below of=id5, yshift=-3mm] {$c$};
\draw [lines1] (id1) -- (br1);
\draw [lines2] (id2) -- (br23);
\draw [lines3] (id3) -- (br23);
\draw [lineS] (br23) -- (br1);
\draw [lineR] (br1) -- (br3);
\draw [lineT] (br3) -- (br12);
\draw [liner1] (br12) -- (id4);
\draw [liner2] (br12) -- (id5);
\draw [liner3] (br3) -- (id6);
\end{scope}
\end{tikzpicture}
\end{equation}

We want to compute the products of generalized Racah-Wigner tensors 
\begin{align}
\tr (U_{\hat R} \, \tilde U_{\hat R}) &\equiv
\sum_{\mu,\rho,\mu',\rho'}
U_{\hat R} \begin{pmatrix}[cccc|cc]
q_1 & q_2 & q_3 & R_{12} & \mu & \rho \\
q'_1 & q'_2 & q'_3 & R_{23} & \mu' & \rho'
\end{pmatrix}_{abc, a'b'c'}  \,
U_{\hat R} \begin{pmatrix}[cccc|cc]
q'_1 & q'_2 & q'_3 & R_{23} & \mu' & \rho' \\
q_1 & q_2 & q_3 & R_{12} & \mu & \rho
\end{pmatrix}_{a'b'c', abc} 
\notag \\[2mm]
\tr (U_{\hat R} \, \tilde U_{\hat R} \, \dbtilde{U}_{\hat R} ) &\equiv
\sum_{\mu,\rho,\mu',\rho',\mu'',\rho''}
U_{\hat R} \begin{pmatrix}[cccc|cc]
q_1 & q_2 & q_3 & R_{12} & \mu & \rho \\
q'_1 & q'_2 & q'_3 & R_{23} & \mu' & \rho'
\end{pmatrix}_{abc, a'b'c'} \ \times
\label{def:trUU} \\[1mm]
&\hspace{5mm} 
U_{\hat R} \begin{pmatrix}[cccc|cc]
q'_1 & q'_2 & q'_3 & R_{23} & \mu' & \rho' \\
q''_1 & q''_2 & q''_3 & R_{23} & \mu'' & \rho''
\end{pmatrix}_{a'b'c', a''b''c''} \,
U_{\hat R} \begin{pmatrix}[cccc|cc]
q''_1 & q''_2 & q''_3 & R_{23} & \mu'' & \rho'' \\
q_1 & q_2 & q_3 & R_{12} & \mu & \rho
\end{pmatrix}_{a''b''c'', abc} 
\notag
\end{align}
which are rewriting of the product of projectors \eqref{cI123 comp},
\begin{equation}
\begin{aligned}
\tr (U_{\hat R} \, \tilde U_{\hat R} )
&= \trb{\, \hat R} \( 
\fP^{\hat R \to \dots \to (q_1,q_2,q_3), \mu\rho,\mu\rho} \,
\tilde \fP^{\hat R \to \dots \to (q'_1,q'_2,q'_3), \mu'\rho',\mu'\rho'} 
\)
\\[1mm]
\tr (U_{\hat R} \, \tilde U_{\hat R} \, \dbtilde{U}_{\hat R} )
&= \trb{\, \hat R} \( 
\fP^{\hat R \to \dots \to (q_1,q_2,q_3), \mu\rho,\mu\rho} \,
\tilde \fP^{\hat R \to \dots \to (q'_1,q'_2,q'_3), \mu'\rho',\mu'\rho'} 
\dbtilde{\fP}^{\hat R \to \dots \to (q''_1,q''_2,q''_3), \mu''\rho'',\mu''\rho''} 
\) .
\end{aligned}
\label{RW as projectors}
\end{equation}
By using $\xi, \xi', \xi''$ in \eqref{ranges xis}, we depict these products as
\begin{equation}
\tr (U_{\hat R} \, \tilde U_{\hat R} ) =
\begin{tikzpicture}[node distance=10mm, baseline=(current bounding box.center)]
\begin{scope}[decoration={markings, mark=at position 0.7 with {\arrow{Straight Barb[]}}}]
\node (br1m) [branch] {$\xi\phantom{{}'\!}$};
\node (br1p) [branch, right of=br1m, xshift=8mm] {$\xi\phantom{{}'\!}$};
\node (br2p) [branch2, below of=br1m, yshift=-12mm] {$\xi'$};
\node (br2m) [branch2, right of=br2p, xshift=8mm] {$\xi'$};
\draw [lineR] (br2p) -- (br1m);
\draw [lineR] (br1p) -- (br2m);
\draw [liner1] (br1m) .. controls (1,.5) .. (br1p);
\draw [lines2] (br1m) -- (br1p);
\draw [liner2] (br1m) .. controls (1,-.5) .. (br1p);
\draw [liner1] (br2m) .. controls (1,-1.7) .. (br2p);
\draw [liner2] (br2m) .. controls (1,-2.7) .. (br2p);
\draw [lines2] (br2m) -- (br2p);
\end{scope}
\end{tikzpicture} 
\hspace{10mm}
\tr (U_{\hat R} \, \tilde U_{\hat R} \, \dbtilde{U}_{\hat R} ) = 
\begin{tikzpicture}[node distance=10mm, baseline=(current bounding box.center)]
\begin{scope}[decoration={markings, mark=at position 0.7 with {\arrow{Straight Barb[]}}}]
\node (proj) {};
\node (u13) {};
\node (u32) {};
\node (u21) {};
\node (br1m) [branch, left of=u13, xshift=-8mm] {$\xi\phantom{{}'\!}$};
\node (br3p) [branch3, right of=u13, xshift=8mm] {$\xi''$};
\node (br3m) [branch3, below right of=u32, xshift=4mm, yshift=-8mm] {$\xi''$};
\node (br2p) [branch2, above left of=u32, xshift=-3mm, yshift=8mm] {$\xi'$};
\node (br2m) [branch2, above right of=u21, xshift=3mm, yshift=8mm] {$\xi'$};
\node (br1p) [branch, below left of=u21, xshift=-4mm, yshift=-8mm] {$\xi\phantom{{}'\!}$};
\draw [lineR] (br1m) -- (br2p);
\draw [lineR] (br2m) -- (br3p);
\draw [lineR] (br3m) -- (br1p);
\draw [liner1] (br1p) .. controls (-1, -.6) .. (br1m);
\draw [liner2] (br1p) .. controls (-1.9, -.9) .. (br1m);
\draw [liner1] (br3p) .. controls (1, -.6) .. (br3m);
\draw [liner2] (br3p) .. controls (1.9, -.9) .. (br3m);
\draw [liner1] (br2p) .. controls (0,1) .. (br2m);
\draw [liner2] (br2p) .. controls (0,2.1) .. (br2m);
\draw [lines2] (br1p) -- (br1m);
\draw [lines2] (br2p) -- (br2m);
\draw [lines2] (br3p) -- (br3m);
\end{scope}
\end{tikzpicture} 
\end{equation}
By grouping pairs of nodes with the same color, we obtain the projector representation \eqref{RW as projectors}.
From the identity of the projectors \eqref{restricted projector relation2}, we get
\begin{equation}
\begin{aligned}
\tr (U_{\hat R} \, \tilde U_{\hat R} ) &=
\( \prod_{i=1}^3 \delta^{q_i q'_i} \, d_{q_i} \)
\delta^{\xi_1 \, \xi_2 } \,
\delta^{\xi_2 \, \xi_1 } 
\\
\tr (U_{\hat R} \, \tilde U_{\hat R}  \, \dbtilde{U}_{\hat R} ) &=
\( \prod_{i=1}^3 \delta^{q_i q'_i} \, \delta^{q_i q''_i} \, d_{q_i} \)
\delta^{\xi_1 \, \xi_2 } \,
\delta^{\xi_2 \, \xi_3 } \,
\delta^{\xi_3 \, \xi_1 } 
\end{aligned}
\label{identities trUUU}
\end{equation}
where we do not sum over the repeated indices ($\xi_i$'s).

The product $\tr (U_{\hat R} \, \tilde U_{\hat R})$ satisfies the following sum rules,
\begin{equation}
\begin{aligned}
\sum_{R_{23}} \sum_{\xi_1,\xi_2} \tr (U_{\hat R} \, \tilde U_{\hat R}) &= 
\( \prod_{i=1}^3 \delta^{q_i q'_i} \, d_{q_i} \)
g(q_1,q_2;R_{12}) \, g(R_{12}, q_3; \hat R)
\\
\sum_{R_{12}} \sum_{\xi_1,\xi_2} \tr (U_{\hat R} \, \tilde U_{\hat R}) &=
\( \prod_{i=1}^3 \delta^{q_i q'_i} \, d_{q_i} \)
g(q_2,q_3;R_{23}) \, g(R_{23}, q_1; \hat R) .
\end{aligned}
\label{constraints trUU}
\end{equation}
We can derive these sum rules by using the identities \eqref{multiple branching identity}, \eqref{product branching2} and \eqref{partial-entire overlap}, as
\begin{equation}
\begin{aligned}
&\sum_{R_{23}} \sum_{\mu,\rho,\mu',\rho'} \ \ 
\begin{tikzpicture}[node distance=10mm, baseline=(current bounding box.center)]
\begin{scope}[decoration={markings, mark=at position 0.8 with {\arrow{Straight Barb[]}}}]
\node (br1) [branch] {$\rho$};
\node (br2) [branch, below right of=br1, xshift=6mm] {$\mu$};
\node (br3) [branch2, right of=br2, xshift=6mm] {$\mu'$};
\node (br4) [branch2, below right of=br3, xshift=6mm] {$\rho'$};
\node (br5) [branch2, right of=br4, xshift=6mm] {$\rho'$};
\node (br6) [branch2, above right of=br5, xshift=6mm] {$\mu'$};
\node (br7) [branch, right of=br6, xshift=6mm] {$\mu$};
\node (br8) [branch, above right of=br7, xshift=6mm] {$\rho$};
\draw [lines1] (br3) .. controls (5,0) .. (br6);
\draw [lines2] (br4) .. controls (5,-1) ..  (br5);
\draw [lines3] (br4) .. controls (5,-2) ..  (br5);
\draw [lineS] (br3) -- (br4);
\draw [lineS] (br5) -- (br6);
\draw [lineR] (br2) -- (br3);
\draw [lineR] (br6) -- (br7);
\draw [lineT] (br1) -- (br2);
\draw [lineT] (br7) -- (br8);
\draw [liner1] (br8) .. controls (5,1) .. (br1);
\draw [liner2] (br8) .. controls (10,1.5) and (0,1.5) .. (br1);
\draw [liner3] (br7) .. controls (9,-3) and (1,-3) .. (br2);
\end{scope}
\end{tikzpicture}
\\
&= \sum_{\mu,\rho,\eta'} \ \ 
\begin{tikzpicture}[node distance=10mm, baseline=(current bounding box.center)]
\begin{scope}[decoration={markings, mark=at position 0.8 with {\arrow{Straight Barb[]}}}]
\node (br1) [branch] {$\rho$};
\node (br2) [branch, below right of=br1, xshift=6mm] {$\mu$};
\node (br3) [branch, right of=br2, xshift=6mm] {$\eta'$};
\node (br6) [branch, right of=br3, xshift=12mm] {$\eta'$};
\node (br7) [branch, right of=br6, xshift=6mm] {$\mu$};
\node (br8) [branch, above right of=br7, xshift=6mm] {$\rho$};
\draw [lines1] (br3) .. controls (4,0) .. (br6);
\draw [lines2] (br3) -- (br6);
\draw [lines3] (br3) .. controls (4,-1.5) ..  (br6);
\draw [lineR] (br2) -- (br3);
\draw [lineR] (br6) -- (br7);
\draw [lineT] (br1) -- (br2);
\draw [lineT] (br7) -- (br8);
\draw [liner1] (br8) .. controls (4,1) .. (br1);
\draw [liner2] (br8) .. controls (8,1.5) and (0,1.5) .. (br1);
\draw [liner3] (br7) .. controls (7,-2) and (1,-2) .. (br2);
\end{scope}
\end{tikzpicture}
\\
&= \delta^{q'_1 q_1} \, \delta^{q'_2 q_2} \, \delta^{q'_3 q_3} \,
d_{q_1} d_{q_2} d_{q_3} \,
g(R_{12}, q_3 ; \hat R) \, g(q_1, q_2 ; R_{12}) \,.
\end{aligned}
\end{equation}
A solution to the equations \eqref{constraints trUU} is
\begin{equation}
\sum_{\xi_1,\xi_2} \tr (U_{\hat R} \tilde U_{\hat R}) \stackrel{?}{=} 
\( \prod_{i=1}^3 \delta^{q_i q'_i} \, d_{q_i} \) 
\frac{g(q_1,q_2;R_{12}) \, g(R_{12}, q_3; \hat R) \, 
g(q_2,q_3;R_{23}) \, g(R_{23}, q_1; \hat R)}{g (q_1, q_2, q_3; \hat R)} \,.
\end{equation}
We conjecture that both sides are equal, and continue the discussion below.
Similarly, we find
\begin{align}
\sum_{R_{31}} \sum_{\xi_1,\xi_2, \xi_3} \tr (U_{\hat R} \tilde U_{\hat R} \dbtilde{U}_{\hat R} ) &=
\( \prod_{i=1}^3 \delta^{q''_i q_i} \, \delta^{q''_i q'_i} \)
\sum_{\mu,\rho,\mu',\rho'} 
U_{\hat R} \begin{pmatrix}[cccc|cc]
q_1 & q_2 & q_3 & R_{12} & \mu & \rho \\
q'_1 & q'_2 & q'_3 & R_{23} & \mu' & \rho'
\end{pmatrix}_{abc, a'b'c'} \ \times
\\
&\hspace{60mm}
U_{\hat R} \begin{pmatrix}[cccc|cc]
q'_1 & q'_2 & q'_3 & R_{23} & \mu' & \rho' \\
q_1 & q_2 & q_3 & R_{12} & \mu & \rho 
\end{pmatrix}_{a'b'c',abc}
\notag \\[2mm]
\sum_{R_{23}} \sum_{\xi_1,\xi_2, \xi_3} \tr (U_{\hat R} \tilde U_{\hat R} \dbtilde{U}_{\hat R} ) &=
\( \prod_{i=1}^3 \delta^{q'_i q_i} \, \delta^{q'_i q''_i} \)
\sum_{\mu,\rho,\mu'',\rho''} 
U_{\hat R} \begin{pmatrix}[cccc|cc]
q_1 & q_2 & q_3 & R_{12} & \mu & \rho \\
q''_1 & q''_2 & q''_3 & R_{31} & \mu'' & \rho''
\end{pmatrix}_{abc, a''b''c''} \ \times
\notag \\
&\hspace{60mm}
U_{\hat R} \begin{pmatrix}[cccc|cc]
q''_1 & q''_2 & q''_3 & R_{31} & \mu'' & \rho'' \\
q_1 & q_2 & q_3 & R_{12} & \mu & \rho 
\end{pmatrix}_{a''b''c'',abc}
\notag \\[2mm]
\sum_{R_{12}} \sum_{\xi_1,\xi_2, \xi_3} \tr (U_{\hat R} \tilde U_{\hat R} \dbtilde{U}_{\hat R} ) &=
\( \prod_{i=1}^3 \delta^{q''_i q_i} \, \delta^{q''_i q'_i} \)
\sum_{\mu',\rho',\mu'',\rho''} 
U_{\hat R} \begin{pmatrix}[cccc|cc]
q''_1 & q''_2 & q''_3 & R_{31} & \mu'' & \rho'' \\
q'_1 & q'_2 & q'_3 & R_{23} & \mu' & \rho'
\end{pmatrix}_{a''b''c'',a'b'c'} \ \times
\notag \\
&\hspace{60mm}
U_{\hat R} \begin{pmatrix}[cccc|cc]
q'_1 & q'_2 & q'_3 & R_{23} & \mu' & \rho' \\
q''_1 & q''_2 & q''_3 & R_{31} & \mu'' & \rho''
\end{pmatrix}_{a'b'c', a''b''c''} .
\notag
\end{align}
A solution to these equations is
\begin{multline}
\sum_{\xi_1,\xi_2, \xi_3} \tr (U_{\hat R} \tilde U_{\hat R} \dbtilde{U}_{\hat R} )
= \( \prod_{i=1}^3 \delta^{q_i q'_i} \, \delta^{q'_i q''_i} \, d_{q_i} \) \ \times
\\
\frac{g(q_1,q_2;R_{12}) \, g(R_{12}, q_3; \hat R) \, 
g(q_2,q_3;R_{23}) \, g(R_{23}, q_1; \hat R) \, 
g(q_3,q_1;R_{31}) \, g(R_{31}, q_2; \hat R)}{g (q_1, q_2, q_3; \hat R)^2} \,.
\label{trUUU conjecture}
\end{multline}

In view of \eqref{identities trUUU}, our conjecture is summarized as
\begin{equation}
\begin{aligned}
\sum_{\xi_1 \in \cM_{12}} \sum_{\xi_2 \in \cM_{23}}
\delta^{\xi_1 \, \xi_2 } \,
\delta^{\xi_2 \, \xi_1 } 
&= \frac{\abs{\cM_{12}} \abs{\cM_{23}} }{\abs{ \cM_{\rm tot}} } 
\\[1mm]
\sum_{\xi_1 \in \cM_{12}} \sum_{\xi_2 \in \cM_{23}} \sum_{\xi_3 \in \cM_{31}} 
\delta^{\xi_1 \, \xi_2 } \,
\delta^{\xi_2 \, \xi_3 } \,
\delta^{\xi_3 \, \xi_1 } 
&= \frac{\abs{\cM_{12}} \abs{\cM_{23}} \abs{\cM_{31}} }{\abs{ \cM_{\rm tot}}^2 } \,.
\end{aligned}
\label{conjecture sum xxx}
\end{equation}

\subsection{Case of $\tilde C^{XYZ}_{\vec h}$}\label{app:CXYZ}

Consider another set of restrictions
\begin{equation}
\begin{aligned}
S_L &\downarrow \Big( 
\Big( ( S_{L_5} \otimes S_{L_6}) \otimes S_{L_1} \otimes S_{L_3} \Big) \otimes (S_{L_2} \otimes S_{L_4}) \Big)
\\
S_L &\downarrow \Big( 
\Big( ( S_{L_3} \otimes S_{L_4}) \otimes S_{L_2} \otimes S_{L_5} \Big) \otimes (S_{L_1} \otimes S_{L_6}) \Big)
\\
S_L &\downarrow \Big( 
\Big( ( S_{L_1} \otimes S_{L_2}) \otimes S_{L_4} \otimes S_{L_6} \Big) \otimes (S_{L_3} \otimes S_{L_5}) \Big)
\end{aligned}
\end{equation}
with $L = \sum_{i=1}^6 L_i$\,, which correspond to the case of $\tilde C^{XYZ}_{\vec h}$ in Section \ref{sec:tilde CXYZ final}.
They induce the irreducible decomposition
\begin{equation}
\begin{aligned}
\hat R &= \bigoplus_{Q, R,T}  \bigoplus_{ \{ q_i \} }
\Bigl\{ g(q_5,q_6; Q) g(Q,q_1,q_3; R) g(q_2, q_4; T) g(R,T;\hat R) \,
\bigotimes_{i=1}^6 q_i \Bigr\}
\\
\hat R &= \bigoplus_{Q', R',T'}  \bigoplus_{ \{ q'_i \} }
\Bigl\{ g(q'_3,q'_4; Q') g(Q',q'_2,q'_5; R') g(q'_1, q'_6; T') g(R',T';\hat R) \,
\bigotimes_{i=1}^6 q'_i \Bigr\}
\\
\hat R &= \bigoplus_{Q'', R'',T''}  \bigoplus_{ \{ q''_i \} }
\Bigl\{ g(q''_1,q''_2; Q'') g(Q'',q''_4,q''_6; R'') g(q''_3, q''_5; T'') g(R'',T'';\hat R) \,
\bigotimes_{i=1}^6 q''_i \Bigr\}.
\end{aligned}
\end{equation}
We fix the representations $(R,Q), (R',Q'), (R'',Q'')$ and the multiplicity labels $\nu,\nu',\nu''$ according to the external operators.
The space of multiplicities run over the spaces
\begin{equation}
\xi \in \cM_{R,Q,\nu} \,, \qquad
\xi' \in \cM_{R',Q',\nu'} \,, \qquad
\xi'' \in \cM_{R'',Q'',\nu''}
\label{ranges xis2}
\end{equation}
where
\begin{equation}
\begin{aligned}
\abs{ \cM_{R,Q,\nu} } &= g(q_5,q_6; Q) g(q_2, q_4; T) g(R,T;\hat R) 
\\
\abs{ \cM_{R',Q',\nu'} } &= g(q'_3,q'_4;Q') g(q'_1, q'_6; T') g(R',T';\hat R)
\\
\abs{ \cM_{R'',Q'',\nu''} } &= g(q''_1,q''_2; Q'') g(q''_3, q''_5; T'') g(R'',T'';\hat R) 
\end{aligned}
\end{equation}
They are subsets of the total multiplicity space
\begin{gather}
\abs{ \cM_{\rm tot} } \equiv g(q_1,q_2,q_3,q_4,q_5, q_6;\hat R) ,
\label{sum over multiplicities} \\[1mm]
\abs{ \cM_{\rm tot} }
= \sum_{R, Q} \sum_{\nu=1}^{g(Q,q_1,q_3; R)} \abs{ \cM_{R,Q,\nu} }
= \sum_{R', Q'} \sum_{\nu'=1}^{g(Q',q'_2,q'_5; R')} \abs{ \cM_{R',Q',\nu'} }
= \sum_{R'', Q''} \sum_{\nu'=1}^{g(Q'',q''_4,q''_6; R'')} \abs{ \cM_{R'',Q'',\nu''} } .
\notag
\end{gather}
Since the restricted Schur characters have two multiplicity labels \eqref{app:restricted character}, we introduce
\begin{equation}
\xi_{\pm} \in \cM_{R_\pm, Q_\pm, \nu_\pm} \,, \qquad
\xi'_{\pm} \in \cM_{R'_\pm, Q'_\pm, \nu'_\pm} \,, \qquad 
\xi''_{\pm} \in \cM_{R''_\pm, Q''_\pm, \nu''_\pm} 
\label{ranges xipms}
\end{equation}
where the $\pm$ signs are correlated.\footnote{Note that $(R_- , R'_-, R''_-) = (R_+, R'_+, R''_+)$ in the main text. We removed these constraints for convenience.}

Let us define the generalized Racah-Wigner tensor by
\begin{equation}
W_{\hat R} \begin{pmatrix}[cccc|cc]
q_1 & q_2 & \dots & q_6 & R_- & \xi_- \\
q'_1 & q'_2 & \dots & q'_6 & R'_+ & \xi'_+
\end{pmatrix}_{ab \dots f, a'b' \dots f'} 
\equiv
\Vev{ \matop{q_1 & q_2 & \dots & q_6}{a & b & \dots & f}{\xi_-} \, \Big| \,
\matop{q'_1 & q'_2 & \dots & q'_6}{a' & b' & \dots & f'}{\xi'_+} }
\label{def:genRW2}
\end{equation}
which is again proportional to $\prod_{i=1}^6 \delta^{q_i q'_i}$.
The RHS depends in $R_- \,, R'_+$ through the multiplicity space $\xi \in \cM_{R_-, Q_-, \nu_-} \,, \xi'_+ \in \cM_{R'_+, Q'_+, \nu'_+}$, as we discussed in \eqref{partial-entire overlap}.
We want to compute their products
\begin{align}
\tr (W_{\hat R} \, \tilde W_{\hat R}) &\equiv
\sum_{\xi_\mp , \xi'_\mp}
W_{\hat R} \begin{pmatrix}[cccc|cc]
q_1 & q_2 & \dots & q_6 & R_- & \xi_- \\
q'_1 & q'_2 & \dots & q'_6 & R'_+ & \xi'_+
\end{pmatrix}_{ab \dots f, a'b' \dots f'} \ \times
\label{def:trWW} \\[1mm]
&\hspace{70mm} 
W_{\hat R} \begin{pmatrix}[cccc|cc]
q'_1 & q'_2 & \dots & q'_6 & R'_- & \xi'_- \\
q_1 & q_2 & \dots & q_6 & R_+ & \xi_+
\end{pmatrix}_{a'b' \dots f',ab \dots f} 
\notag \\[2mm]
\tr (W_{\hat R} \, \tilde W_{\hat R} \, \dbtilde{W}_{\hat R} ) &\equiv
\sum_{\xi_\mp, \xi'_\mp, \xi''_\mp}
W_{\hat R} \begin{pmatrix}[cccc|cc]
q_1 & q_2 & \dots & q_6 & R_- & \xi_- \\
q'_1 & q'_2 & \dots & q'_6 & R'_+ & \xi'_+
\end{pmatrix}_{ab \dots f, a'b' \dots f'} \ \times
\label{def:trWWW} \\[1mm]
&\hspace{-15mm} 
W_{\hat R} \begin{pmatrix}[cccc|cc]
q'_1 & q'_2 & \dots & q'_6 & R'_- & \xi'_- \\
q''_1 & q''_2 & \dots & q''_6 & R''_+ & \xi''_+
\end{pmatrix}_{a'b' \dots f',a''b'' \dots f''} \,
W_{\hat R} \begin{pmatrix}[cccc|cc]
q''_1 & q''_2 & \dots & q''_6 & R''_- & \xi''_- \\
q_1 & q_2 & \dots & q_6 & R_+ & \xi_+
\end{pmatrix}_{a''b'' \dots f'', ab \dots f} \,.
\notag
\end{align}
They are identical to the product of projectors \eqref{cI123 comp},
\begin{align}
\tr (W_{\hat R} \, \tilde W_{\hat R}) &=
\trb{\, \hat R} \( 
\fP_{\hat I_1 \hat I_2}^{\hat R \to \dots \to (q_1 , q_2, \dots , q_6), \xi_{-}, \xi_{+} } \,
\fP_{\hat I_2 \hat I_1}^{\hat R \to \dots \to (q'_1 , q'_2, \dots , q'_6), \xi'_{-}, \xi'_{+} } 
\)
\\
\tr (W_{\hat R} \, \tilde W_{\hat R} \, \dbtilde{W}_{\hat R} )
&= \trb{\, \hat R} \( 
\fP_{\hat I_1 \hat I_2}^{\hat R \to \dots \to (q_1 , q_2, \dots , q_6), \xi_{-}, \xi_{+} } \,
\fP_{\hat I_2 \hat I_3}^{\hat R \to \dots \to (q'_1 , q'_2, \dots , q'_6), \xi'_{-}, \xi'_{+} } \,
\fP_{\hat I_3 \hat I_1}^{\hat R \to \dots \to (q''_1 , q''_2, \dots , q''_6), \xi''_{-}, \xi''_{+} }
\) .
\notag 
\end{align}
These products are depicted as
\begin{equation}
\tr (W_{\hat R} \, \tilde W_{\hat R} ) =
\begin{tikzpicture}[node distance=10mm, baseline=(current bounding box.center)]
\begin{scope}[decoration={markings, mark=at position 0.7 with {\arrow{Straight Barb[]}}}]
\node (br1m) [branch] {$\xi_-\phantom{{}'\!}$};
\node (br1p) [branch, right of=br1m, xshift=8mm] {$\xi_+\phantom{{}'\!}$};
\node (br2p) [branch2, below of=br1m, yshift=-12mm] {$\xi'_+$};
\node (br2m) [branch2, right of=br2p, xshift=8mm] {$\xi'_-$};
\draw [lineR] (br2p) -- (br1m);
\draw [lineR] (br1p) -- (br2m);
\draw [liner1] (br1m) .. controls (1,.5) .. (br1p);
\draw [lines2] (br1m) -- (br1p);
\draw [liner2] (br1m) .. controls (1,-.5) .. (br1p);
\draw [liner1] (br2m) .. controls (1,-1.7) .. (br2p);
\draw [liner2] (br2m) .. controls (1,-2.7) .. (br2p);
\draw [lines2] (br2m) -- (br2p);
\end{scope}
\end{tikzpicture} 
\hspace{10mm}
\tr (W_{\hat R} \, \tilde W_{\hat R} \, \dbtilde{W}_{\hat R} ) = 
\begin{tikzpicture}[node distance=10mm, baseline=(current bounding box.center)]
\begin{scope}[decoration={markings, mark=at position 0.7 with {\arrow{Straight Barb[]}}}]
\node (proj) {};
\node (u13) {};
\node (u32) {};
\node (u21) {};
\node (br1m) [branch, left of=u13, xshift=-8mm] {$\xi_-\phantom{{}'\!}$};
\node (br3p) [branch3, right of=u13, xshift=8mm] {$\xi''_+$};
\node (br3m) [branch3, below right of=u32, xshift=4mm, yshift=-8mm] {$\xi''_-$};
\node (br2p) [branch2, above left of=u32, xshift=-3mm, yshift=8mm] {$\xi'_+$};
\node (br2m) [branch2, above right of=u21, xshift=3mm, yshift=8mm] {$\xi'_-$};
\node (br1p) [branch, below left of=u21, xshift=-4mm, yshift=-8mm] {$\xi_+\phantom{{}'\!}$};
\draw [lineR] (br1m) -- (br2p);
\draw [lineR] (br2m) -- (br3p);
\draw [lineR] (br3m) -- (br1p);
\draw [liner1] (br1p) .. controls (-1, -.6) .. (br1m);
\draw [liner2] (br1p) .. controls (-1.9, -.9) .. (br1m);
\draw [liner1] (br3p) .. controls (1, -.6) .. (br3m);
\draw [liner2] (br3p) .. controls (1.9, -.9) .. (br3m);
\draw [liner1] (br2p) .. controls (0,1) .. (br2m);
\draw [liner2] (br2p) .. controls (0,2.1) .. (br2m);
\draw [lines2] (br1p) -- (br1m);
\draw [lines2] (br2p) -- (br2m);
\draw [lines2] (br3p) -- (br3m);
\end{scope}
\end{tikzpicture} 
\end{equation}
As a corollary of the identity of the projectors \eqref{restricted projector relation2}, we find that
\begin{equation}
\begin{aligned}
\tr (W_{\hat R} \, \tilde W_{\hat R} ) &=
\( \prod_{i=1}^6 \delta^{q_i q'_i} \, d_{q_i} \)
\delta^{\xi_- \, \xi'_+} 
\delta^{\xi'_- \, \xi_+} \,
\\[1mm]
\tr (W_{\hat R} \, \tilde W_{\hat R}  \, \dbtilde{W}_{\hat R} ) &=
\( \prod_{i=1}^6 \delta^{q_i q'_i} \, \delta^{q_i q''_i} \, d_{q_i} \)
\delta^{\xi_- \, \xi'_+ } \,
\delta^{\xi'_- \, \xi''_+ } \,
\delta^{\xi''_- \, \xi_+ } \,.
\end{aligned}
\label{identities trWWW}
\end{equation}
By summing $\{ \xi_\mp, \xi'_\mp, \xi''_\mp \}$ over the ranges $\{ \cM_{R_\mp,Q_\mp,\nu_\mp} , \cM_{R'_\mp,Q'_\mp,\nu'_\mp}, \cM_{R''_\mp,Q''_\mp,\nu''_\mp} \}$, we discover the overlap
\begin{equation}
\sum_{\xi_- \in \cM_{R_-,Q_-,\nu_-} } \ \sum_{\xi'_+ \in \cM_{R'_+,Q'_+,\nu'_+}} 
\delta^{\xi_- \, \xi'_+ } 
= \abs{ \cM_{R_-,Q_-,\nu_-} \cap \cM_{R'_+,Q'_+,\nu'_+} } \ .
\end{equation}
The overlap satisfies the sum rules
\begin{equation}
\begin{gathered}
\sum_{R_-,Q_-,\nu_-} \ \sum_{R'_+,Q'_+,\nu'_+}
\abs{ \cM_{R_-,Q_-,\nu_-} \cap \cM_{R'_+,Q'_+,\nu'_+} } 
= \abs{\cM_{\rm tot}} 
\\
\sum_{R_-,Q_-,\nu_-} 
\abs{ \cM_{R_-,Q_-,\nu_-} \cap \cM_{R'_+,Q'_+,\nu'_+} } 
= \abs{\cM_{R'_+,Q'_+,\nu'_+}} 
\\
\sum_{R'_+,Q'_+,\nu'_+} 
\abs{ \cM_{R_-,Q_-,\nu_-} \cap \cM_{R'_+,Q'_+,\nu'_+} } 
= \abs{\cM_{R_-,Q_-,\nu_-}} \ .
\end{gathered}
\end{equation}
As a solution to the sum rules, we conjecture that
\begin{equation}
\abs{ \cM_{R_-,Q_-,\nu_-} \cap \cM_{R'_+,Q'_+,\nu'_+} } 
= \oldelta^{\, \nu_- \, \nu'_+ } \,
\frac{\abs{\cM_{R_-,Q-,\nu_-}} \abs{\cM_{R'_+,Q'_+,\nu'_+}} }{\abs{\cM_{\rm tot}}} 
\end{equation}
where $\oldelta^{\nu \nu'}$ should be understood as the intersection inside $\cM_{\rm tot}$
\begin{equation}
\oldelta^{\, \nu_+ \, \nu'_- } = 
\begin{cases}
1 &\qquad 
\( \cM_{R_-,Q_-,\nu_-} \cap \cM_{R'_+,Q'_+,\nu'_+} \neq \emptyset \)
\\[1mm]
0 &\qquad \( {\rm otherwise} \).
\end{cases}
\label{def:oldelta}
\end{equation}
It follows that
\begin{align}
\sum_{\xi_\mp, \xi'_\mp} \tr (W_{\hat R} \, \tilde W_{\hat R} ) &=
\( \prod_{i=1}^6 \delta^{q_i q'_i} \, d_{q_i} \)
\oldelta^{\, \nu_- \, \nu'_+ } \, \oldelta^{\, \nu'_- \, \nu_+ } \ \times
\label{trWW conjecture} \\
&\hspace{-10mm} 
\frac{\abs{\cM_{R_-,Q-,\nu_-}} \abs{\cM_{R_+,Q+,\nu_+}} \abs{\cM_{R'_-,Q'_-,\nu'_-}}\abs{\cM_{R'_+,Q'_+,\nu'_+}} }{\abs{\cM_{\rm tot}}^2 } 
\notag \\[1mm]
\sum_{\xi_\mp, \xi'_\mp, \xi''_\mp} \tr (W_{\hat R} \, \tilde W_{\hat R}  \, \dbtilde{W}_{\hat R} ) &=
\( \prod_{i=1}^6 \delta^{q_i q'_i} \, \delta^{q_i q''_i} \, d_{q_i} \)
\oldelta^{\, \nu_- \, \nu'_+ } \,
\oldelta^{\, \nu'_- \, \nu''_+ } \,
\oldelta^{\, \nu''_- \, \nu_+ } \ \times
\label{trWWW conjecture}  \\
&\hspace{-10mm} 
\frac{\abs{\cM_{R_-,Q-,\nu_-}} \abs{\cM_{R_+,Q+,\nu_+}} \abs{\cM_{R'_-,Q'_-,\nu'_-}}\abs{\cM_{R'_+,Q'_+,\nu'_+}} \abs{\cM_{R''_-,Q''_-,\nu''_-}}\abs{\cM_{R''_+,Q''_+,\nu''_+}} }{\abs{\cM_{\rm tot}}^3 } .
\notag
\end{align}

\subsection{Restricted Littlewood-Richardson coefficients}\label{app:resLR}

Let us compute the restricted Littlewood-Richardson coefficients in \cite{Bhattacharyya:2008xy} in our method.
We will find the perfect agreement. However, they considered multiplicity-free cases only.
Thus, this agreement does not provide non-trivial checks of our conjectured formula.

\bigskip
We define the restricted Littlewood-Richardson coefficients by
\begin{equation}
\begin{gathered}
F^{\{ 3 \}}_{\{1 \} \{2 \}} =
\frac{1}{L_1! \, L_2!} \, \sum_{\sigma_1 \in S_{L_1}} \sum_{\sigma_2 \in S_{L_2}}
\chi^{\bsR_1} (\sigma_1) \, \chi^{\bsR_2} (\sigma_2) \, \chi^{\bsR_3} (\sigma_1 \circ \sigma_2)
\\[1mm]
L_i =m_i + n_i \,, \quad
\bsR_i = \pare{ R_i, (r_i, s_i), (\nu_{i-} \,, \nu_{i+} ) }.
\end{gathered}
\label{def:resLR ch}
\end{equation}
The definition used in \cite{Bhattacharyya:2008xy} is
\begin{equation}
f^{\{ 3 \}}_{\{1 \} \{2 \}} =
\frac{1}{m_1! n_1! m_2 ! n_2!} \, \frac{m_3! n_3!}{L_3!} \, \frac{d_{R_3}}{d_{r_3} \, d_{s_3}} 
\sum_{\sigma_1 \in S_{L_1}} \sum_{\sigma_2 \in S_{L_2}}
\chi^{\bsR_1} (\sigma_1) \, \chi^{\bsR_2} (\sigma_2) \, \chi^{\bsR_3} (\sigma_1 \circ \sigma_2) .
\end{equation}
The two definitions are related by
\begin{equation}
F^{\{ 3 \}}_{\{1 \} \{2 \}} =\frac{m_1! n_1! m_2 ! n_2!}{m_3! n_3!} \,
\frac{L_3!}{L_1! L_2!} \, 
\frac{d_{r_3} \, d_{s_3}}{d_{R_3}} \,
f^{\{ 3 \}}_{\{1 \} \{2 \}} \,.
\end{equation}

The restricted Littlewood-Richardson coefficients $F^{\{ 3 \}}_{\{1 \} \{2 \}}$ can be computed as follows.
First, consider the restriction $S_{L_3} \downarrow (S_{L_1} \otimes S_{L_2})$, which gives
\begin{equation}
R_3 = \bigoplus_{T_1, T_2} g (T_1, T_2 ; R_3) \( T_1 \otimes T_2 \) .
\end{equation}
The restricted character in \eqref{def:resLR ch} becomes
\begin{multline}
\chi^{\bsR_3} (\sigma_1 \circ \sigma_2)
=
\sum_{T_1, T_2} \sum_{\mu=1}^{g (T_1, T_2 ; R_3)}
D^{T_1}_{h_1 h'_1} (\sigma_1) \, D^{T_2}_{h_2 h'_2} (\sigma_2) \, 
\tilde B^{R_3 \to (T_1, T_2) \mu}_{I \to (h_1 h_2)} \,
(\tilde B^T)^{R_3 \to (T_1, T_2) \mu}_{I' \to (h'_1 h'_2)} \ \times
\\
B^{R_3 \to (r_3, s_3), \nu_{3-}}_{I \to (i,j)} \,
\brT^{R_3 \to (r_3, s_3), \nu_{3+}}_{I' \to (i,j)} \,.
\end{multline}
In the quiver notation, we can depict this equation as
\begin{equation}
\chi^{R_3 (r_3 , s_3), (\nu_{3-}, \nu_{3+})} (\sigma_1 \circ \sigma_2) = \ 
\begin{tikzpicture}[node distance=10mm, baseline=(current bounding box.center)]
\begin{scope}[decoration={markings, mark=at position 0.7 with {\arrow{Straight Barb[]}}}]
\node (sig2) [perm] {$\sigma_1 \circ \sigma_2$};
\node (br1) [branch, above of=sig2, yshift=4mm] {$\nu_{3-}$};
\node (br2) [branch, below of=sig2, yshift=-4mm] {$\nu_{3+}$}; 
\draw [lineR] (br1) -- (sig2);
\draw [lineR] (sig2) -- (br2);
\draw [liner1] (br2.west) .. controls (-1.5,-0.5) and (-1.5,0.5) .. (br1.west);
\draw [liner2] (br2.east) .. controls (1.5,-0.5) and (1.5,0.5) .. (br1.east);
\end{scope}
\end{tikzpicture}
\ = \sum_{T_1, T_2, \mu} \ 
\begin{tikzpicture}[node distance=10mm, baseline=(current bounding box.center)]
\begin{scope}[decoration={markings, mark=at position 0.7 with {\arrow{Straight Barb[]}}}]
\node (sig1) [perm] {$\sigma_1$};
\node (sig2) [perm, right of=sig1, xshift=5mm] {$\sigma_2$};
\node (br3) [branch2, above right of=sig1, yshift=3mm] {$\mu$};
\node (br4) [branch2, below right of=sig1, yshift=-3mm] {$\mu$};
\node (br1) [branch, above of=br3, yshift=4mm] {$\nu_{3-}$};
\node (br2) [branch, below of=br4, yshift=-4mm] {$\nu_{3+}$}; 
\draw [lineR] (br1) -- (br3);
\draw [lineR] (br4) -- (br2);
\draw [liner1] (br2.west) .. controls (-1.5,-0.5) and (-1.5,0.5) .. (br1.west);
\draw [liner2] (br2.east) .. controls (3,-0.5) and (3,0.5) .. (br1.east);
\draw [lines1] (br3.west) -- (sig1.north);
\draw [lines1] (sig1.south) -- (br4.west);
\draw [lines2] (br3.east) -- (sig2.north);
\draw [lines2] (sig2.south) -- (br4.east);
\end{scope}
\end{tikzpicture}
\label{restricted ch decomp}
\end{equation}
By summing over $\sigma_1$ and $\sigma_2$ in \eqref{def:resLR ch}, we get $\delta^{T_1 ,R_1} \, \delta^{T_2 ,R_2}$ and another sets of branching coefficients in place of $\sigma_1 \,, \sigma_2$ in \eqref{restricted ch decomp}, giving us
\begin{equation}
\begin{tikzpicture}[node distance=10mm, baseline=(current bounding box.center)]
\begin{scope}[decoration={markings, mark=at position 0.8 with {\arrow{Straight Barb[]}}}]
\node (br1) [branch] {$\nu_{3-}$};
\node (br2) [branch2, right of=br1, xshift=6mm] {$\mu$};
\node (br3) [branch, above right of=br2, xshift=8mm] {$\nu_{1-}$};
\node (br4) [branch, below right of=br2, xshift=8mm] {$\nu_{2-}$};
\node (br5) [branch, right of=br3, xshift=8mm] {$\nu_{1+}$};
\node (br6) [branch, right of=br4, xshift=8mm] {$\nu_{2+}$};
\node (br7) [branch2, below right of=br5, xshift=8mm] {$\mu$};
\node (br8) [branch, right of=br7, xshift=6mm] {$\nu_{3+}$};
\draw [liner1] (br8) .. controls (9,-2.5) and (0,-2.5) .. (br1);
\draw [liner2] (br8) .. controls (9,2.5) and (0,2.5) .. (br1);
\draw [lineR] (br1) -- (br2);
\draw [lines1] (br2) -- (br3);
\draw [lines2] (br2) -- (br4);
\draw [liner3] (br3.north east) -- (br5.north west);
\draw [liner1] (br3.south east) -- (br5.south west);
\draw [lines3] (br4.north east) -- (br6.north west);
\draw [liner2] (br4.south east) -- (br6.south west);
\draw [lines1] (br5) -- (br7);
\draw [lines2] (br6) -- (br7);
\draw [lineR] (br7) -- (br8);
\end{scope} 
\end{tikzpicture}
= \tr ( \proj \, \tilde{\proj} ) .
\end{equation}
The restricted Littlewood-Richardson coefficient \eqref{def:resLR ch} becomes
\begin{equation}
F^{\{ 3 \}}_{\{1 \} \{2 \}} = \frac{1}{d_{R_1} d_{R_2}} \, \sum_\mu \,
\tr \Bigl( 
\proj^{R_3 \to (r_3 , s_3), (\nu_{3-} \nu_{3+})}
\tilde{\proj}^{R_3 \to (R_1 , R_2), \mu \to (r_1 , s_1 , r_2 , s_2), (\mu, (\nu_{1+} , \nu_{2+}), (\nu_{1-} , \nu_{2-}) )}
\Bigr).
\end{equation}

To evaluate the projectors, we introduce the permutations on the fully-split space
\begin{equation}
S_{\rm FS} = S_{m_1} \otimes S_{m_2} \otimes S_{n_1} \otimes S_{n_2} 
\end{equation}
and consider sub-projectors.
The total multiplicity space for the restriction $S_{L_3} \downarrow S_{\rm FS}$ is
\begin{equation}
\abs{ \cM_{\rm tot} } = g(r_1 , r_2 , s_1 , s_2 ; R_3). 
\end{equation}
The multiplicity space for the first projector $\proj^{R_3 \to (r_3 , s_3), (\nu_{3-} \nu_{3+})}$ is
\begin{equation}
\begin{gathered}
\abs{ \cM_{r_3, s_3, \nu_{3\mp}} } = g(r_1 , r_2 ; r_3) g(s_1 , s_2 ; s_3),
\\[1mm]
\sum_{r_3, s_3} \sum_{\nu_{3-} = 1}^{g(r_3, s_3; R_3)} \abs{ \cM_{r_3, s_3, \nu_{3-}} } = 
\sum_{r_3, s_3} \sum_{\nu_{3+} = 1}^{g(r_3, s_3; R_3)} \abs{ \cM_{r_3, s_3, \nu_{3+}} } = 
\abs{ \cM_{\rm tot} }.
\end{gathered}
\end{equation}
The multiplicity space for the second projector $\tilde{\proj}^{R_3 \to \dots \to (r_1 , s_1 , r_2 , s_2), (\mu, \nu_{1\mp} , \nu_{2\mp})}$ is
\begin{equation}
\begin{gathered}
\abs{ \cM_{R_1, R_2, \nu_{1\mp} , \nu_{2\mp}} } = g(R_1 , R_2 ; R_3)
\\[1mm]
\sum_{R_1, R_2} \sum_{\nu_{1-} = 1}^{g(r_1, s_1; R_1)} \sum_{\nu_{2-} = 1}^{g(r_2, s_2; R_2)}
\abs{ \cM_{R_1, R_2, \nu_{1-} , \nu_{2-}} } = 
\sum_{R_1, R_2} \sum_{\nu_{1+} = 1}^{g(r_1, s_1; R_1)} \sum_{\nu_{2+} = 1}^{g(r_2, s_2; R_2)}
\abs{ \cM_{R_1, R_2, \nu_{1+} , \nu_{2+}} } = 
\abs{ \cM_{\rm tot} }.
\end{gathered}
\end{equation}
From the identity of the projector \eqref{restricted projector relation2}, we obtain
\begin{equation}
\tr ( \proj \, \tilde{\proj} ) =
\oldelta^{\, \nu_{3+} \, (\nu_{1+}, \nu_{2+}) } \,
\oldelta^{\, (\nu_{1-}, \nu_{2-}) \, \nu_{3-} } \,
d_{r_1} d_{r_2} d_{s_1} d_{s_2} \, \cG_{\rm LR}
\end{equation}
where we grouped $(\nu_{1\mp}, \nu_{2\mp})$ so that they can be compared with $\nu_{3\mp}$.
Just like before, we conjecture that
\begin{equation}
\begin{aligned}
\cG_{\rm LR} &= \frac{\abs{ \cM_{r_3, s_3, \nu_{3-}} } \, \abs{ \cM_{r_3, s_3, \nu_{3+}} }
\abs{ \cM_{R_1, R_2, \nu_{1-} , \nu_{2-}} } \, \abs{ \cM_{R_1, R_2, \nu_{1+} , \nu_{2+}} } }{\abs{ \cM_{\rm tot} }^2 }
\\[1mm]
&= \( \frac{g(R_1 , R_2 ; R_3) g(r_1 , r_2 ; r_3) g(s_1 , s_2 ; s_3)}{g(r_1 , r_2 , s_1 , s_2 ; R_3)} \)^2 .
\end{aligned}
\end{equation}
In summary, we get
\begin{equation}
F^{\{ 3 \}}_{\{1 \} \{2 \}} = \oldelta^{\, \nu_{3+} \, (\nu_{1+}, \nu_{2+}) } \,
\oldelta^{\, (\nu_{1-}, \nu_{2-}) \, \nu_{3-} } \,
\frac{ d_{r_1} d_{r_2} d_{s_1} d_{s_2} }{ d_{R_1} d_{R_2} } \
\( \frac{g(R_1 , R_2 ; R_3) \, g(r_1 , r_2; r_3) \, g(s_1 , s_2; s_3)}{g(r_1, r_2, s_1, s_2 ; R_3)} \)^2.
\label{formula:resLR}
\end{equation}

\bigskip
Three cases have been considered in \cite{Bhattacharyya:2008xy}.
The first case is the antisymmetric representations,
\begin{align}
(R_i, r_i, s_i) = \( [1^{m_i +n _i}], [1^{m_i}] , [1^{n_i}] \) 
\end{align}
and the second case is the symmetric representations,
\begin{align}
(R_i, r_i, s_i) = \( [m_i +n _i], [m_i] , [n_i] \) .
\end{align}
In both cases, all representations are one-dimensional and multiplicity-free.
Therefore $F^{\{ 3 \}}_{\{1 \} \{2 \}} = 1$, which means
\begin{equation}
f^{\{ 3 \}}_{\{1 \} \{2 \}} = \frac{m_3! \, n_3! \, L_1! \, L_2!}{m_1! \, n_1! \, m_2 ! \, n_2! \, L_3!} \,.
\end{equation}
The last case is $r_2 = s_1 = \emptyset$, implying that
\begin{equation}
R_1 = r_1 = r_3 \,, \quad
R_2 = s_2 = s_3 \,, \quad
F^{\{ 3 \}}_{\{1 \} \{2 \}} = 1
\end{equation}
and hence
\begin{equation}
f^{\{ 3 \}}_{\{1 \} \{2 \}} = \delta^{R_1 , r_3} \, \delta^{R_2 , s_3} \, 
\frac{L_1! L_2!}{L_3!} \, 
\frac{d_{R_3}}{d_{r_3} \, d_{s_3}} \,.
\end{equation}
All the results agree with \cite{Bhattacharyya:2008xy}.

\bibliographystyle{utphys}
\bibliography{bibmix-ex}{}

\end{document}